\documentclass[lettersize,journal]{IEEEtran}
\usepackage{amsmath,amsfonts}
\usepackage{algorithmic}
\usepackage{algorithm}
\usepackage{mathtools}
\usepackage{hyperref}
\usepackage{amsmath}
\usepackage{bm}
\usepackage{array}
\usepackage[caption=false,font=normalsize,labelfont=sf,textfont=sf]{subfig}
\usepackage{textcomp}
\usepackage{stfloats}
\usepackage{url}
\usepackage{verbatim}
\usepackage{graphicx}
\usepackage{color}
\usepackage{natbib}
\usepackage{amssymb}
\usepackage{subfloat}
\usepackage{subcaption}
\usepackage{multirow}
\usepackage{amsthm}
\usepackage{booktabs}
\usepackage{gensymb}
\usepackage{caption}
\usepackage[table]{xcolor} 

\usepackage{multirow}
\usepackage{makecell}
\usepackage{color, colortbl}
\definecolor{LightGray}{gray}{0.85}
\definecolor{White}{gray}{1.0}
\definecolor{Celadon}{RGB}{175, 225, 175}
\definecolor{lightpink}{RGB}{220, 182, 120}
\definecolor{skyblue}{HTML}{D9EAD3}
\definecolor{minired}{HTML}{F4CCCC}
\newtheorem{definition}{Definition}[section]  

\def\red#1{\textcolor{red}{#1}}

\newcommand{\etal}{\textit{et~al.}}

\newcommand{\ie}{\textit{i.e.}}
\newcommand{\eg}{\textit{e.g.}}

\newcommand{\triangledComment}[1]{\hfill $\triangleleft$ \textcolor{black}{#1}}

\def\vx{{\bm{x}}}
\def\vt{\bm{t}}
\def\vm{{\bm{m}}}
\def\va{{\bm{a}}}
\def\vmu{{\bm{\mu}}}
\def\vsigma{{\bm{\sigma}}}
\def\vr{\bm{r}}
\def\gD{\mathcal{D}}

\def\gM{\mathcal{M}}
\begin{document}

\title{FLARE: Toward Universal Dataset Purification against Backdoor Attacks}


\author{IEEE Publication Technology,~\IEEEmembership{Staff,~IEEE,}
\thanks{This paper was produced by the IEEE Publication Technology Group. They are in Piscataway, NJ.}
\thanks{Manuscript received April 19, 2021; revised August 16, 2021.}}

\markboth{IEEE Transactions on Information Forensics and Security}%
{IEEE Transactions on Information Forensics and Security}


\author{
Linshan Hou, Wei Luo, Zhongyun Hua, Songhua Chen, Leo Yu Zhang, and Yiming Li

\thanks{This work was supported by by Guangdong Basic and Applied Basic Research Foundation under Grant 2024A1515012299.}
\thanks{Linshan Hou is with the School of Computer Science and Technology, Harbin Institute of Technology, Shenzhen, Guangdong 518055, China (email: lizzieandland@gmail.com).}
\thanks{Wei Luo is with the School of Information Technology, Deakin University, Australia. (email: wei.luo@deakin.edu.au).}
\thanks{Zhongyun Hua is with the School of Computer Science and Technology, Harbin Institute of Technology, Shenzhen, Guangdong 518055, China. (e-mail: huazhongyun@hit.edu.cn).}
\thanks{Songhua Chen. Independent Researcher. (email: frederichen01@gmail.com)}
\thanks{Leo Yu Zhang is with the School of Information and Communication Technology, Griffith University, Southport, Gold Coast, QLD 4215, Australia (email: leo.zhang@griffith.edu.au).}
\thanks{
Yiming Li is with College of Computing and Data Science, Nanyang Technological University, Singapore 639798. (email: liyiming.tech@gmail.com)
}
\thanks{Corresponding Author(s): Zhongyun Hua and Yiming Li.}
}



\maketitle

\begin{abstract}
Deep neural networks (DNNs) are susceptible to backdoor attacks, where adversaries poison datasets with adversary-specified triggers to implant hidden backdoors, enabling malicious manipulation of model predictions. Dataset purification serves as a proactive defense by removing malicious training samples to prevent backdoor injection at its source. We first reveal that the current advanced purification methods rely on a latent assumption that the backdoor connections between triggers and target labels in backdoor attacks are simpler to learn than the benign features. We demonstrate that this assumption, however, does not always hold, especially in all-to-all (A2A) and untargeted (UT) attacks. As a result, purification methods that analyze the separation between the poisoned and benign samples in the input-output space or the final hidden layer space are less effective. We observe that this separability is not confined to a single layer but varies across different hidden layers. Motivated by this understanding, we propose FLARE, a universal purification method to counter various backdoor attacks. FLARE aggregates abnormal activations from all hidden layers to construct representations for clustering. To enhance separation, FLARE develops an adaptive subspace selection algorithm to isolate the optimal space for dividing an entire dataset into two clusters. FLARE assesses the stability of each cluster and identifies the cluster with higher stability as poisoned. Extensive evaluations on benchmark datasets demonstrate the effectiveness of FLARE against 22 representative backdoor attacks, including all-to-one (A2O), all-to-all (A2A), and untargeted (UT) attacks, and its robustness to adaptive attacks. Codes are available at \href{https://github.com/THUYimingLi/BackdoorBox}{BackdoorBox} and \href{https://github.com/vtu81/backdoor-toolbox}{backdoor-toolbox}.

\end{abstract}

\section{Introduction}

Deep neural networks (DNNs) are widely deployed in mission-critical applications, including autonomous driving~\cite{kong2020physgan,grigorescu2020survey}, and face recognition~\cite{li2016mutual,li2014common}. Currently, due to the complicated structure and large parameter scale of modern DNNs, training these models usually relies on large-scale datasets, typically from external sources (\eg, data markets). 

However, recent studies reveal that using such external datasets introduces security risks~\cite{li2023detecting,zhang2024badrobot,10795257,zhang2025improving,song2025pb}. In particular, backdoor attacks have been identified as a significant threat~\cite{gu2017badnets,li2022backdoor,wei2024pointncbw,gao2024backdoor,luo2025just}. Specifically, the backdoor adversaries poison a small subset of training data by embedding a predefined trigger (\eg, a subtle white square) and re-assigning their labels as the adversary-specified target label(s). As a result, all DNNs trained on these poisoned samples will learn a hidden backdoor, \ie, the latent and malicious connection between the triggers and the target labels. During the inference process of the backdoored model, the adversaries can activate its backdoor by implanting trigger patterns to maliciously change the predictions of any samples to the target labels. This attack is highly stealthy since the attacked model behaves normally on benign samples so that users cannot easily detect it simply based on the results of their local validation samples.

\begin{figure}[!t]
	\centering
    \includegraphics[width=0.9\linewidth]{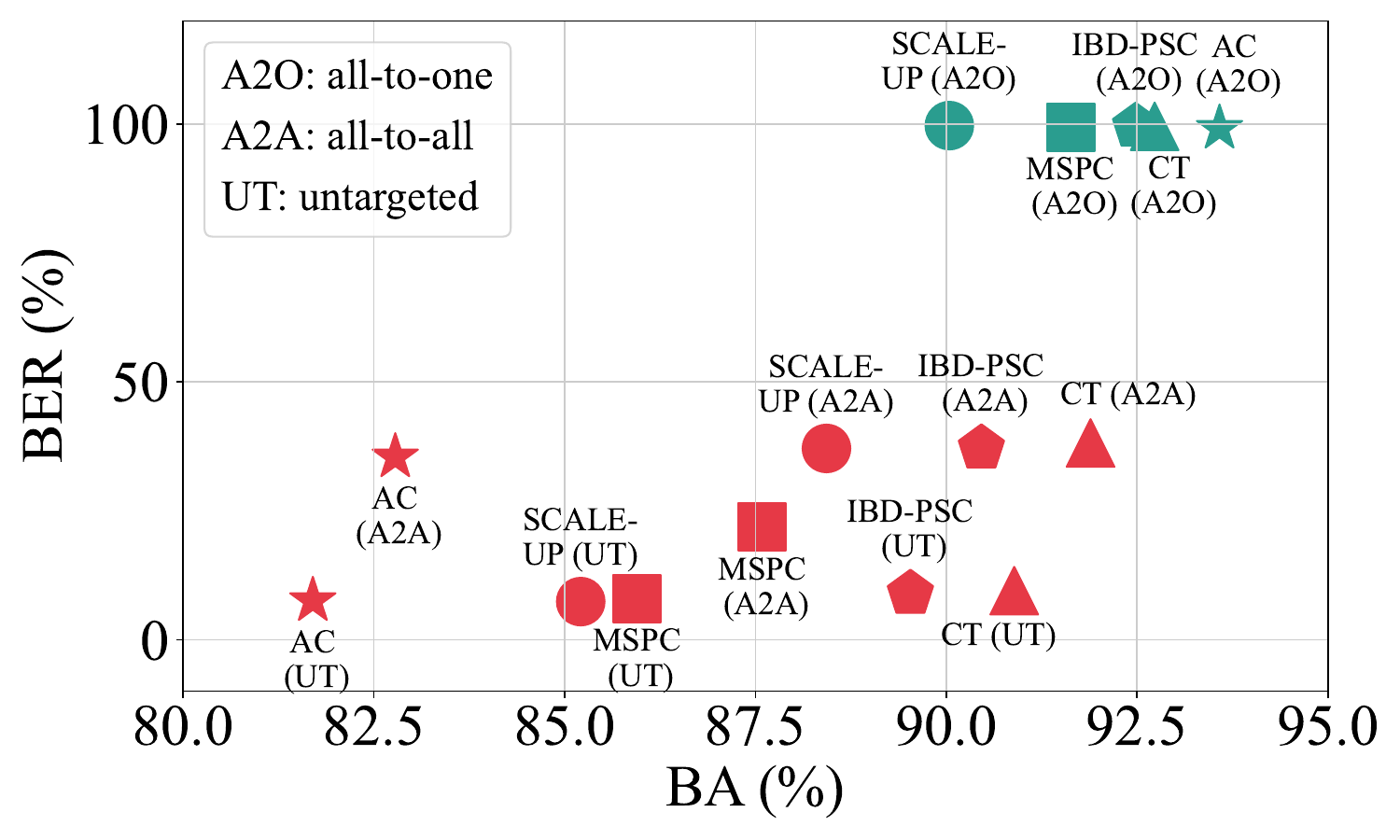}
\vspace{-0.8em}
 \caption{The benign accuracy (BA) and backdoor elimination rate (BER) of models trained on purified dataset. In these cases, BER is calculated as $100\% -$ ASR, where ASR (attack success rate) measures the ratio of poisoned samples misclassified to the target labels by the backdoored model.} 
 \label{fig:intro}
  \vspace{-1.7em}
 \end{figure}
 
Currently, researchers have investigated five representative defensive strategies against backdoor attacks. These strategies include: \textbf{(1)} dataset purification~\cite{qi2023towards,yao2024reverse,hou2024ibd}, \textbf{(2)} poison suppression~\cite{huang2022backdoor,gao2023backdoor,tang2023setting}, \textbf{(3)} model-level backdoor detection~\cite{wang2019neural,xiang2023umd,wang2024mm}, \textbf{(4)} input-level backdoor detection~\cite{guo2023scale,pal2024backdoor,hou2024ibd}, and \textbf{(5)} backdoor mitigation~\cite{liu2018fine,zeng2022adversarial,xu2024towards}. Among these strategies, poison suppression modifies the training process to reduce the impact of poisoned samples; model-level backdoor detection and mitigation identify or remove embedded backdoors in post-development; and input-level detection prevents backdoor activation by identifying malicious inputs during inference. In contrast, dataset purification acts as a proactive defense, identifying and removing poisoned samples from the dataset before training begins. This paper focuses on dataset purification, aiming to prevent backdoor creation at its source and precisely trace malicious samples to their origins.

In this paper, we first revisit existing advanced dataset purification methods. We reveal that their effectiveness relies on an implicit assumption: the connections between triggers and target labels in backdoor attacks are inherently simpler to learn than the benign features. This assumption enables these methods to achieve promising results by focusing primarily on input-output relationships. Specifically, \textbf{(1)} one line of work~\cite{li2021anti,zhang2023backdoor} suggests that DNNs converge significantly faster on poisoned samples, indicating that the backdoor connections are easier to learn; \textbf{(2)} studies such as~\cite{chou2020sentinet,huang2023distilling} observe that poisoned samples often exhibit highly localized and small saliency regions. This suggests that backdoor triggers function as shortcut features, allowing the model to bypass learning complex, distributed features in benign data and thus simplifying the backdoor establishment; \textbf{(3)} research from~\cite{guo2023scale,hou2024ibd} leverages the scaled prediction consistency of poisoned samples, implying that DNNs overfit on distinctive, artificially crafted triggers, thereby simplifying backdoor formation. This assumption generally holds for typical all-to-one (A2O) backdoor attacks, as backdoor triggers are often simple and easy to learn. However, this assumption does not always hold for modern complicated backdoor attacks, especially for all-to-all (A2A) and untargeted (UT) backdoor attacks (as shown in Figure~\ref{fig:intro}). As such, an intriguing question arises:

\emph{Shall we design a universal dataset purification method that is effective against various types of backdoor attacks?}

The answer is in the positive! Inspired by the finding that DNN memorization is distributed across neurons in multiple hidden layers~\cite{maini2023can}, we explore the distinctions between poisoned and benign samples throughout the model instead of simply the input-output relationships. Our analysis of A2A and UT backdoor attacks reveals a critical observation: poisoned and benign samples do not consistently separate within specific layers; instead, distinctions emerge across different hidden layers, varying with the attack types. Building on this insight, we propose a universal method for dataset purification, termed \textbf{F}ull-spectrum \textbf{L}earning \textbf{A}nalysis for \textbf{R}emoving \textbf{E}mbedded poisoned samples (FLARE). FLARE comprises two main stages: latent representation extraction and poisoned sample detection. In the first stage, FLARE constructs a comprehensive latent representation for each training sample by consolidating the abnormal values from all hidden layers’ feature maps. Specifically, FLARE first aligns all feature maps to a uniform scale (\eg, [0,1]) by leveraging the statistics of Batch Normalization (BN) layers. FLARE then extracts an abnormally large or small value from each feature map and consolidates these values across all hidden layers to construct the latent representation. In the second stage, FLARE detects poisoned samples through cluster analysis. In general, FLARE splits the entire dataset into two distinct clusters and identifies the cluster with higher \textit{cluster stability} as poisoned. Specifically, FLARE first applies dimensionality reduction to reduce the computation consumption and improve clustering efficiency. FLARE then selects a stable subspace by adaptively excluding category-specific features from the last few hidden layers, isolating an optimal subspace where benign samples from various classes are still close together. FLARE evaluates \textit{cluster stability} as the `density' difference between the density level at which the cluster first appears and that where the cluster divides into smaller sub-clusters. FLARE finally identifies the cluster exhibiting higher stability as poisoned since poisoned samples tend to form compact clusters due to the sharing of the same trigger-related features. 


In conclusion, our main contributions are three-fold. \textbf{(1)} We reveal that the underlying assumption of existing advanced dataset purification methods, \ie, the backdoor connections are easier to learn than benign ones, does not always hold, particularly under all-to-all and untargeted attacks. We also demonstrate that poisoned and benign samples do not consistently separate within particular layers across various types of attacks. 
\textbf{(2)} Based on our intriguing findings, we develop a universal dataset purification method (dubbed FLARE). It separates poisoned and benign samples throughout the model instead of simply the input-output relationship. \textbf{(3)} We conduct extensive experiments on benchmark datasets, verifying the effectiveness of our FLARE against 22 representative backdoor attacks (including A2O, A2A, and UT ones) and its resistance to potential adaptive attacks.


\section{Related Work}

DNNs are inherently vulnerable to various security threats due to their high model complexity, large parameter space, and tendency to overfit. These vulnerabilities make DNNs susceptible to a range of attacks, including adversarial attacks~\cite{gao2025pixels,zhang2024does}, membership inference~\cite{he2025labelonly,li2024yes}, and backdoor attacks~\cite{gu2017badnets,gao2020backdoor}. Among them, backdoor attacks represent an emerging but highly threatening security vulnerability~\cite{gu2017badnets,li2022backdoor}. In such attacks, an adversary implants a malicious connection, known as a backdoor, between a predefined trigger and an adversary-specified target label. This backdoor remains dormant for benign inputs and only activates when the trigger appears, making backdoor attacks stealthy.


\subsection{Poison-only Backdoor Attacks}

\vspace{0.3em} 
\noindent\textbf{Targeted Attacks.} Gu~\etal~\cite{gu2017badnets} proposed the first poison-only backdoor attack, BadNets, where an adversary embeds an adversary-specified trigger into a few training samples and modifies their labels. This attack supports two primary modes: \textbf{(1)} all-to-one (A2O), where all poisoned samples are assigned a single target label, and \textbf{(2)} all-to-all (A2A), where poisoned samples from a class $i$ are relabeled as a different class label, typically the next consecutive class (\ie, $i+1$). Models trained on the poisoned dataset establish a backdoor connection between the trigger and target labels. Subsequent research aimed to make backdoor attacks more stealthy by developing invisible triggers~\cite{nguyen2021wanet} and employing label-consistent poisoning~\cite{turner2019label}, where only samples from the target class are poisoned, thus avoiding label modification and bypassing manual inspection. Follow-up research further introduced more sophisticated triggers, including sample-specific triggers~\cite{li2021invisible}, sparse triggers~\cite{gao2024backdoor}, horizontal triggers~\cite{ma2023horizontal}, and asymmetric triggers~\cite{qi2023revisiting}, crafted to evade defenses.


\vspace{0.3em} 
\noindent\textbf{Untargeted Attacks.} Targeted attacks assign poisoned samples to adversary-specific labels, ensuring that the backdoored model consistently misclassifies the poisoned samples into particular classes. Recently, some pioneering work has discussed untargeted attacks~\cite{li2022untargeted,xue2023untargeted}, where the goal is to make the predictions deviate from the true labels instead of approaching particular ones (\ie, target labels). For example, poisoned samples are reassigned with random incorrect labels \cite{li2022untargeted}, causing the backdoored model to misclassify the poisoned samples into incorrect classes. As a result, the predictions for all poisoned samples approximate a uniform distribution, making the attack more difficult to learn and detect.




\subsection{Backdoor Defenses}
Based on the stage at which they occur, existing defenses can be divided into five main categories: \textbf{(1)} dataset purification~\cite{qi2023towards,yao2024reverse,gao2025try}, which focuses on detecting and removing poisoned samples from a given suspicious dataset before model training, \textbf{(2)} poison suppression~\cite{huang2022backdoor,gao2023backdoor,tang2023setting}, which modifies the training process to limit the impact of poisoned samples, \textbf{(3)} model-level backdoor detection~\cite{wang2019neural,xiang2023umd,wang2024mm}, which assesses whether a suspicious model contains hidden backdoors; \textbf{(4)} input-level backdoor detection~\cite{gao2019strip,li2023ntd,luo2025just}, which identifies malicious inputs at inference; \textbf{(5)} backdoor mitigation~\cite{zeng2022adversarial,li2024purifying,xu2024towards}, which directly removes backdoors after model development. This paper primarily focuses on dataset purification since it reduces backdoor threats from the source and can be easily used to mitigate model backdoors. We hereby provide an overview of the defences related to dataset purification and backdoor mitigation.

\vspace{0.3em} 
\noindent\textbf{Dataset Purification.} Existing strategies can be divided into four types: \textbf{(1)} purification via latent separability, \textbf{(2)} purification via early convergence, \textbf{(3)} purification via dominant trigger effects, \textbf{(4)} purification via perturbation consistency.

Specifically, latent separability-based defenses exploited the detectable traces left by poisoned samples in the feature space. For example, Chen~\etal~\cite{chen2018detecting} observed that, in the feature space of the final hidden layer, samples from the target class form two distinct clusters, with the smaller cluster identified as poisoned. Ma~\etal~\cite{ma2022beatrix} utilized high-order statistics (\ie, Gram matrix) to analyze the differences between poisoned and benign samples; Early convergence-based defenses relied on the observation that DNNs converge on poisoned samples more rapidly than on benign ones. During the early stages of training, the losses for poisoned samples quickly drop to near zero, while those for benign samples remain relatively high. For example, researchers in~\cite{li2021anti,zhang2023backdoor} traped samples whose losses decreased more rapidly by using the local gradient ascent technique during the initial five training epochs; Dominant trigger-based defenses assume that backdoor triggers play a dominant role in DNN predictions. For example, Chou~\etal~\cite{chou2020sentinet} utilized model interpretability techniques (\eg, Grad-CAM~\cite{selvaraju2017grad}) to visualize salient regions of an input image, identifying highly localized and small regions as potential trigger areas. Similarly, Huang~\etal~\cite{huang2023distilling} distilled minimal patterns from input images that influenced the model's predictions and identified images with abnormally small patterns as poisoned; Perturbation consistency-based defenses assumed that poisoned samples are resistant to perturbations. For instance, Guo~\etal~\cite{guo2023scale} observed that poisoned samples exhibited prediction consistency under pixel-level amplification and proposed analyzing this consistency to distinguish poisoned samples. To address constraints in pixel values and insensitivity to amplification, Pal~\etal~\cite{pal2024backdoor} optimized a mask for selective pixel amplification; Qi~\etal~\cite{qi2023towards} analyzed the prediction consistency by unlearning the benign connections of the backdoored models; and Hou~\etal~\cite{hou2024ibd} examined prediction consistency under unbounded weights.

However, in this paper, we find that all existing methods suffer from poor performance in some cases. Specifically, the first type of purification method generally only utilizes information at a specific layer (\eg, the last hidden layer) and can be easily bypassed by advanced attacks \cite{qi2023revisiting}. We will also show that the last three types of methods, however, relied on an underlying assumption that does not always hold, especially under A2A and UT attacks. Most recently, a concurrent work (\ie, Telltale~\cite{gao2025try}), offered a novel and effective solution for detecting poisoned samples in the training process. It gathered loss trajectories of samples after model convergence and applied clustering analysis to identify poisoned clusters. It is applicable to non-classification tasks, mitigating the reliance on clean data and addressing inconsistencies in defense effectiveness. However, it focuses mainly on the A2O and clean-label attack modes, overlooking other attack modes such as A2A and UT. How to design an effective and comprehensive dataset purification method is still an important open question.


\vspace{0.3em} 
\noindent\textbf{Backdoor Mitigation.} This defense occurs in the post-development phase, aiming to remove implanted backdoors from attacked models. For instance, Li~\cite{liu2018fine} pruned dormant neurons with benign inputs; Wu~\cite{wu2021adversarial} pruned neurons that are sensitive to adversarial perturbations; Chai~\cite{chai2022one} applied weight-level pruning, which is more precise than neuron-level pruning, thereby preserving benign task performance; Li~\cite{li2021neural} adopted knowledge distillation to guide the fine-tuning of backdoored models. Instead of direct removal, researchers in~\cite{wang2019neural, zeng2022adversarial, xu2024towards} reverse-engineered triggers and decouple them from target labels through unlearning and fine-tuning.

Existing backdoor mitigation defenses have shown effectiveness against different attacks; however, Zhu~\etal~\cite{zhu2024breaking} demonstrated that the injected backdoors can persist and reactivate during inference, even after backdoor mitigation. This underscores the urgent need for a dataset purification method capable of defending against a broader range of attacks, including A2O, A2A, and UT attacks, to prevent backdoor creation at its source.

\section{Revisiting Dataset Purification}
\label{sec:revisiting}

\subsection{Preliminaries}

\noindent\textbf{The Main Pipeline of Poison-only Backdoor Attack.} Let $ \mathcal{D} = \{ (\vx_i, y_i) \}_{i=1}^{N} $ denotes a training set composed of $ N $ \textit{i.i.d.} samples. Each sample $(\vx, y)$ is characterized by $\vx \in \mathcal{X} = [0,1]^{d_C \times d_W \times d_H}$ and $y \in \mathcal{Y} = \{1, 2, \ldots, K\}$. An adversary can generate a poisoned dataset $\hat{\mathcal{D}}$ by modifying a subset of benign samples (\ie, $\mathcal{D}_s$), \ie, $\hat{\mathcal{D}} = \mathcal{D}_p \cup \mathcal{D}_b$, where $\mathcal{D}_b = \mathcal{D} - \mathcal{D}_s \subset \mathcal{D}$, $\mathcal{D}_p=\{(\hat{\vx}, \hat{y})\mid \hat{\vx}=\mathcal{G}_X(\vx)), \hat{y}=\mathcal{G}_Y(y), (\vx, y)\in \mathcal{D}_s \}$, and $\rho = |\mathcal{D}_s| / |\hat{\mathcal{D}}|$ is the poisoning rate. $\mathcal{G}_X: \mathcal{X} \rightarrow \mathcal{X}$, $\mathcal{G}_Y: \mathcal{Y} \rightarrow \mathcal{Y}$ are adversary-specified poisoned image generator and target label generator, respectively. For instance, in the BadNets attack~\cite{gu2017badnets}, $\mathcal{G}_X(x)=(\bm{1}-\vm)\odot\vx+\vm\odot \vt$, where $\vm\in\{0,1\}^{d_C \times d_W \times d_H}, \vt\in \mathcal{X}$ is the malicious trigger. As for the $\mathcal{G}_Y(y)$, there are two primary attack paradigms: \textbf{(1)} targeted attacks and \textbf{(2)} untargeted attacks. Specifically, for the targeted attacks, poisoned images are re-labeled to the designated target labels: $\mathcal{G}_Y(y)=y_t, y_t\in \mathcal{Y}$ in A2O attacks, and $\mathcal{G}_Y(y)=(y+1) \pmod{K}$ in A2A attacks. For the untargeted attacks, each poisoned image is re-labeled to a random label uniformly sampled from \(\mathcal{Y}\) (\ie, $G_Y(y)\sim U(1, K)$). 

\begin{figure}[!t]
\vspace{-1.5em}
\centering
 \begin{minipage}{0.99\linewidth}
 \centering
    \includegraphics[width=1\linewidth]{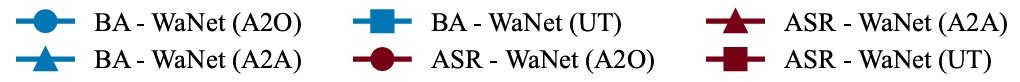}
\end{minipage}
 \begin{minipage}{0.99\linewidth}
    \begin{minipage}{0.49\linewidth}
        \centering
         \includegraphics[width=1\linewidth]{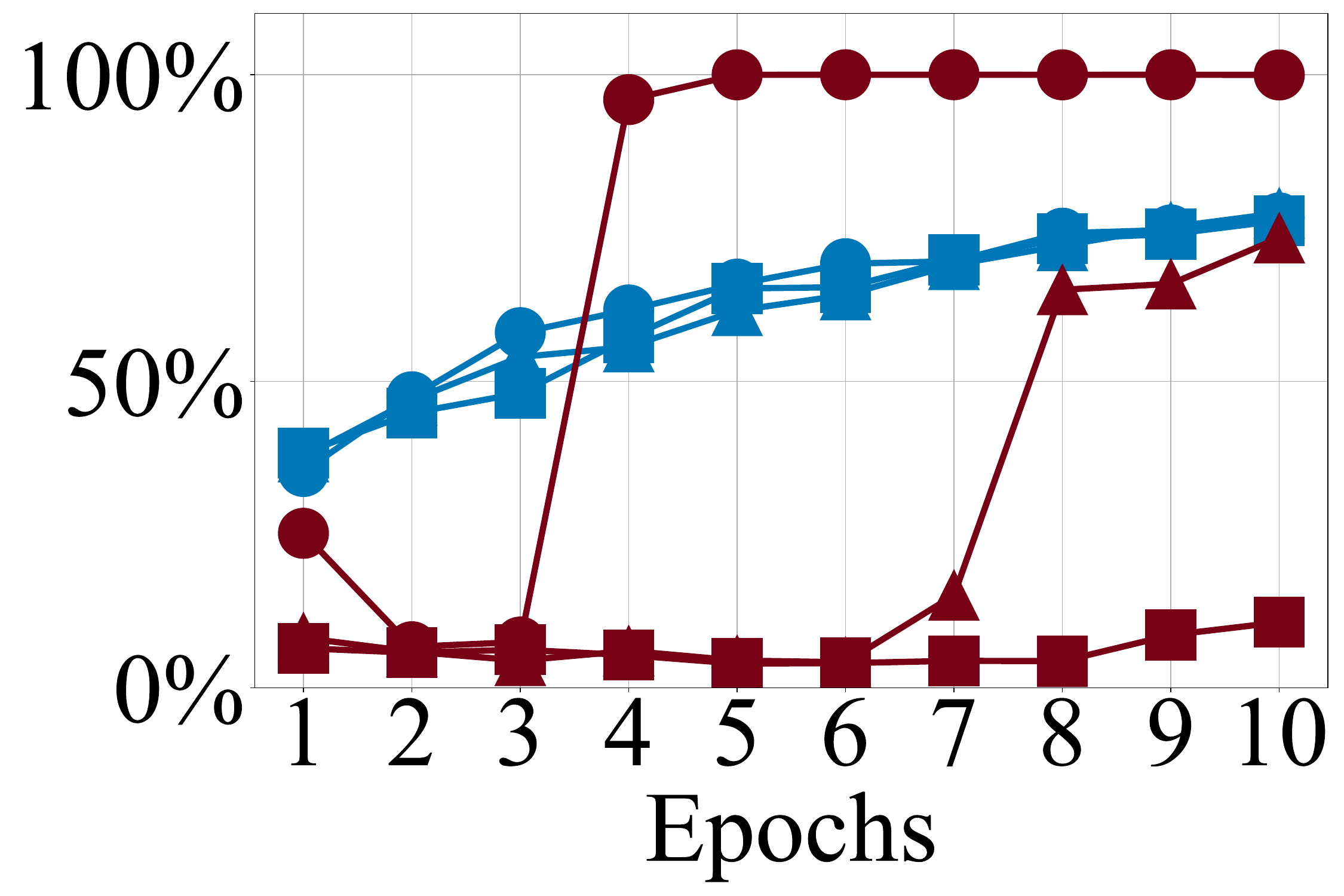}
            \centerline{(a) BadNets}
    \end{minipage}
   \begin{minipage}{0.49\linewidth}
        \centering
         \includegraphics[width=1\linewidth]{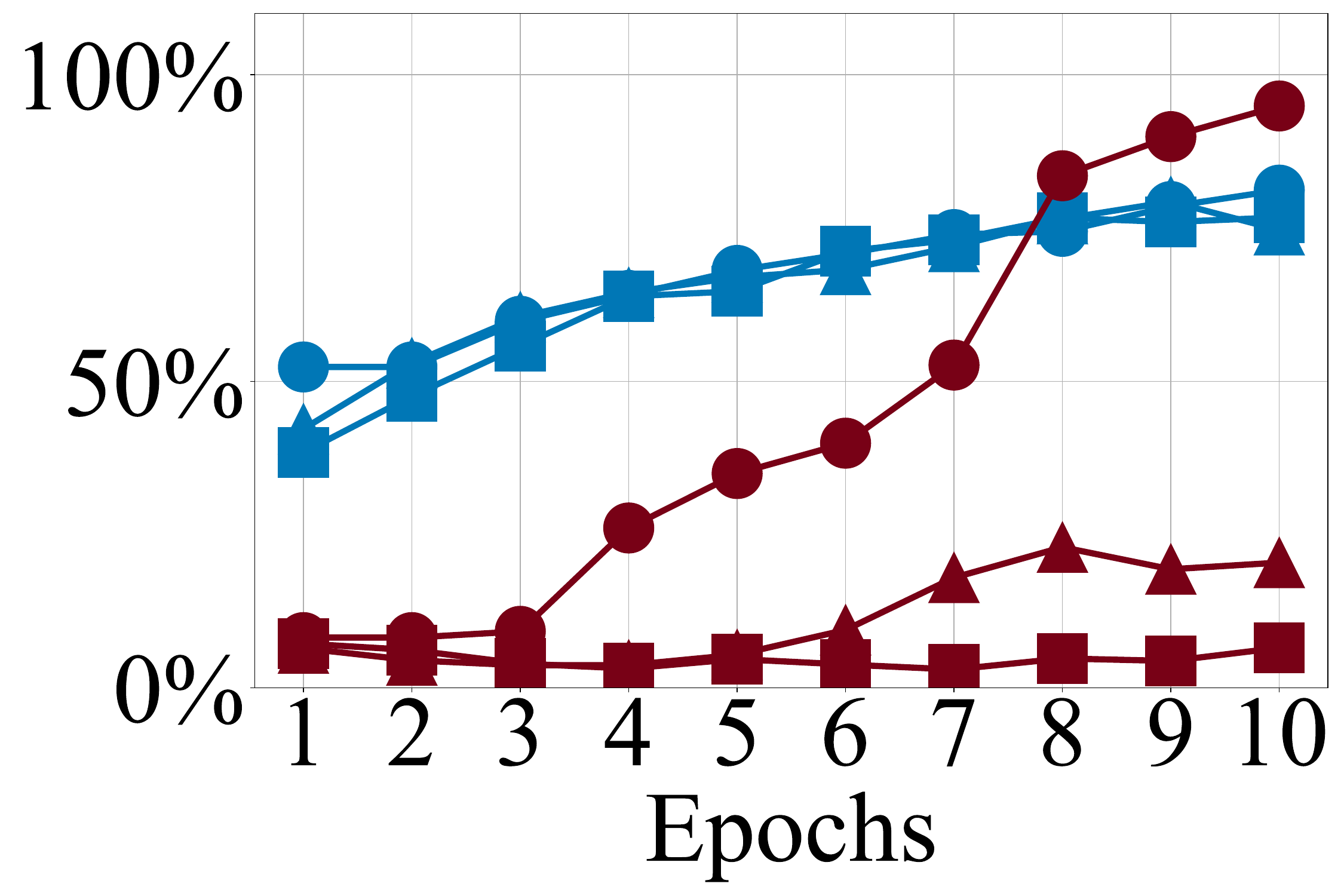}
            \centerline{(b) WaNet}
    \end{minipage}
\end{minipage}
\vspace{-0.5em}
\caption{The benign accuracy (BA) and attack success rate (ASR) during the initial ten epochs of model training. These curves show the model's convergence behavior, with rapid increases in ASR indicating that the corresponding backdoor connection is easier to learn, while slower rises suggest more challenging ones, especially under A2A and UT attacks.}
 \label{fig:5epoch}
  \vspace{-1em}
 \end{figure}


\vspace{0.3em} 
\noindent\textbf{Threat Model of Dataset Purification.} Similar to existing studies~\cite{chen2018detecting,pal2024backdoor}, we assume that adversaries can freely poison the training dataset but can not manipulate the training process. Defenders aim to identify and filter out poisoned samples from a given third-party dataset. This is achieved through two main goals: effectiveness and generalizability. \textbf{Effectiveness} ensures that all poisoned samples are detected while minimizing the incorrect removal of benign samples. \textbf{Generalizability} guarantees that the defense is effective against various attack types, including A2O, A2A, and UT attacks. We assume that defenders have access to the poisoned training dataset and full control over the training process, but they have no knowledge of the backdoor attacks. In particular, our method requires less capacity for defenders compared to classical methods~\cite{qi2023towards,hou2024ibd} since it does not require additional benign local samples.


\subsection{Revisiting Strategy 1 (Early Convergence)}
\label{sec:early}
In general, the purification methods in~\cite{li2021anti,huang2022backdoor,zhang2023backdoor} are designed based on the observation that DNNs converge faster on poisoned samples during the early stages of training, suggesting that the backdoor connections are easier to learn. 


\vspace{0.3em} 
\noindent\textbf{Settings.} To validate the effectiveness of this purification strategy across different attack types, we conduct two representative backdoor attacks, \ie, BadNets~\cite{gu2017badnets} and WaNet~\cite{nguyen2021wanet}, on the CIFAR-10 dataset using ResNet-18. They are the representatives of sample-agnostic and sample-specific attacks. Each attack includes three variants: A2O, A2A, and UT. For each attack, we calculate the benign accuracy (BA) and the attack success rate (ASR) during the initial ten epochs.

\vspace{0.3em} 
\noindent\textbf{Results.} As shown in Figure~\ref{fig:5epoch}, the ASR for the BadNets (A2O) and WaNet (A2O) attacks rapidly approaches 100\%. However, the ASRs for the other attacks remain lower than the BAs, with UT attacks even nearing 0\%. These results suggest that in A2A and UT attacks, DNNs do not converge quickly on poisoned samples, indicating that the assumption that backdoor connections are simpler to learn than the benign ones does not hold for A2A and UT attacks.




\subsection{Revisiting Strategy 2 (Dominant Trigger Effects)}
\label{sec:local}
Researchers in~\cite{chou2020sentinet,huang2023distilling} observed that poisoned samples often exhibit highly localized and small saliency regions, suggesting that trigger-related features can be regarded as `short-cut' that are more easily learned by DNNs compared to benign features.

\vspace{0.3em} 
\noindent\textbf{Settings.} We use BadNets and WaNet for our discussions, following the same experimental settings as described in Section~\ref{sec:early}, with WaNet results limited to the A2A setting for brevity. We compare saliency regions from two perspectives: visualization and quantitative measurement. Specifically, we utilize Grad-CAM~\cite{selvaraju2017grad} to visualize the saliency regions of both benign and poisoned samples. We then calculate the $\ell^1$ norms of these regions to quantify their size.

\begin{figure}[!t]
  \vspace{-1.8em}
	\centering
    \includegraphics[width=0.95\linewidth]{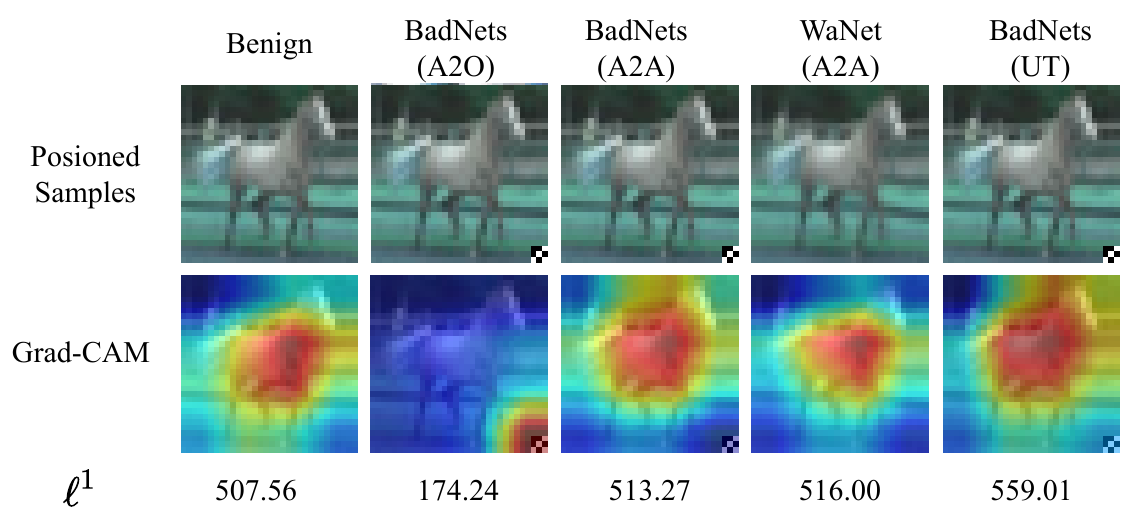}
\vspace{-0.5em}
 \caption{Grad-CAM visualization of saliency regions for benign and poisoned samples. The $\ell^1$ norm values of these regions are presented to quantify the amount of information the model relies upon for predictions. Larger $\ell^1$ norm values indicate that the model's predictions depend on more input features.}
\label{fig:gradcam}
\vspace{-1.5em}
\end{figure}

\vspace{0.3em} 
\noindent\textbf{Results.} As shown in Figure~\ref{fig:gradcam}, the saliency region of poisoned samples under the BadNets (A2O) attack is concentrated around the trigger, with its $\ell^1$ norm substantially smaller than that of the corresponding benign sample. However, the saliency regions for other attacks shift from the trigger area to benign regions, accompanied by high $\ell^1$ norms exceeding those of benign samples. These results suggest that, under A2A and UT attacks, the predictions of poisoned samples rely on distributed features, instead of simply trigger-related features. Consequently, the assumption that backdoor connections are inherently simpler does not hold for A2A and UT attacks.

\begin{figure*}[!t]
  \vspace{-1.8em}
\centering
 \begin{minipage}{0.8\linewidth}
 \begin{minipage}{0.02\linewidth}
            \rotatebox{90}{\hspace{40pt} CT  \hspace{40pt} IBD-PSC \hspace{40pt}  MSPC \hspace{40pt} SCALE-UP
            \vspace{20pt}}
\end{minipage}
 \begin{minipage}{0.95\linewidth}
    \centering
    \begin{minipage}{0.24\linewidth}
    \includegraphics[width=1\linewidth]{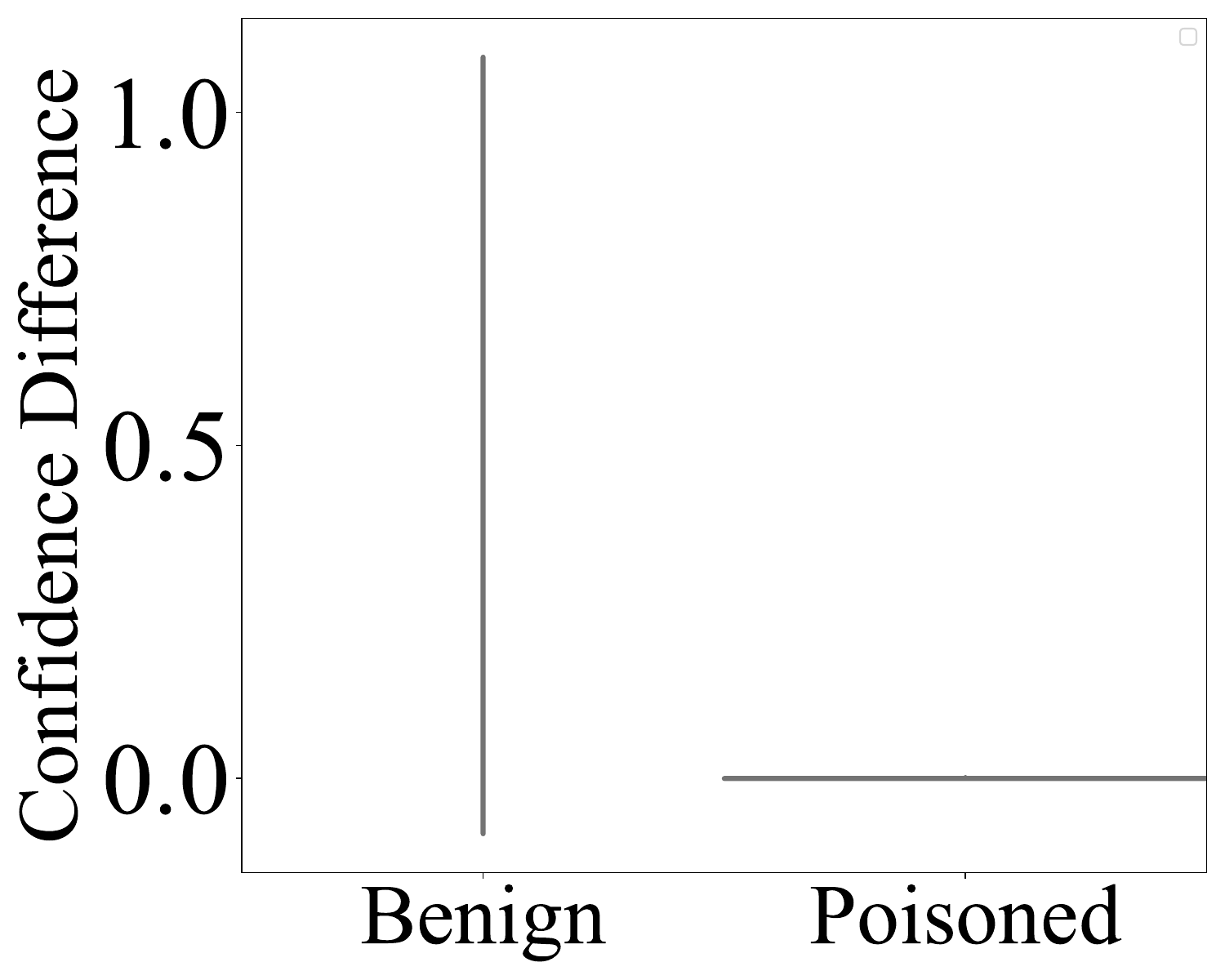}
    \includegraphics[width=1\linewidth]{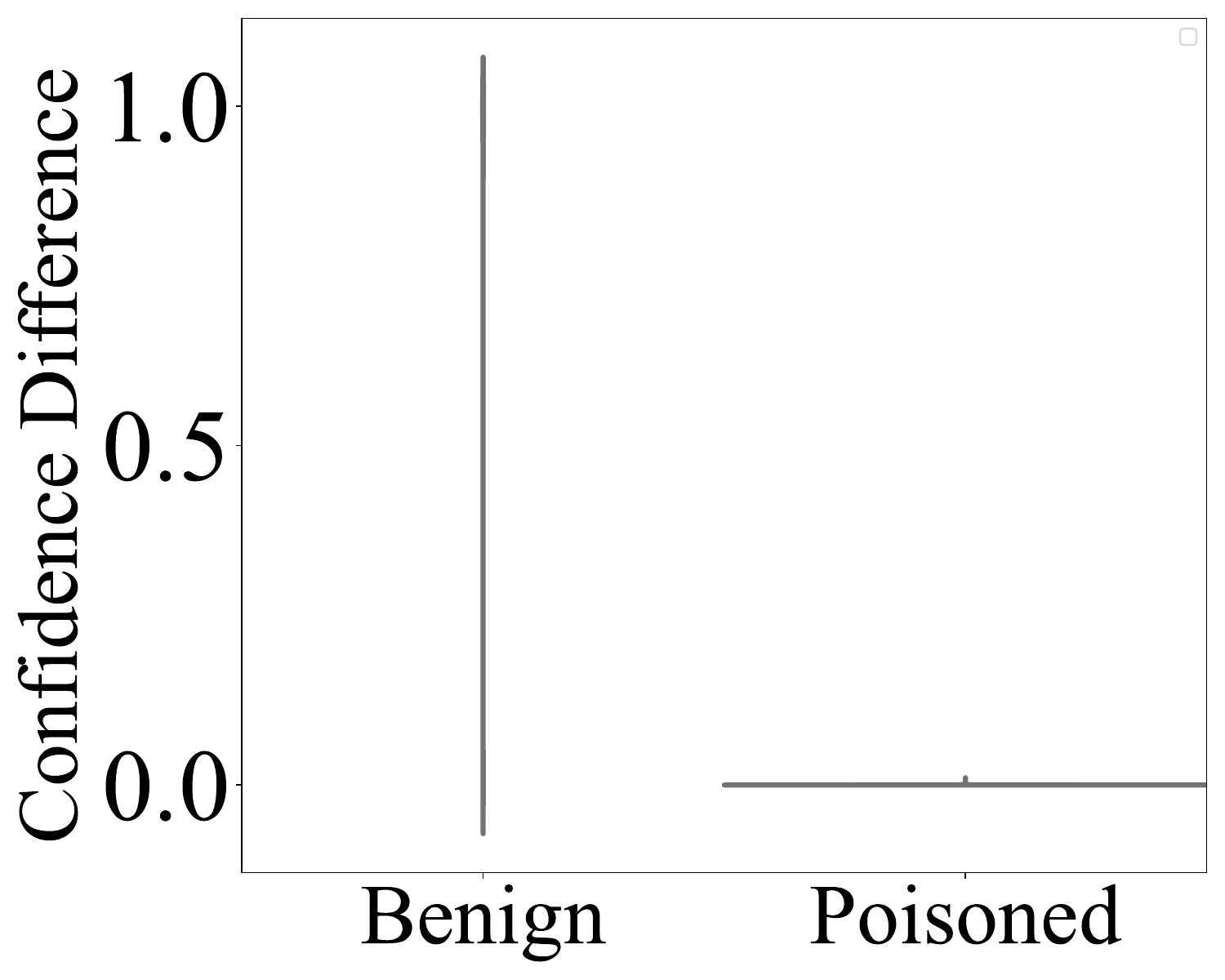}
            \includegraphics[width=1\linewidth]{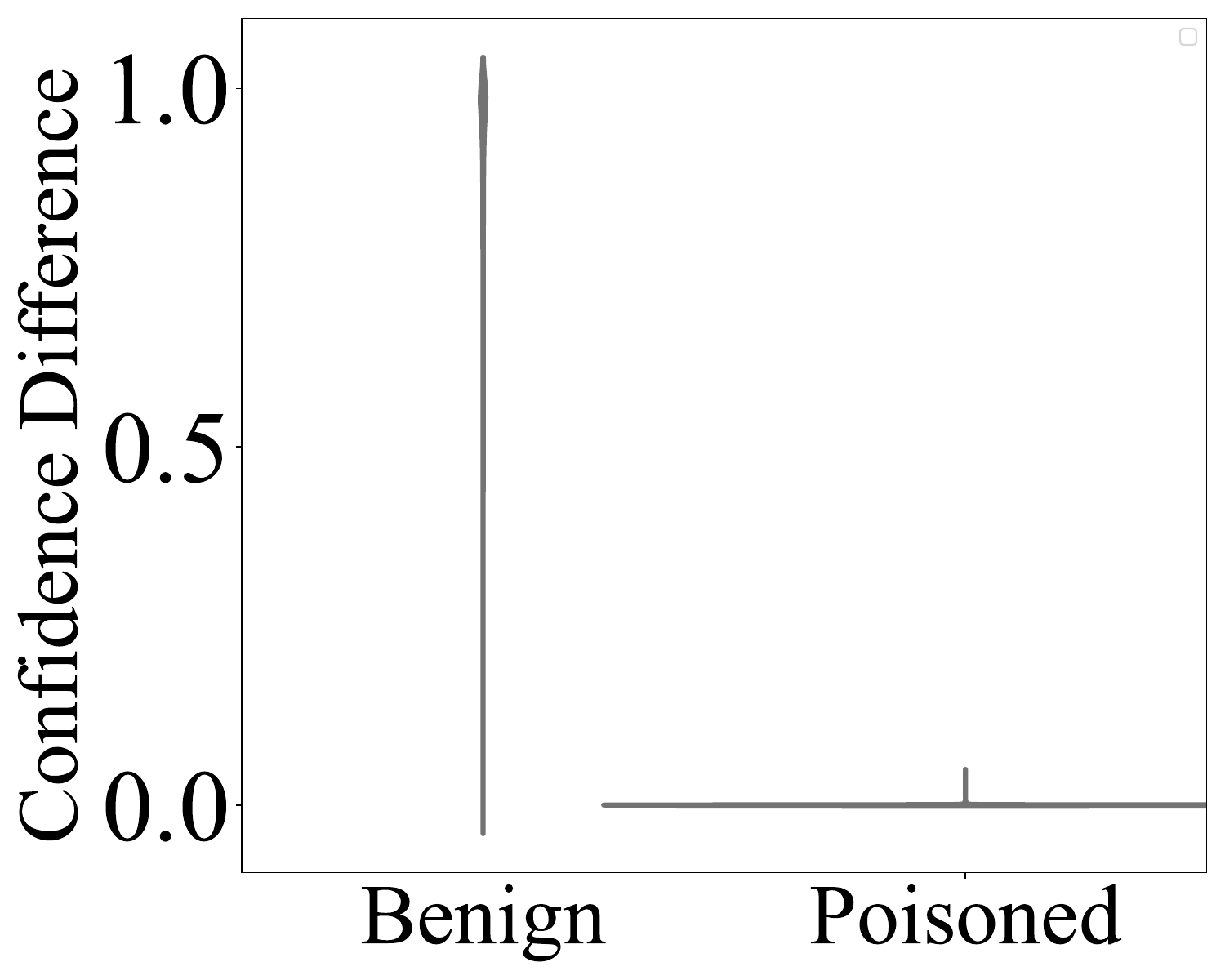}
            \includegraphics[width=1\linewidth]{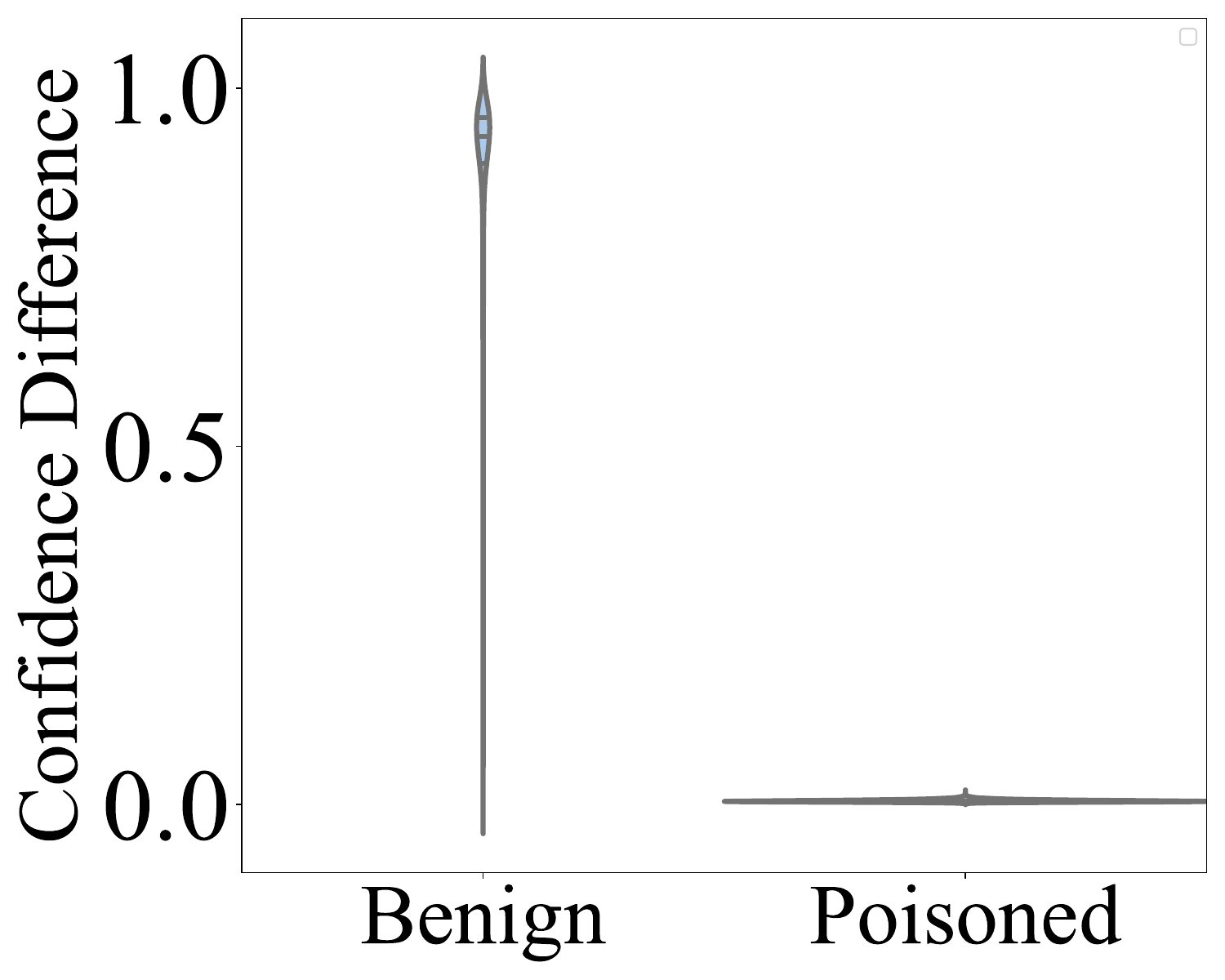}
            \centerline{(a) BadNets (A2O)}
    \end{minipage}    
    \begin{minipage}{0.24\linewidth}
            \includegraphics[width=1\linewidth]{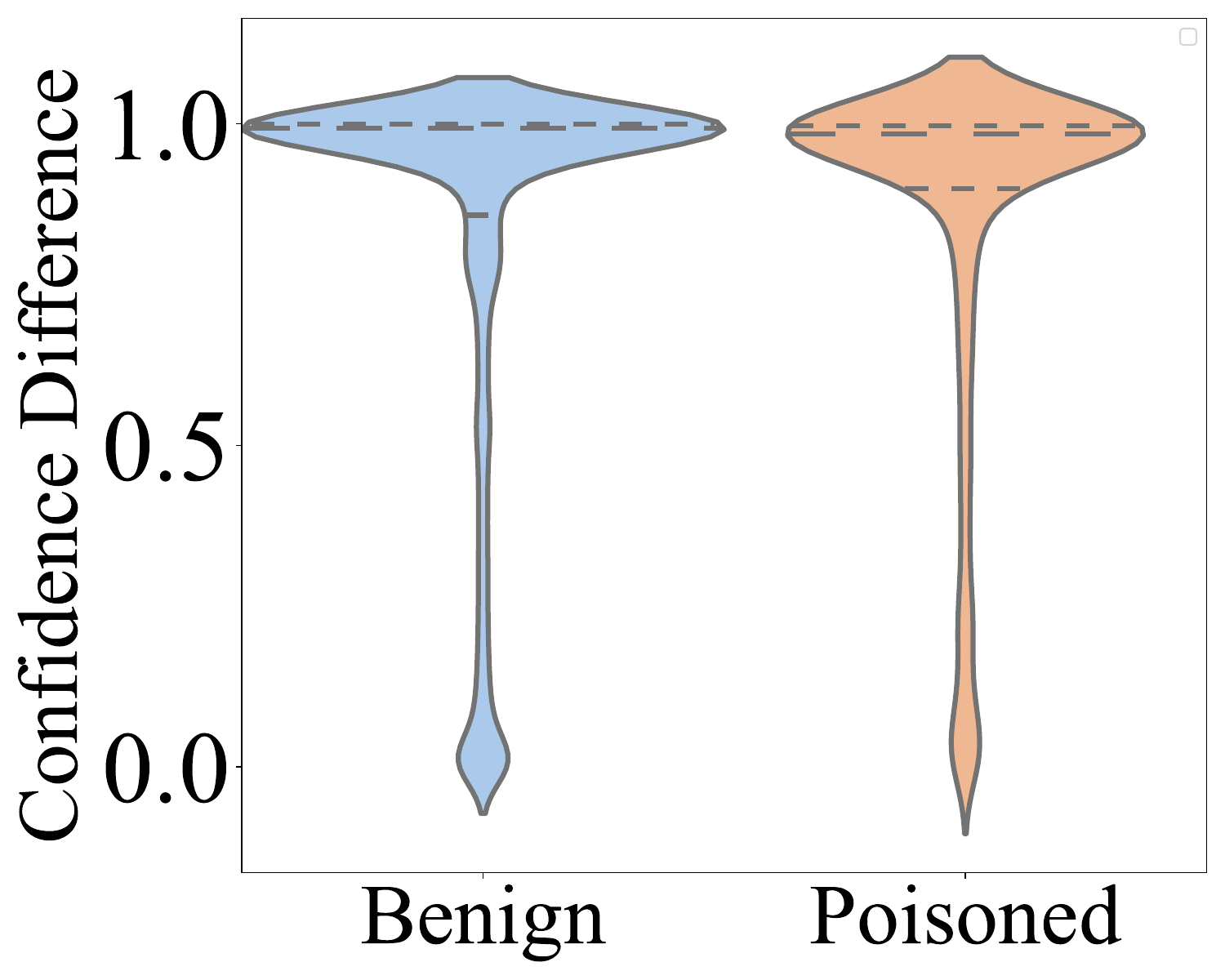}
            \includegraphics[width=1\linewidth]{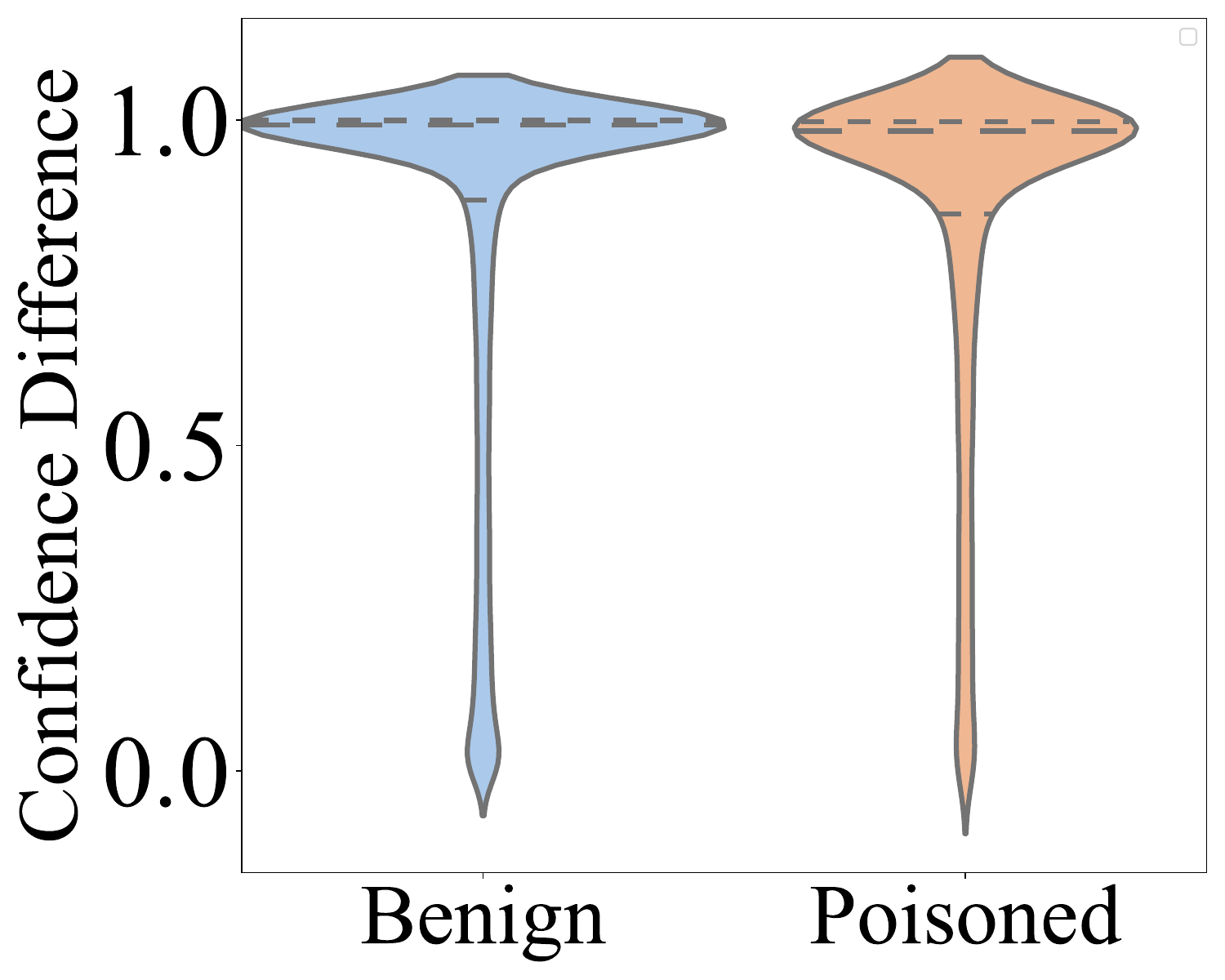}
            \includegraphics[width=1\linewidth]{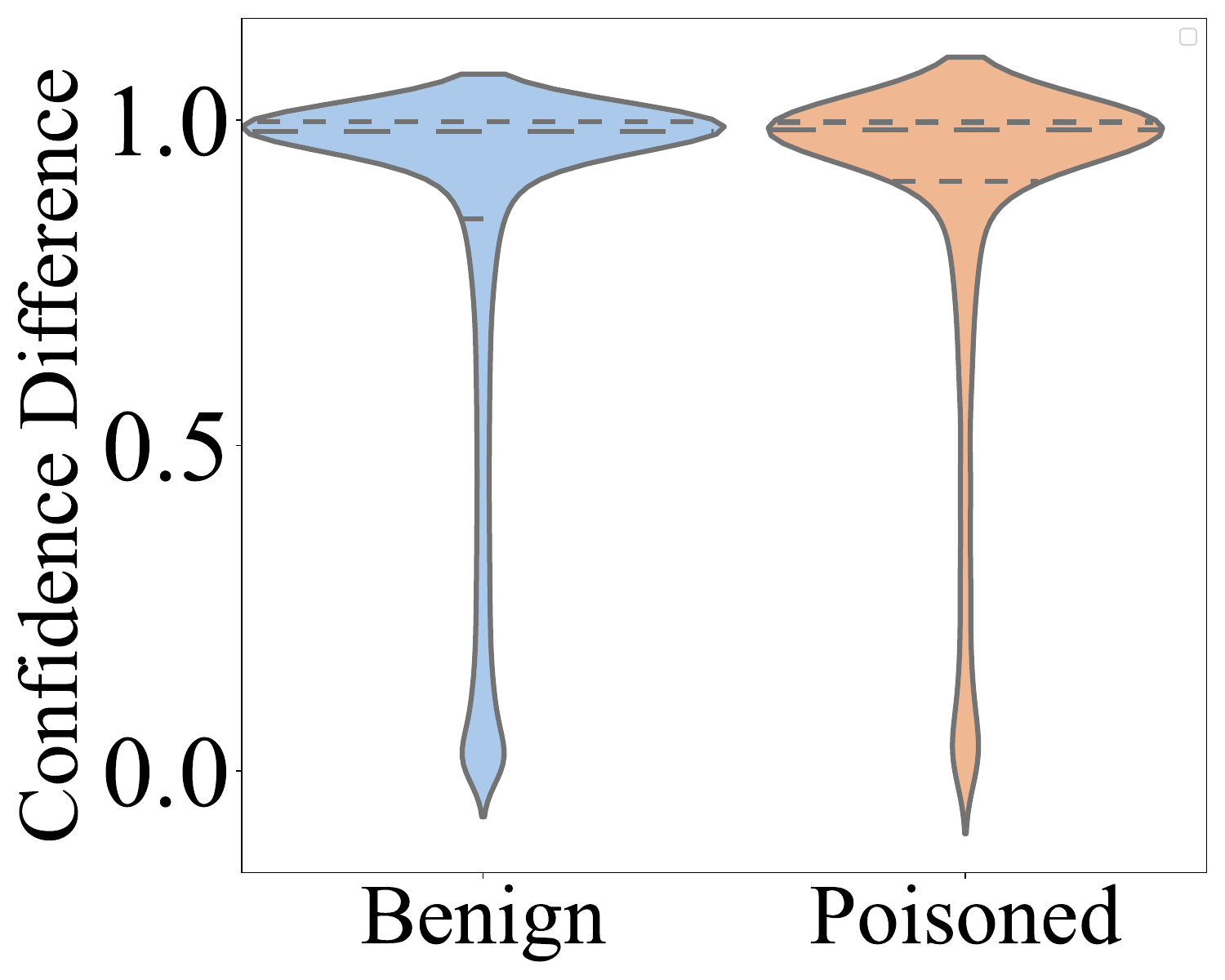}
            \includegraphics[width=1\linewidth]{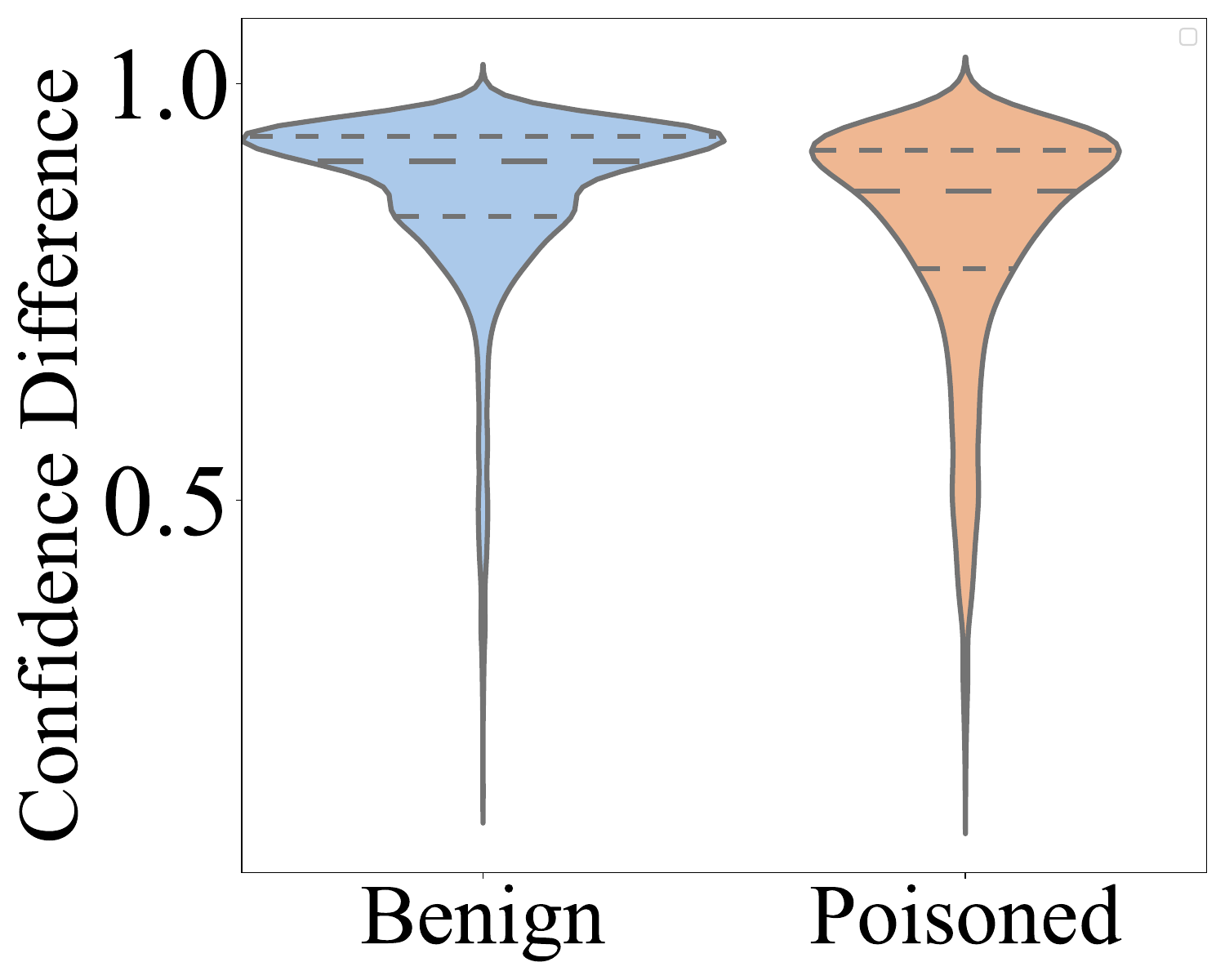}
            \centerline{(b) BadNets (A2A)}
    \end{minipage}
         \begin{minipage}{0.24\linewidth}
            \includegraphics[width=1\linewidth]{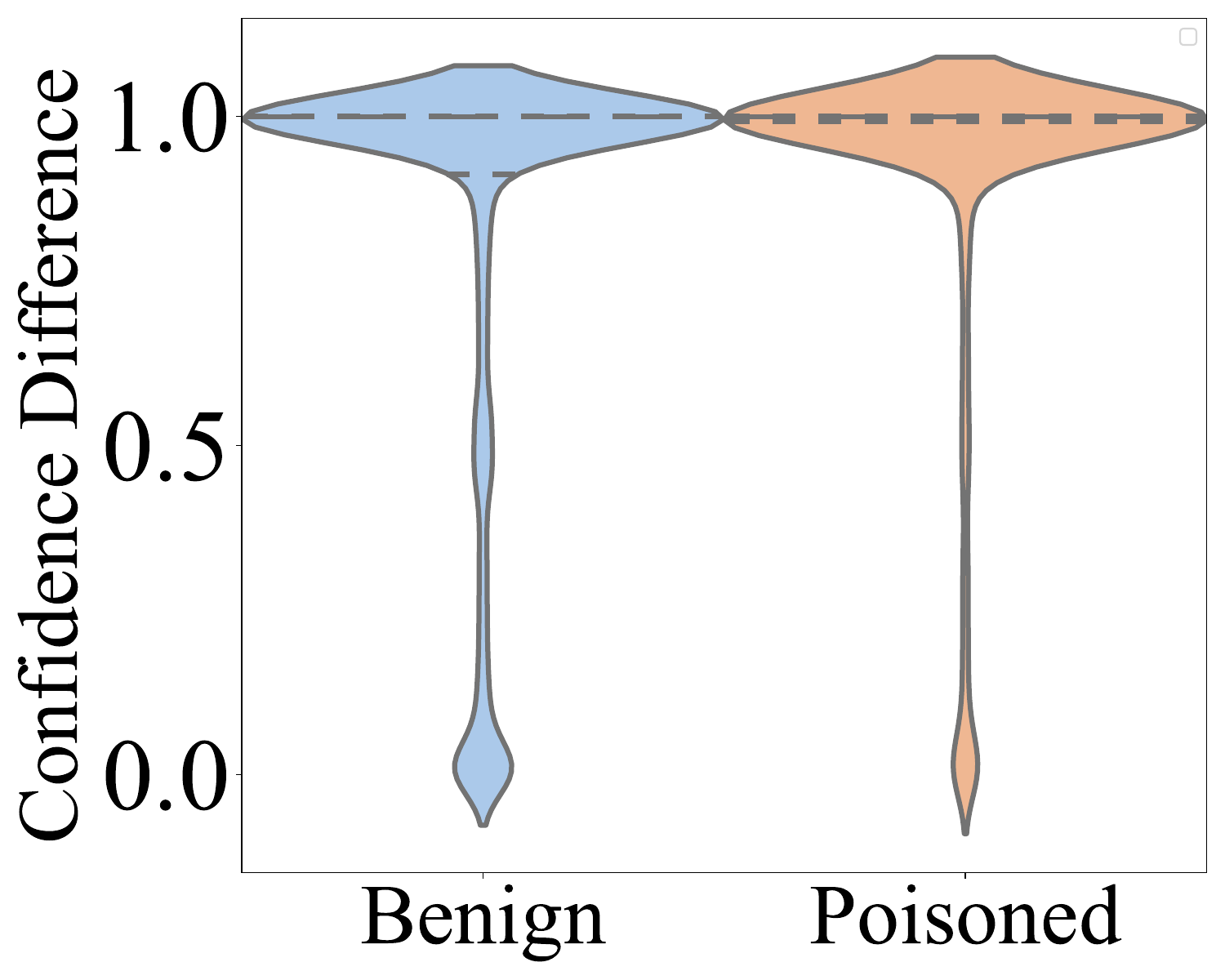}
            \includegraphics[width=1\linewidth]{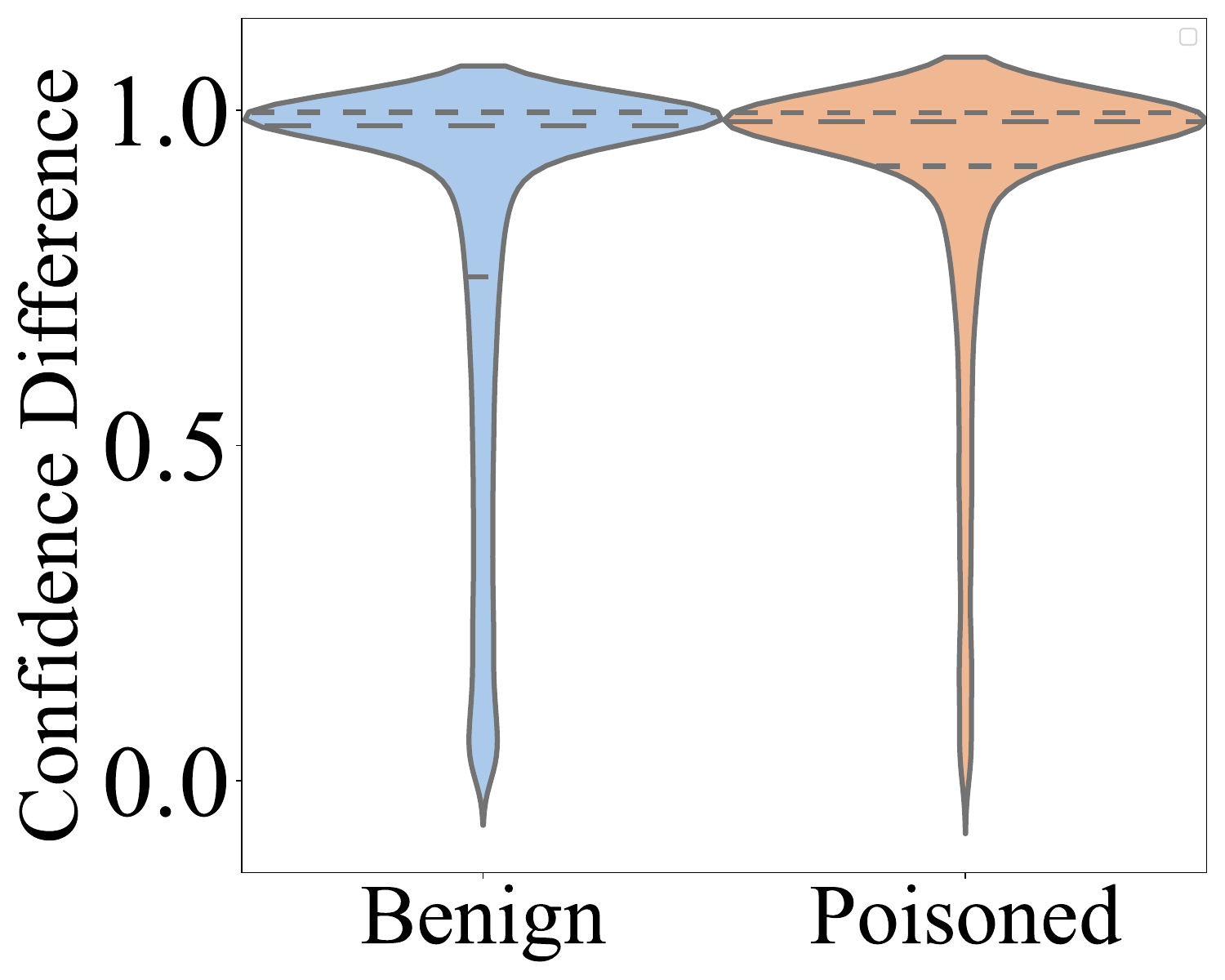}
             \includegraphics[width=1\linewidth]{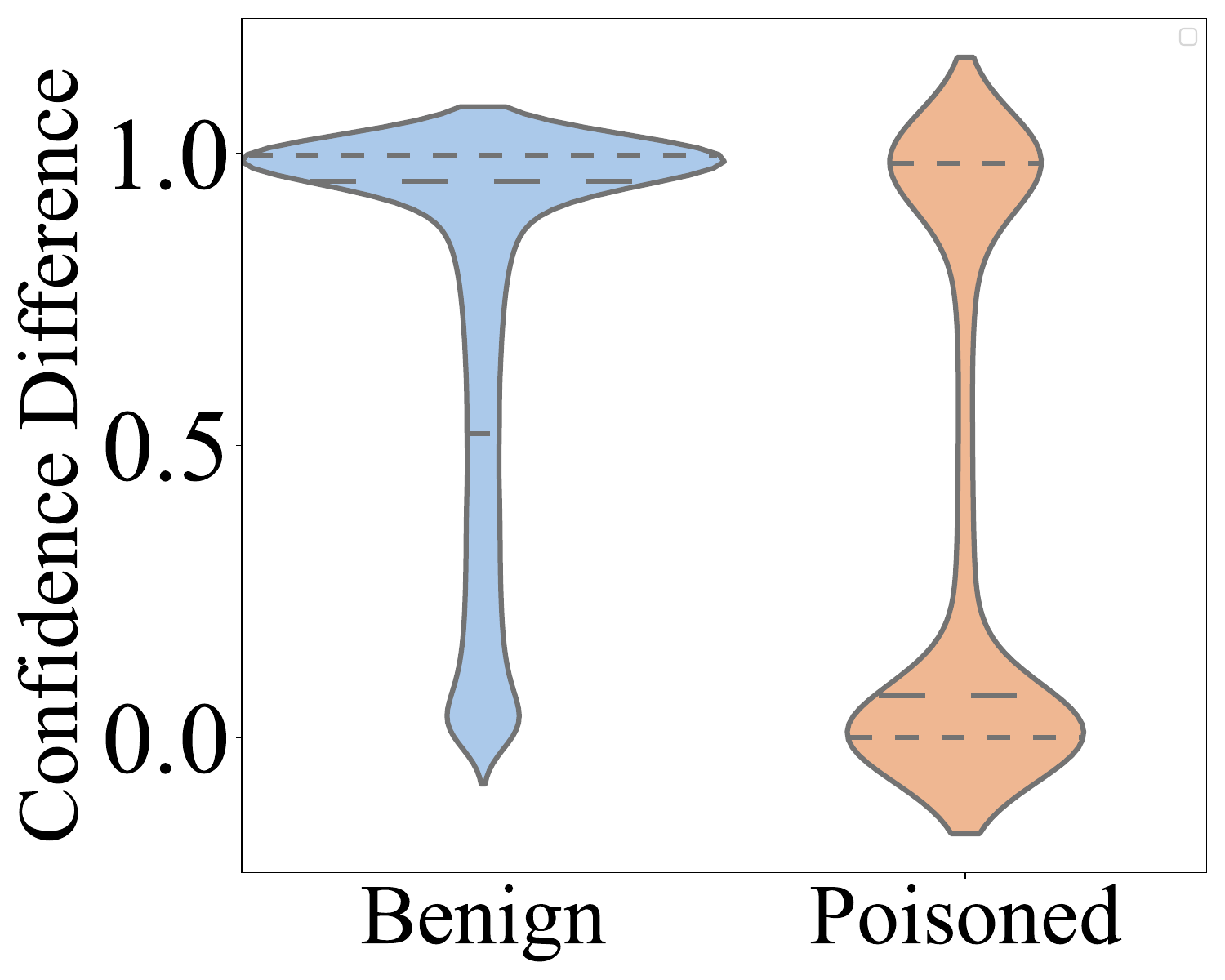}
            \includegraphics[width=1\linewidth]{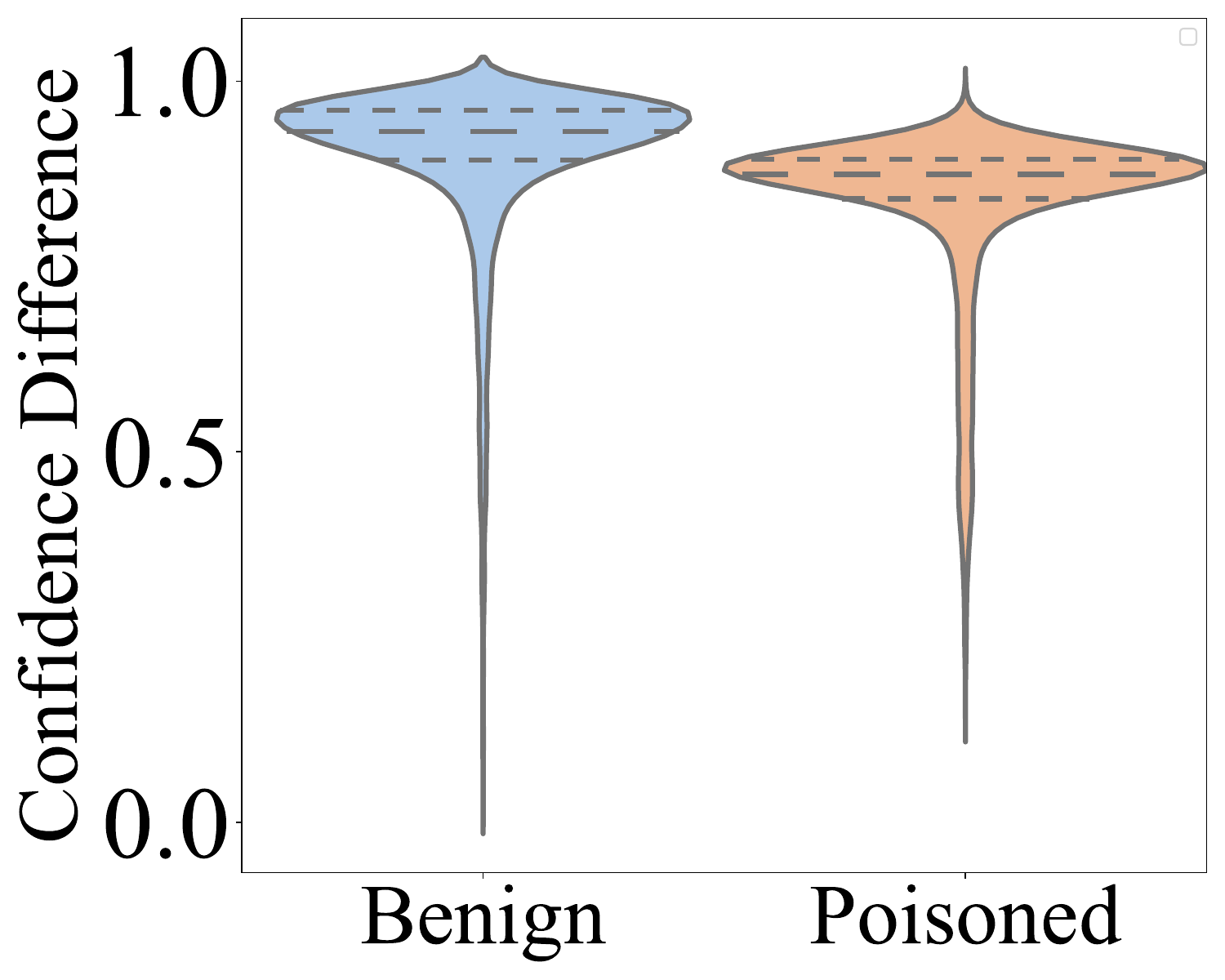}
            \centerline{(c) WaNet (A2A)}
    \end{minipage}
    \begin{minipage}{0.24\linewidth}
        \includegraphics[width=1\linewidth]{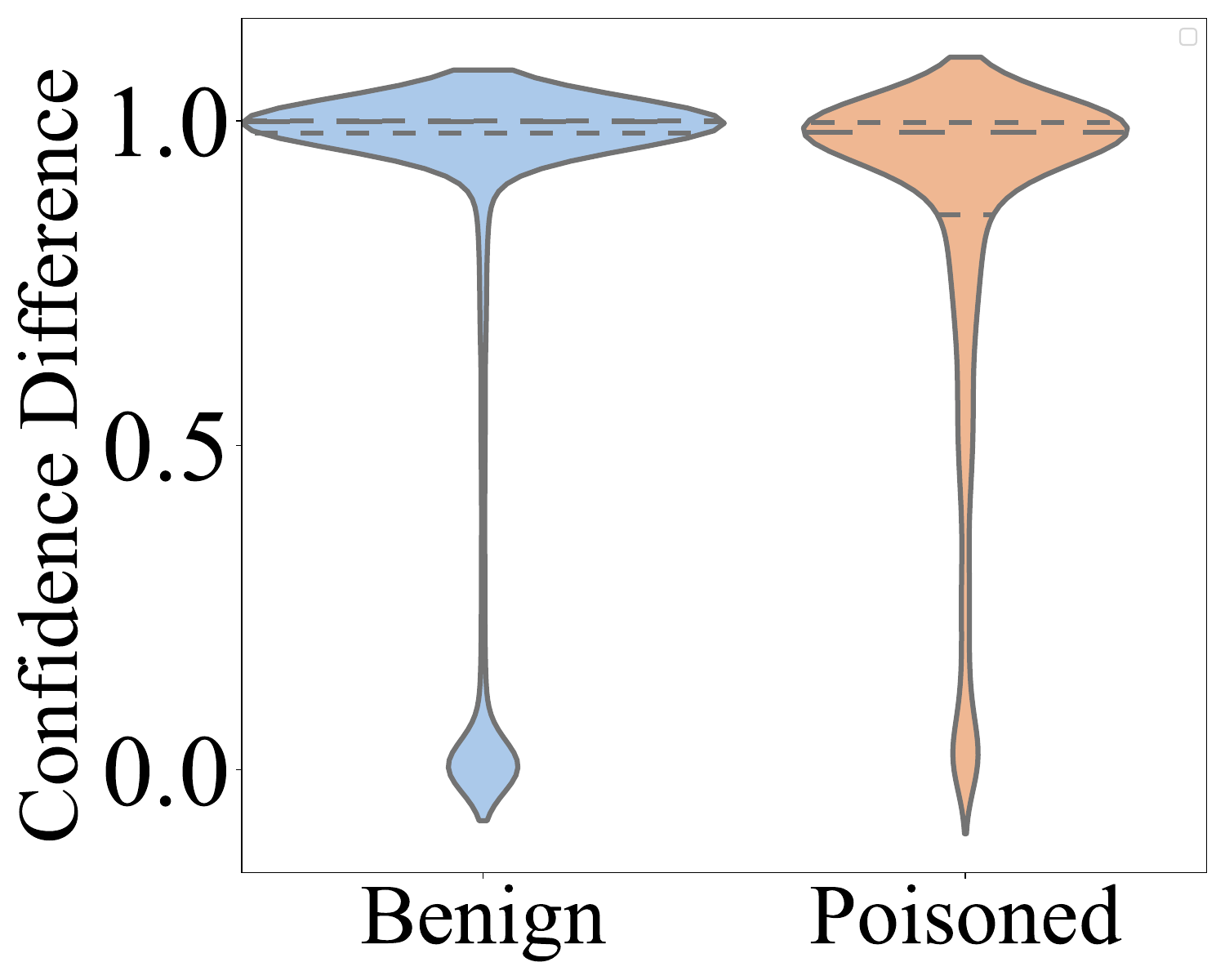}
         \includegraphics[width=1\linewidth]{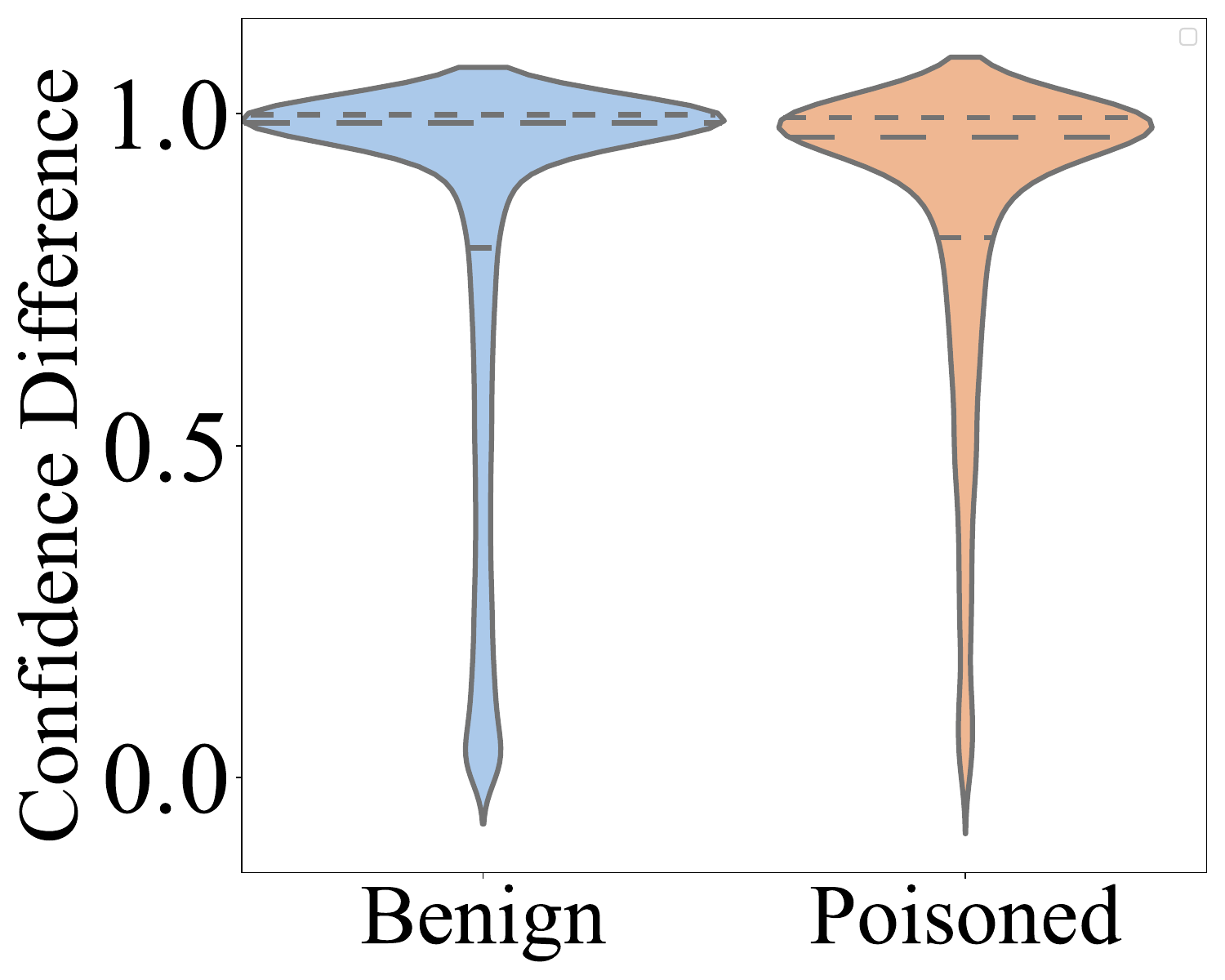}
        \includegraphics[width=1\linewidth]{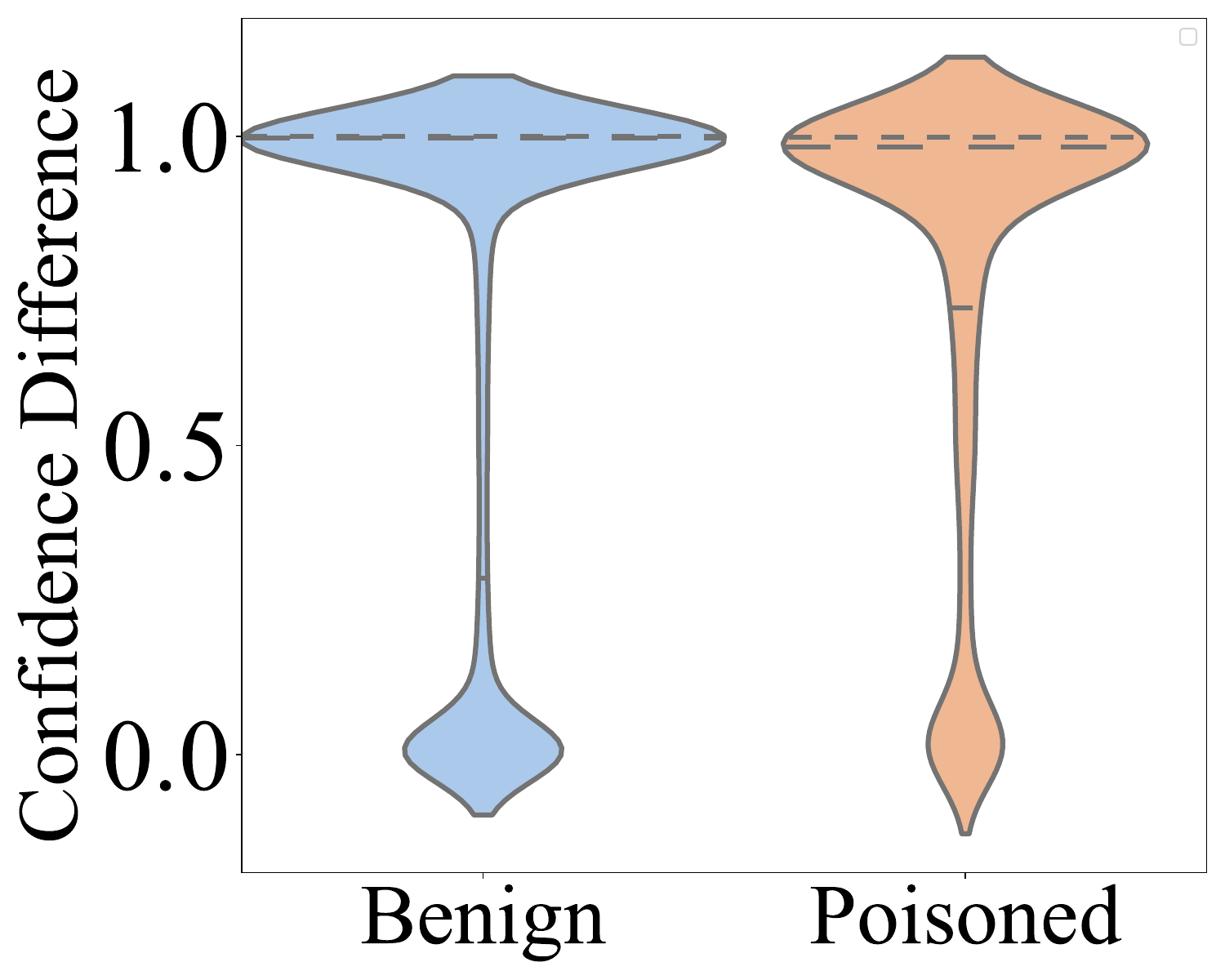}
            \includegraphics[width=\linewidth]{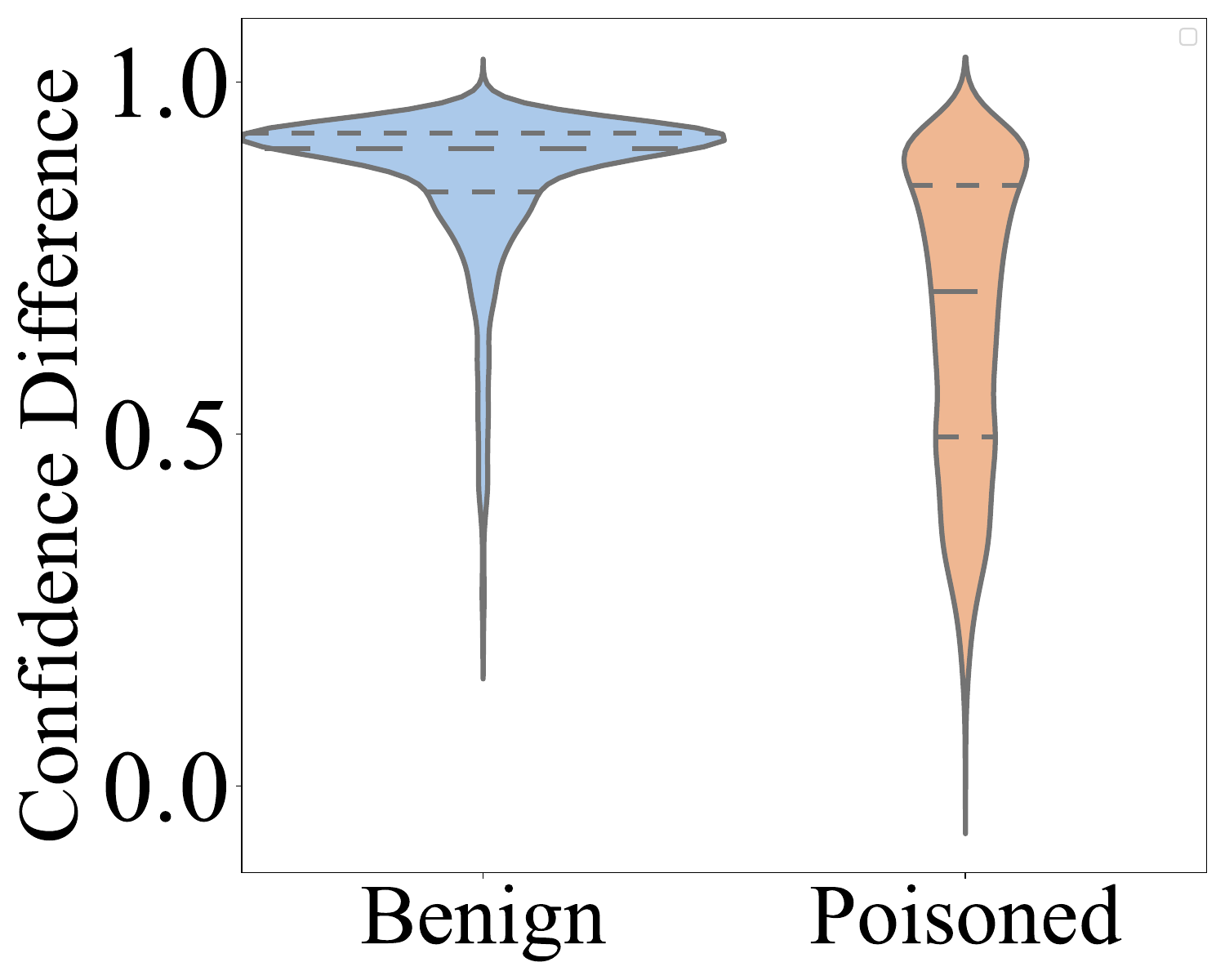}
            \centerline{(d) BadNets (UT)}
    \end{minipage}
    \end{minipage}
\end{minipage}
\vspace{-0.2em}
 \caption{Difference in prediction confidences for benign and poisoned samples on CIFAR-10 under input-level and weight-level perturbations, calculated by comparing prediction confidences on the original label before and after perturbation. A significant difference in confidence patterns indicates effective detection, as poisoned samples are distinguishable from benign ones.}
 \label{fig:conf_diff}
 \vspace{-0.5em}
\end{figure*}

\begin{figure*}[!t]
\centering
 \begin{minipage}{0.8\linewidth}
    \begin{minipage}{0.02\linewidth}
            \vspace{-25pt}
            \rotatebox{90}{\small BadNets (UT)   \hspace{10pt} \small WaNet (A2A)  \hspace{12pt} \small BadNets (A2A) }  
    \end{minipage}
    \begin{minipage}{0.97\linewidth}
    \centering
     \begin{minipage}{0.193\linewidth}
     \includegraphics[width=1\linewidth]{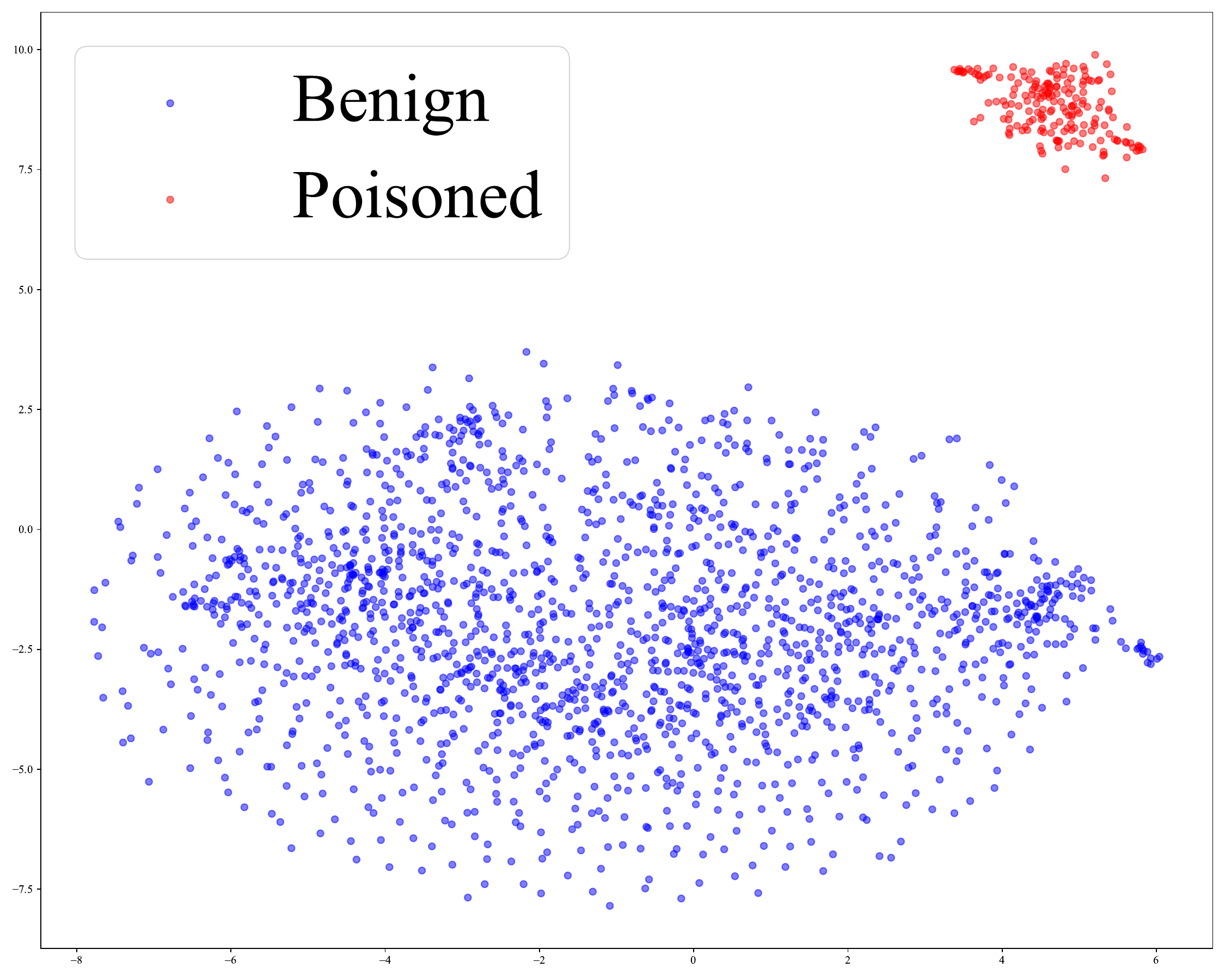}
          \includegraphics[width=1\linewidth]{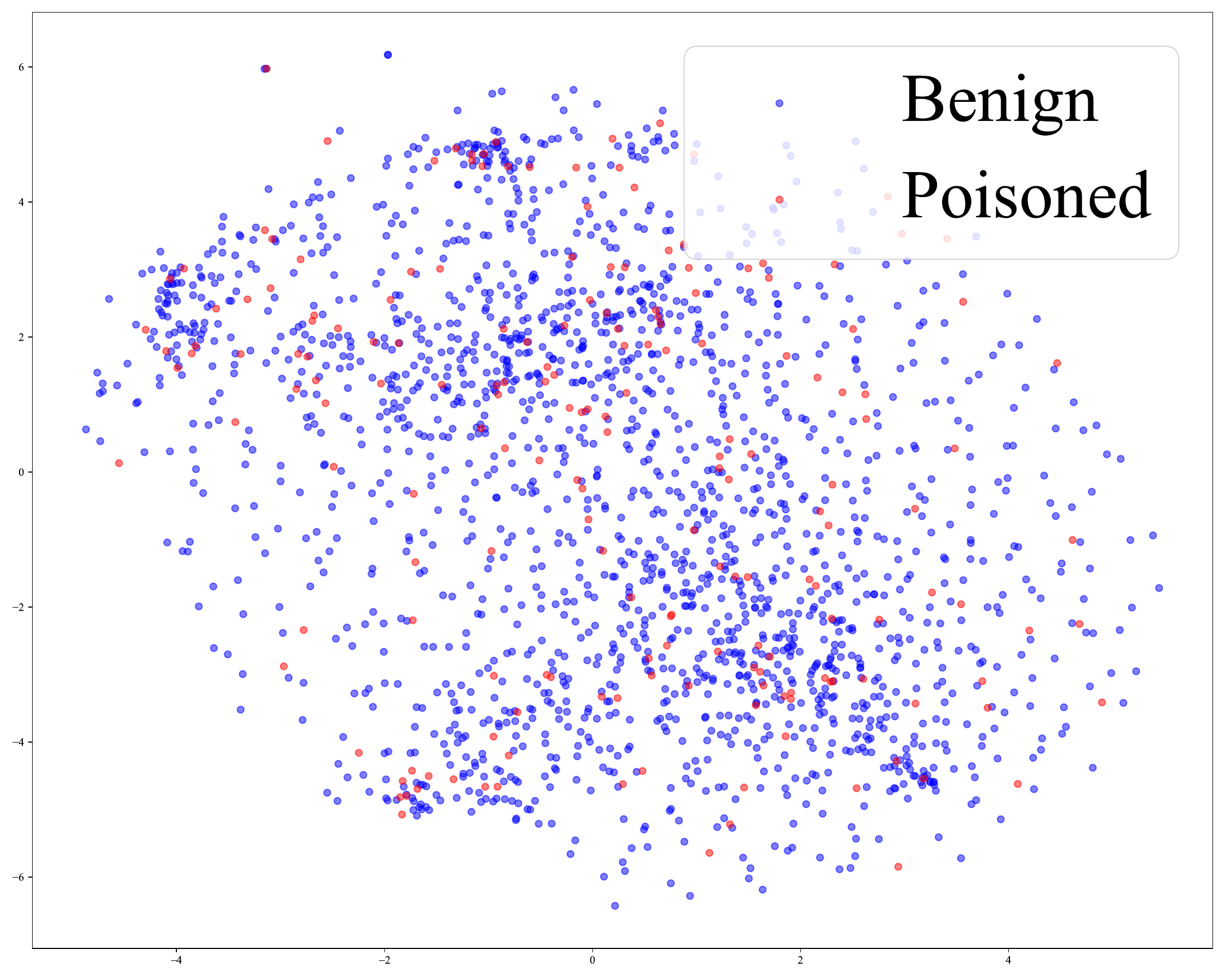}
            \includegraphics[width=1\linewidth]{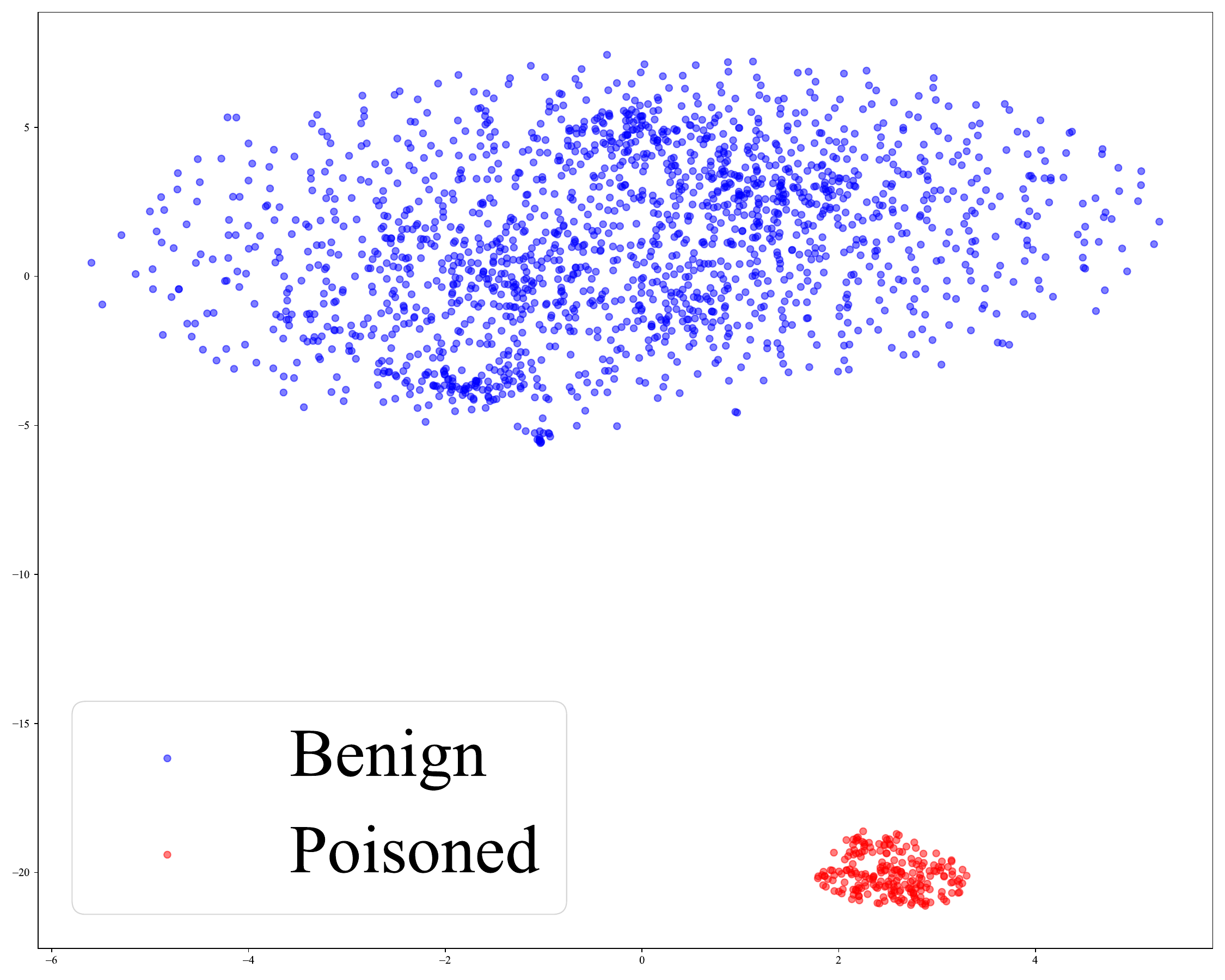}
            \centerline{(a) Layer 2}
    \end{minipage}
     \begin{minipage}{0.193\linewidth}
     \includegraphics[width=1\linewidth]{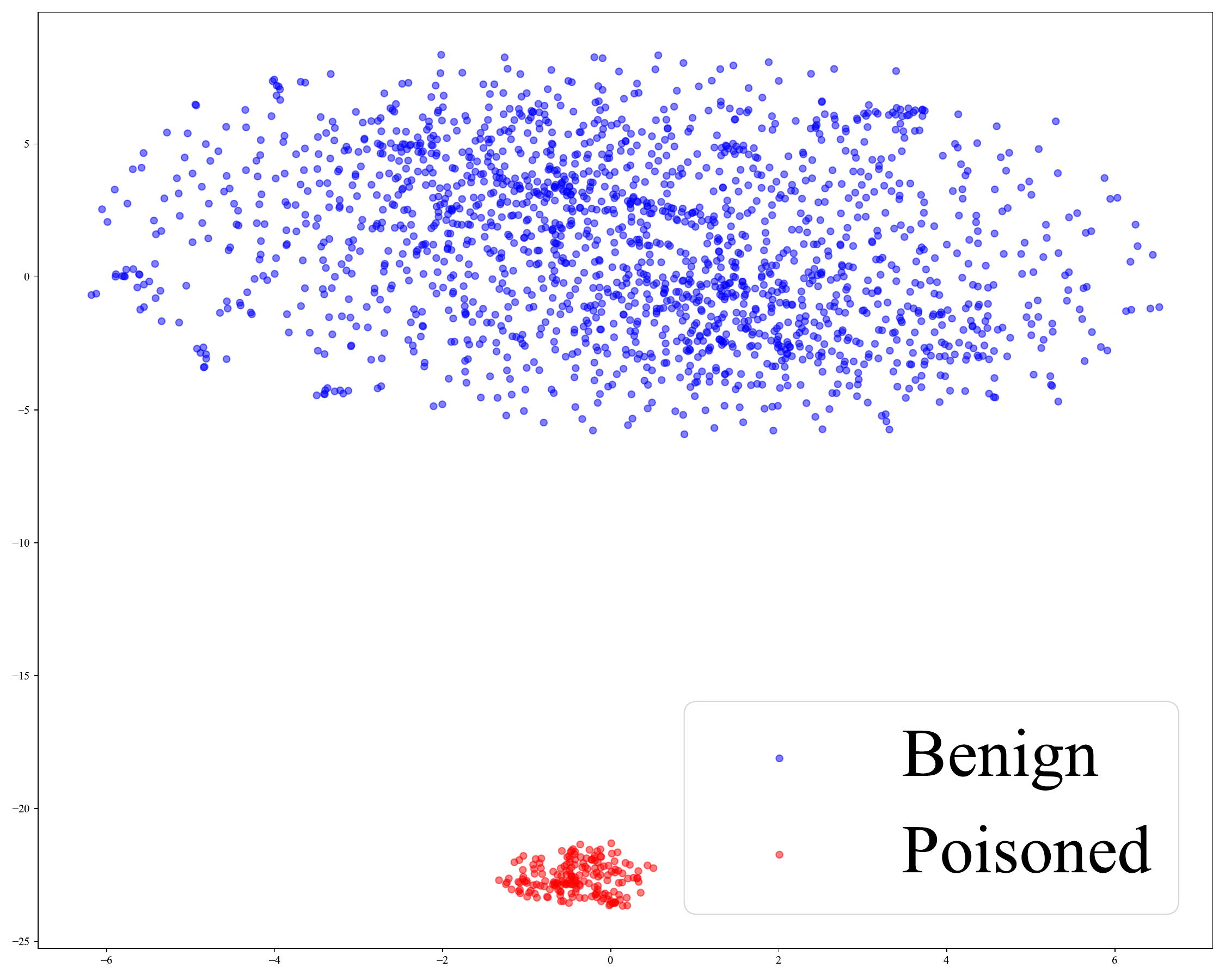}
     \includegraphics[width=1\linewidth]{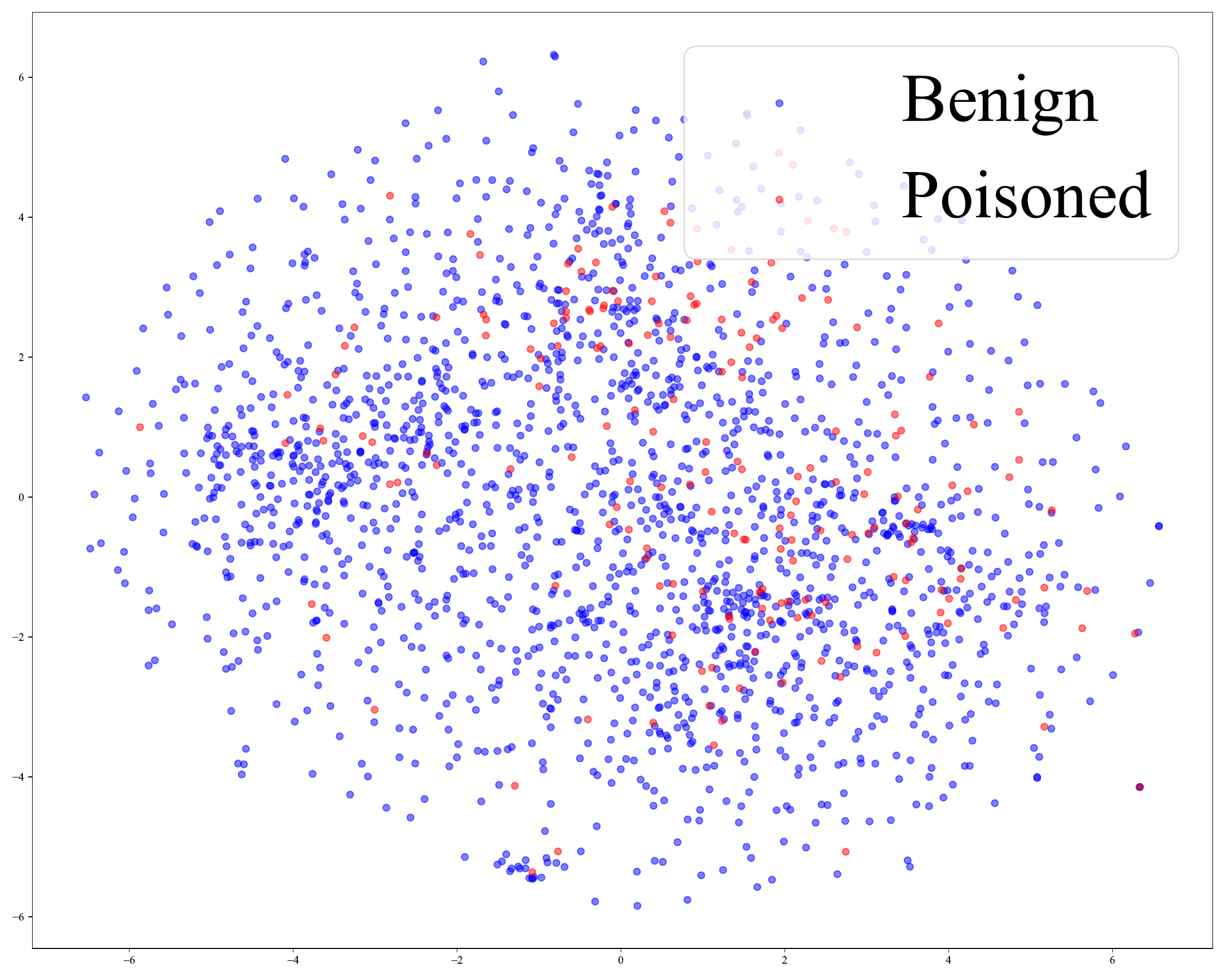}
            \includegraphics[width=1\linewidth]{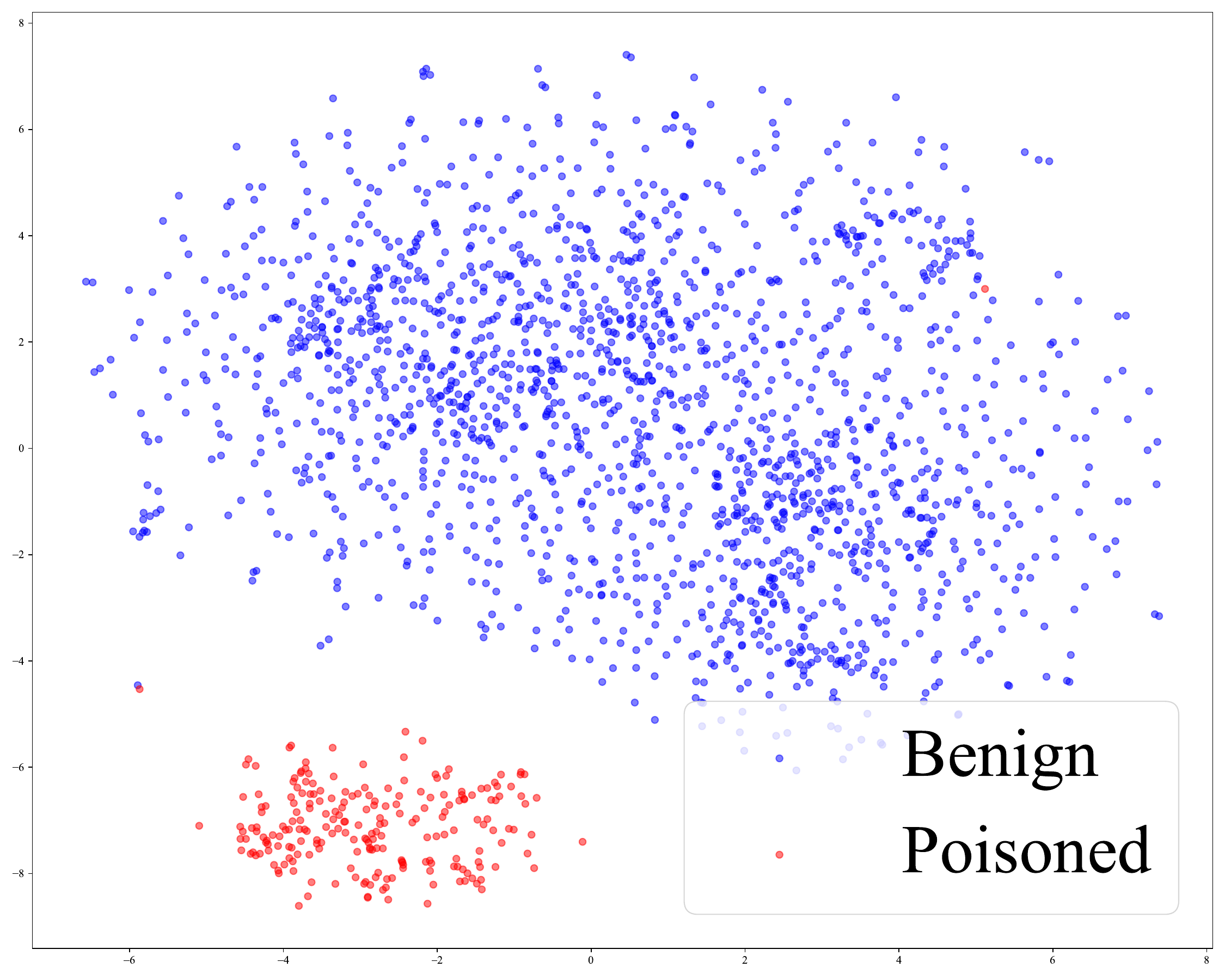}
            \centerline{(b) Layer 6}
    \end{minipage}
     \begin{minipage}{0.193\linewidth}
     \includegraphics[width=1\linewidth]{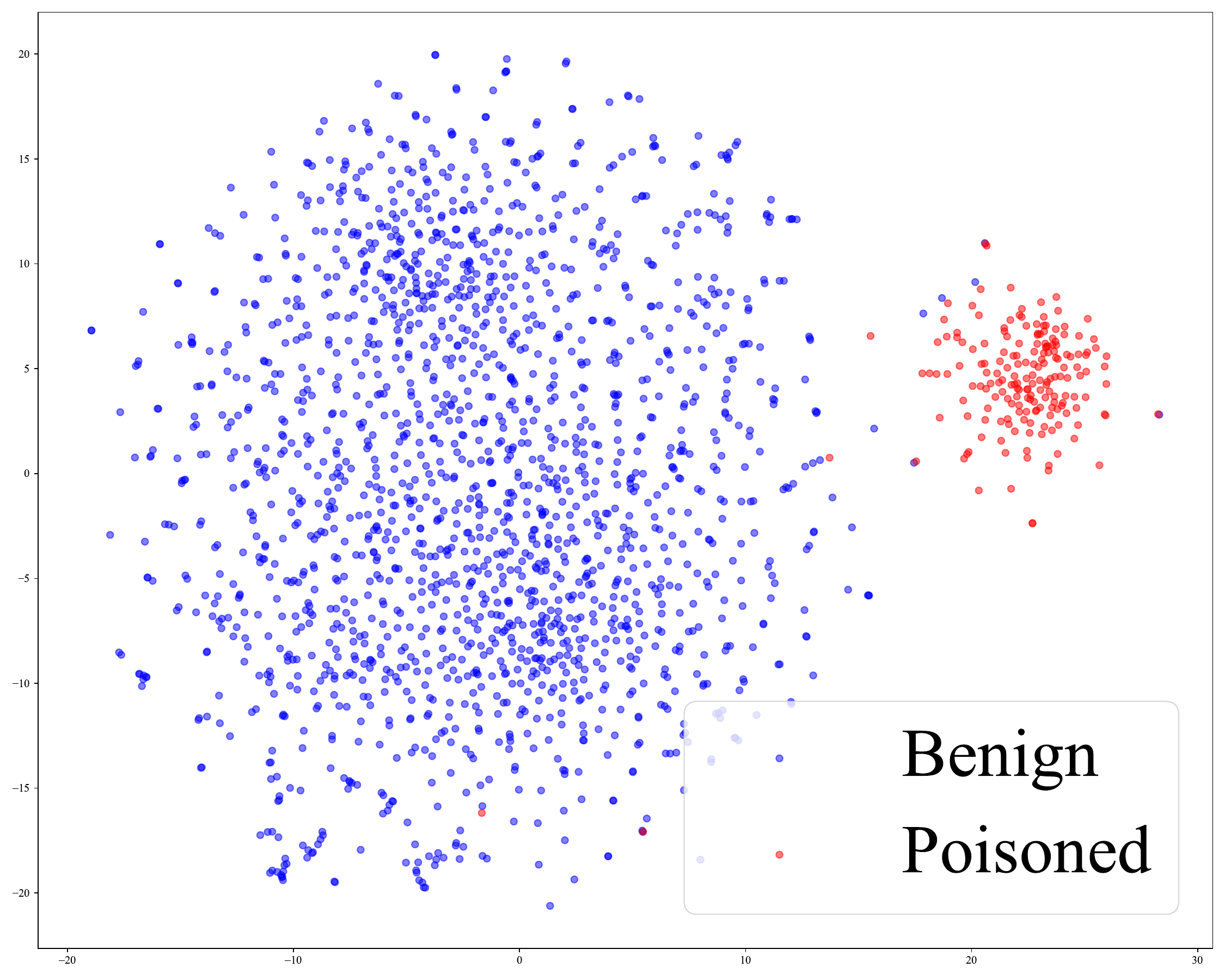}
     \includegraphics[width=1\linewidth]{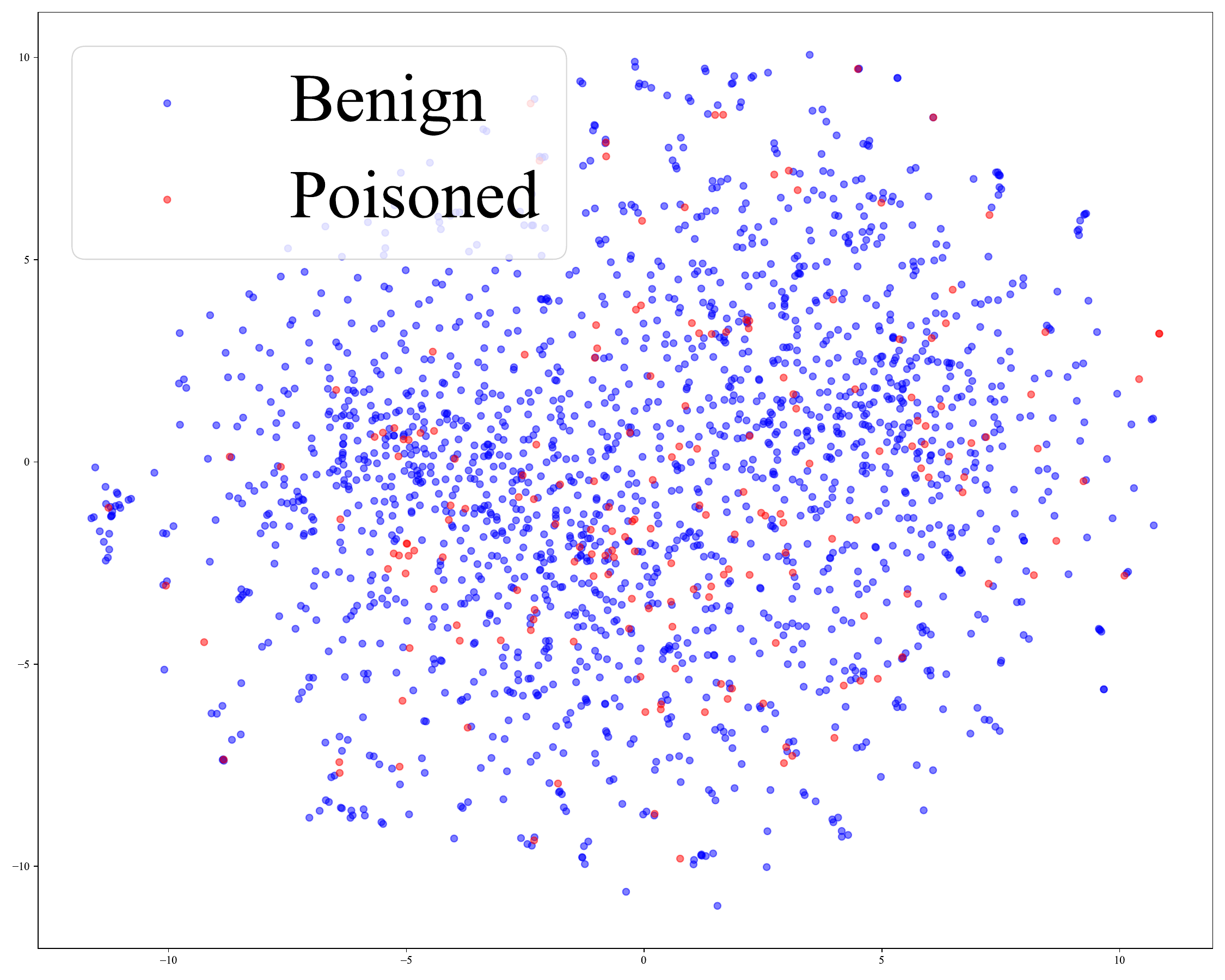}
            \includegraphics[width=1\linewidth]{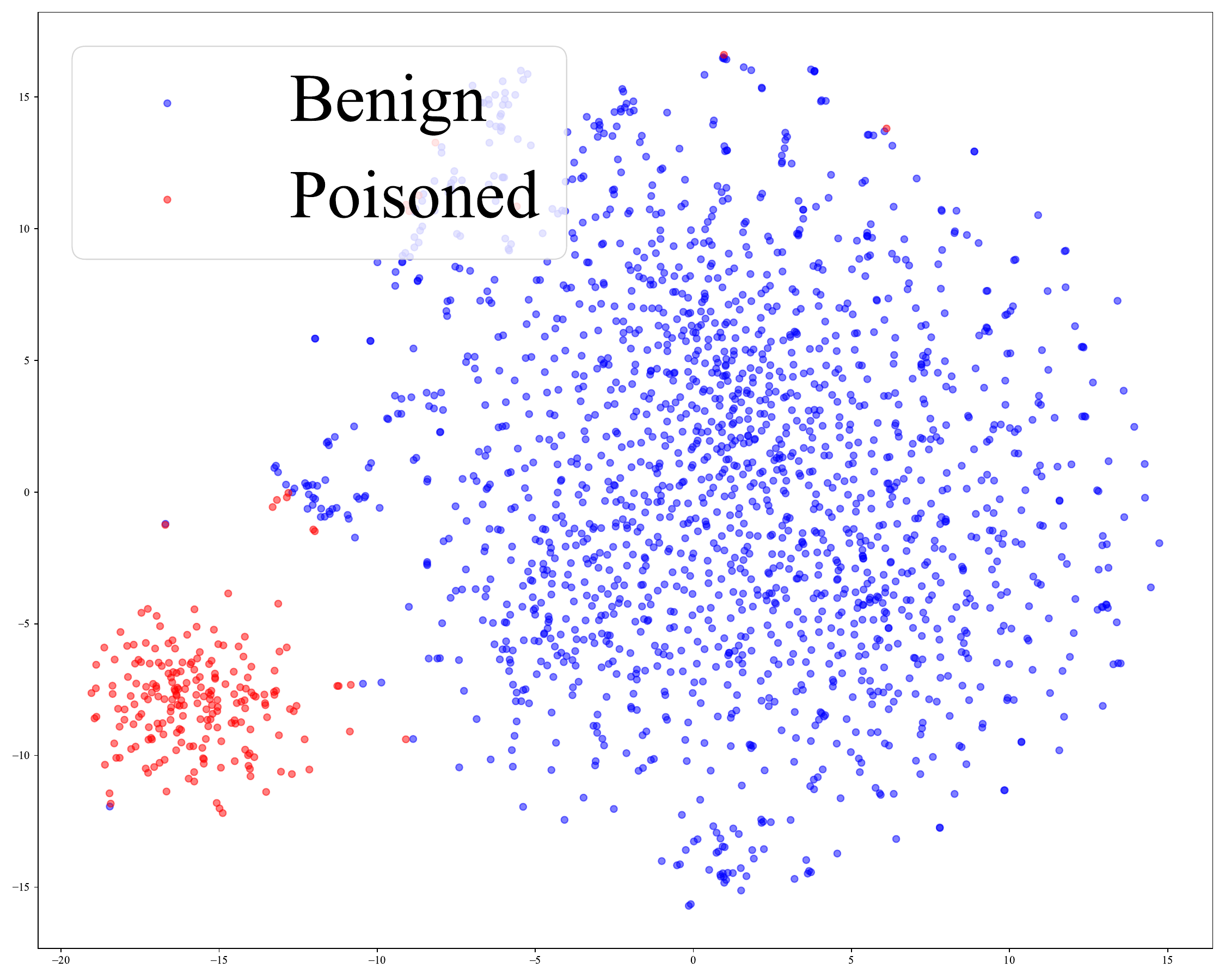}
            \centerline{(c) Layer 10}
    \end{minipage}
     \begin{minipage}{0.193\linewidth}
    \includegraphics[width=1\linewidth]{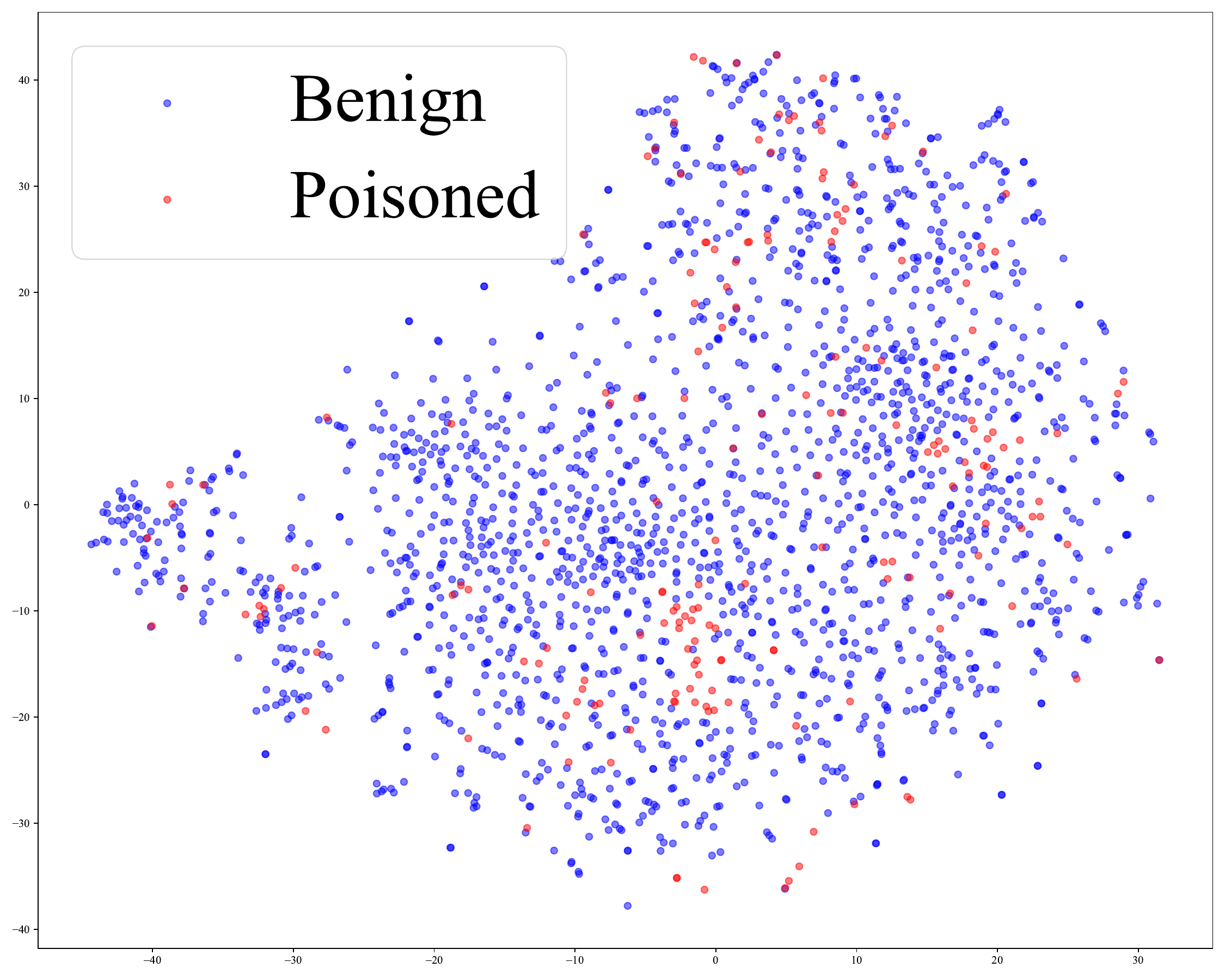}
    \includegraphics[width=1\linewidth]{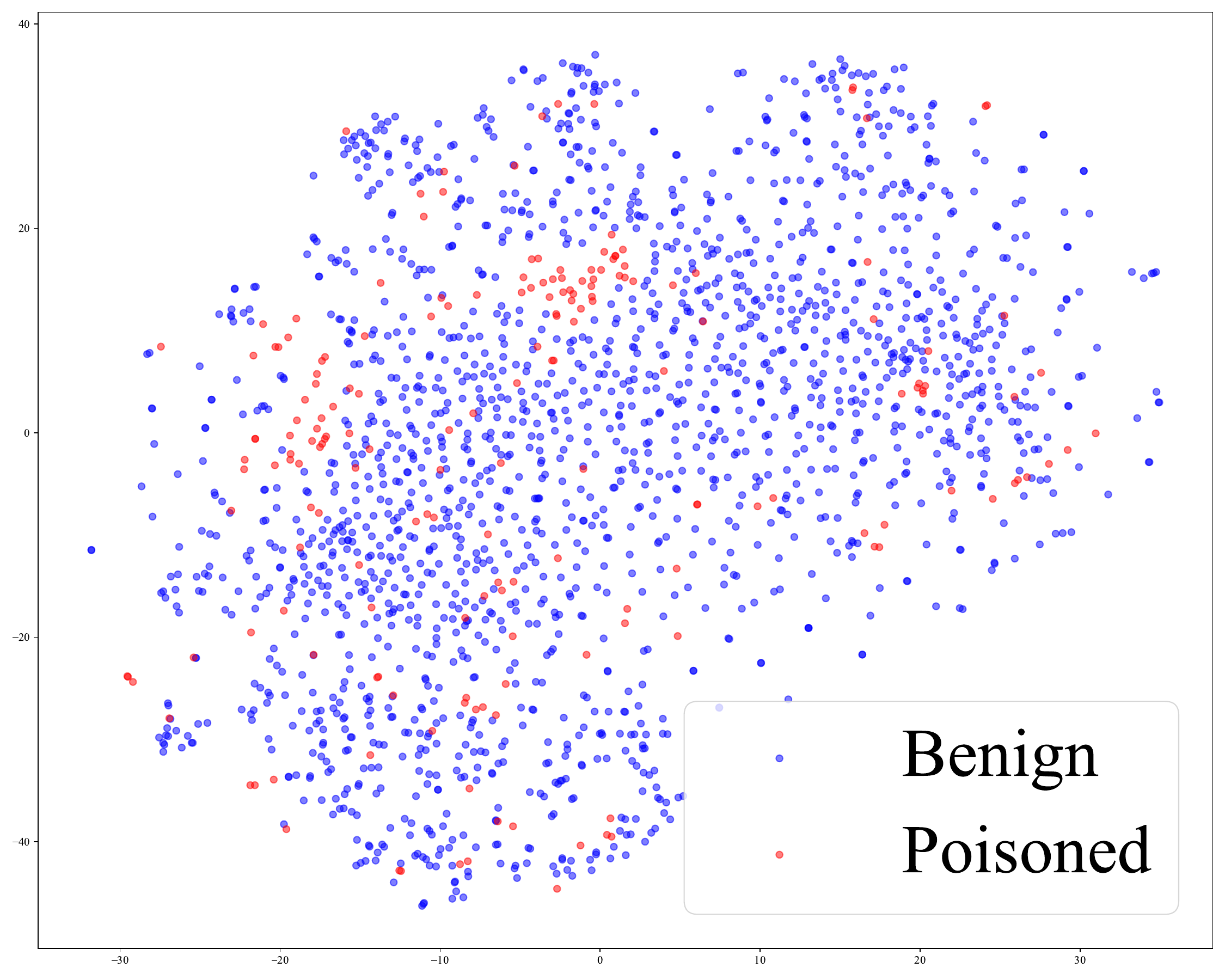}
            \includegraphics[width=1\linewidth]{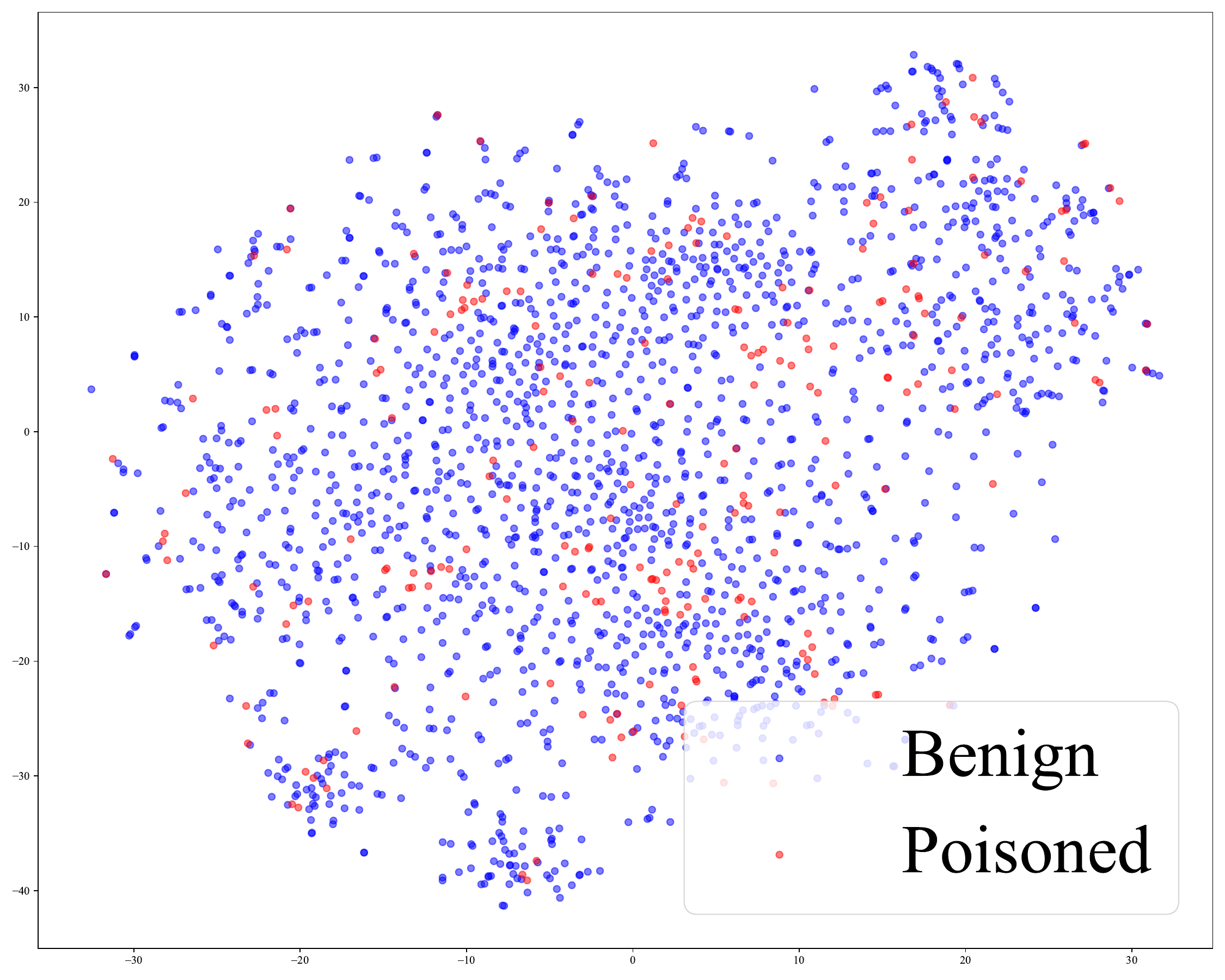}
            \centerline{(d) Layer 14}
    \end{minipage}
    \begin{minipage}{0.193\linewidth}
    \includegraphics[width=1\linewidth]{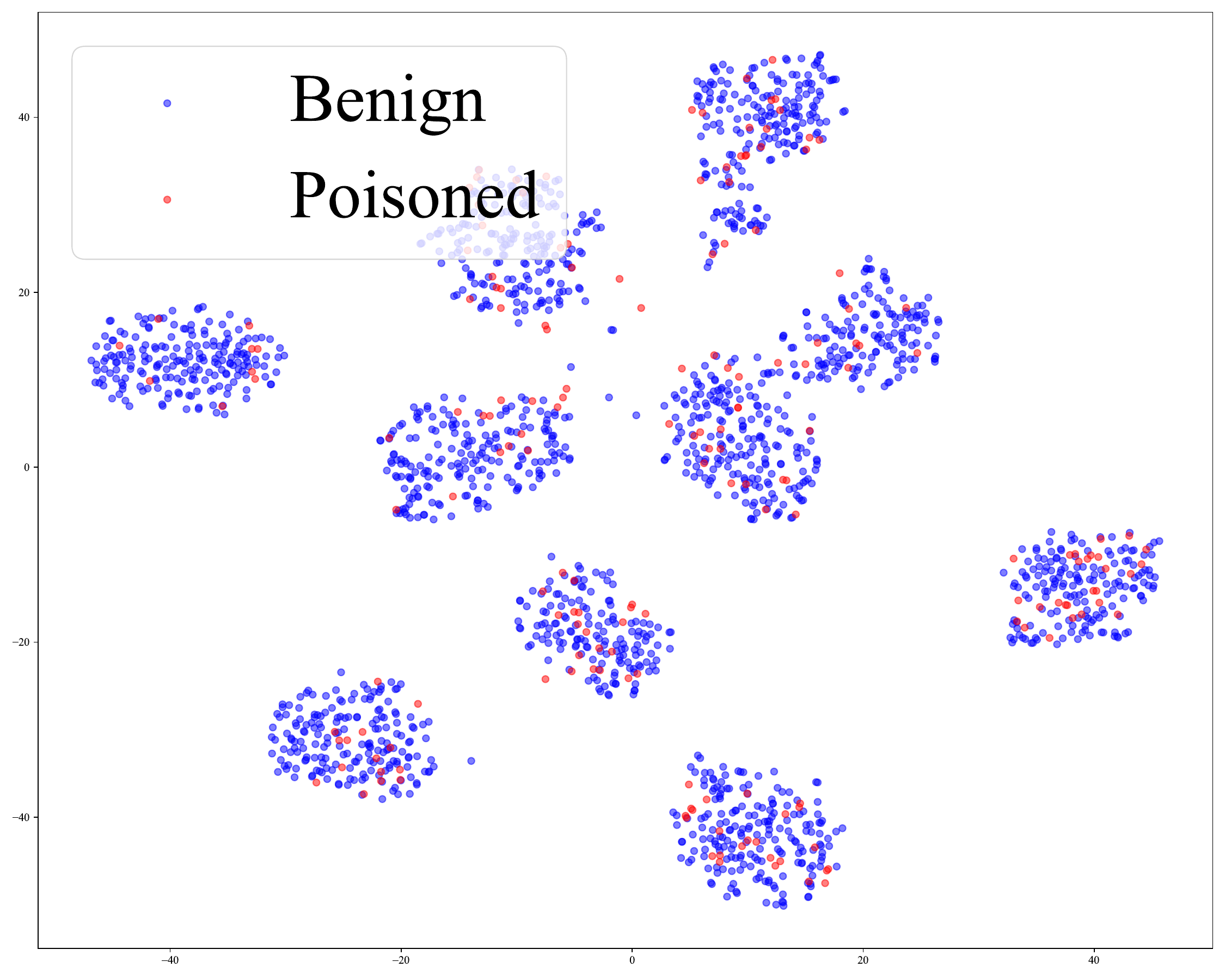}
        \includegraphics[width=1\linewidth]{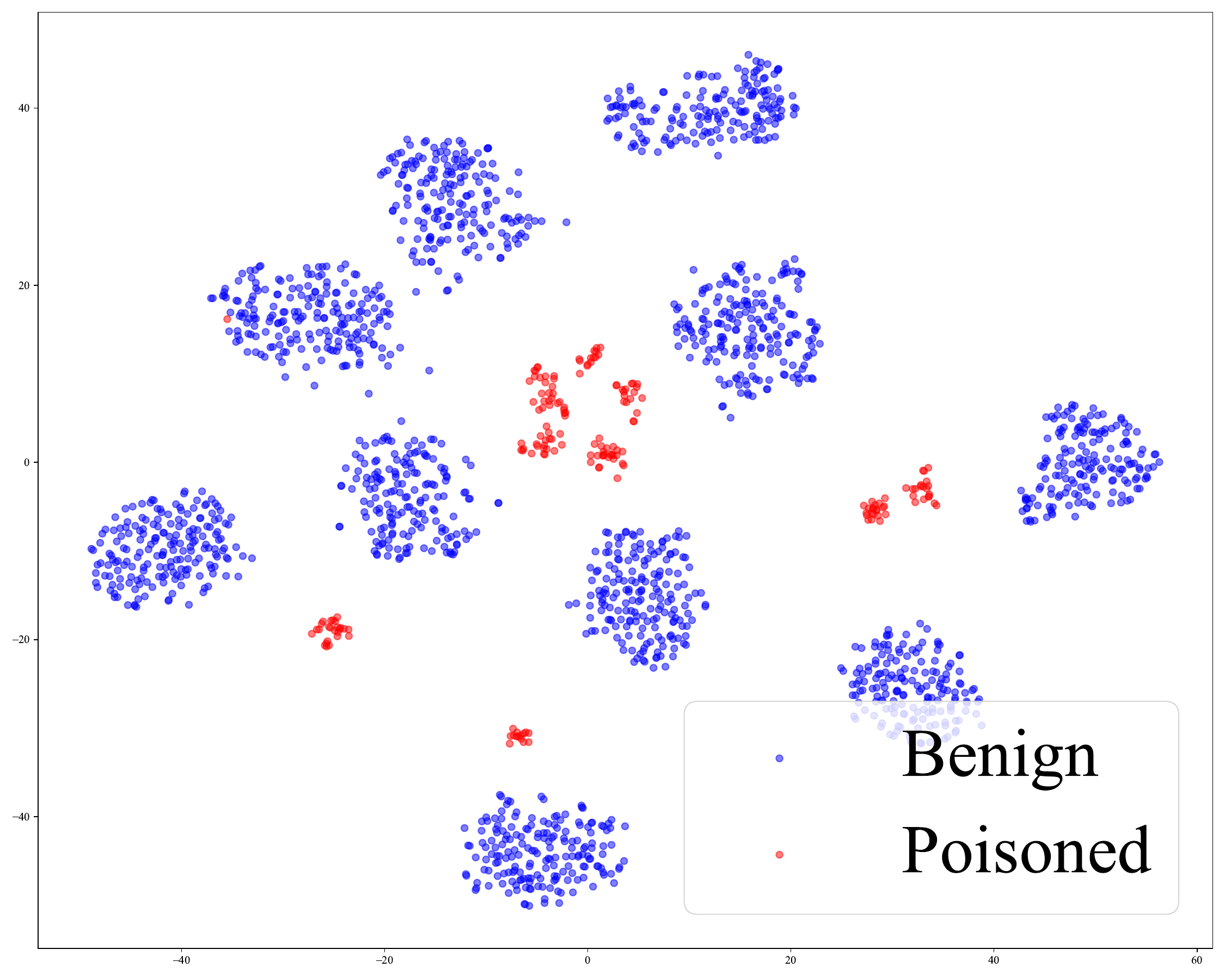}
            \includegraphics[width=1\linewidth]{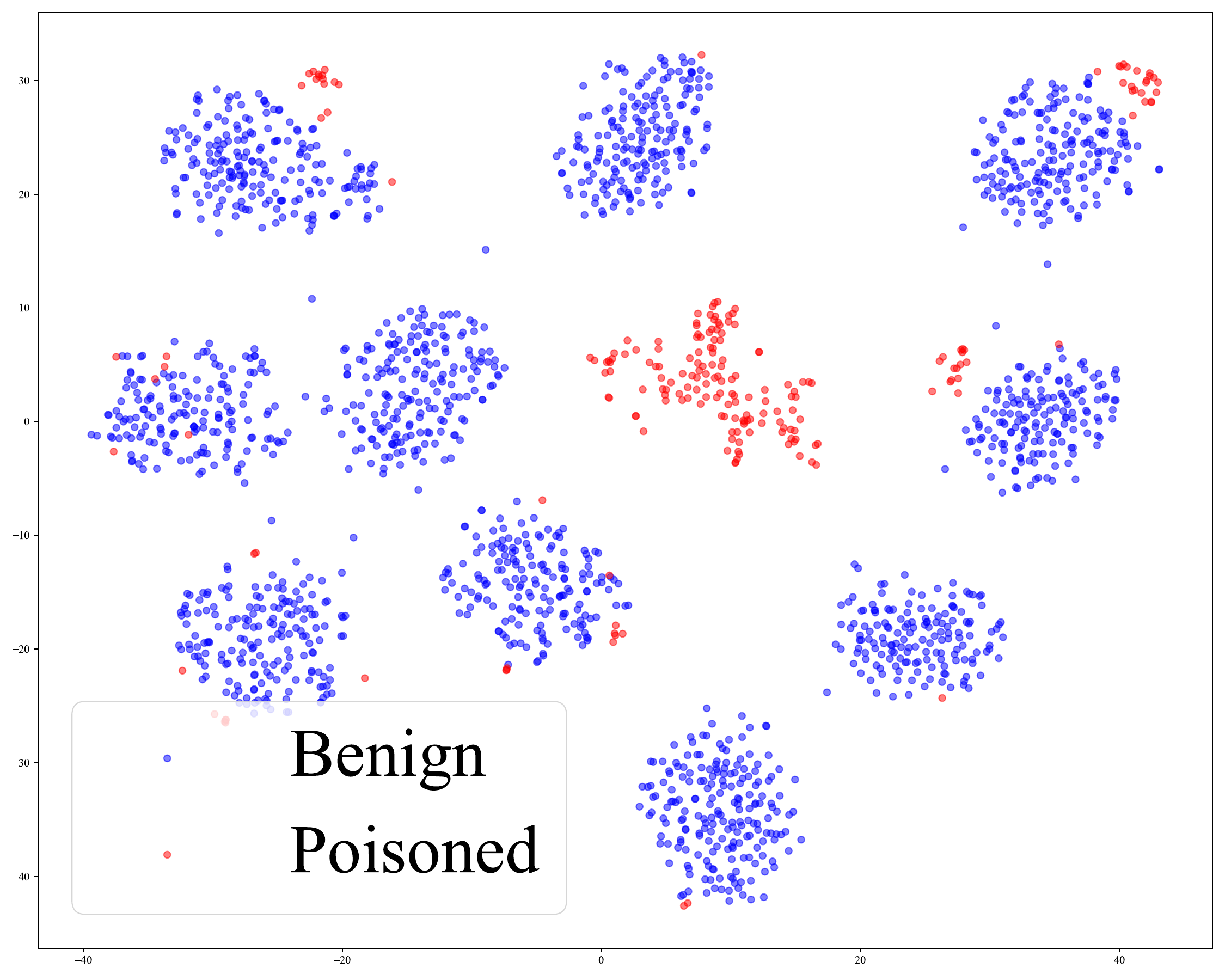}
            \centerline{(e) Layer 18}
    \end{minipage}
   \end{minipage}
\end{minipage}
\vspace{-0.2em}
 \caption{T-SNE visualization of latent representations across different hidden layers on the CIFAR-10 dataset. A total of 2,000 samples are randomly sampled from the training set, including both benign and poisoned samples, with a poisoning rate of 0.1. Each point corresponds to a training sample with poisoned samples marked in red and benign samples in blue.}
 \label{fig:tsne_diff}
 \vspace{-1.5em} 
\end{figure*}

\subsection{Revisiting Strategy 3 (Perturbation Consistency)}
Researchers also observed that poisoned samples exhibit greater prediction consistency than those of benign samples under pixel-level amplification~\cite{guo2023scale,pal2024backdoor} or weight-level alterations~\cite{qi2023towards,hou2024ibd}. This implies that DNNs tend to overfit on triggers instead of learning semantic features.

\begin{figure*}[!t]
 \vspace{-1em}
	\centering
    \includegraphics[width=0.8\linewidth]{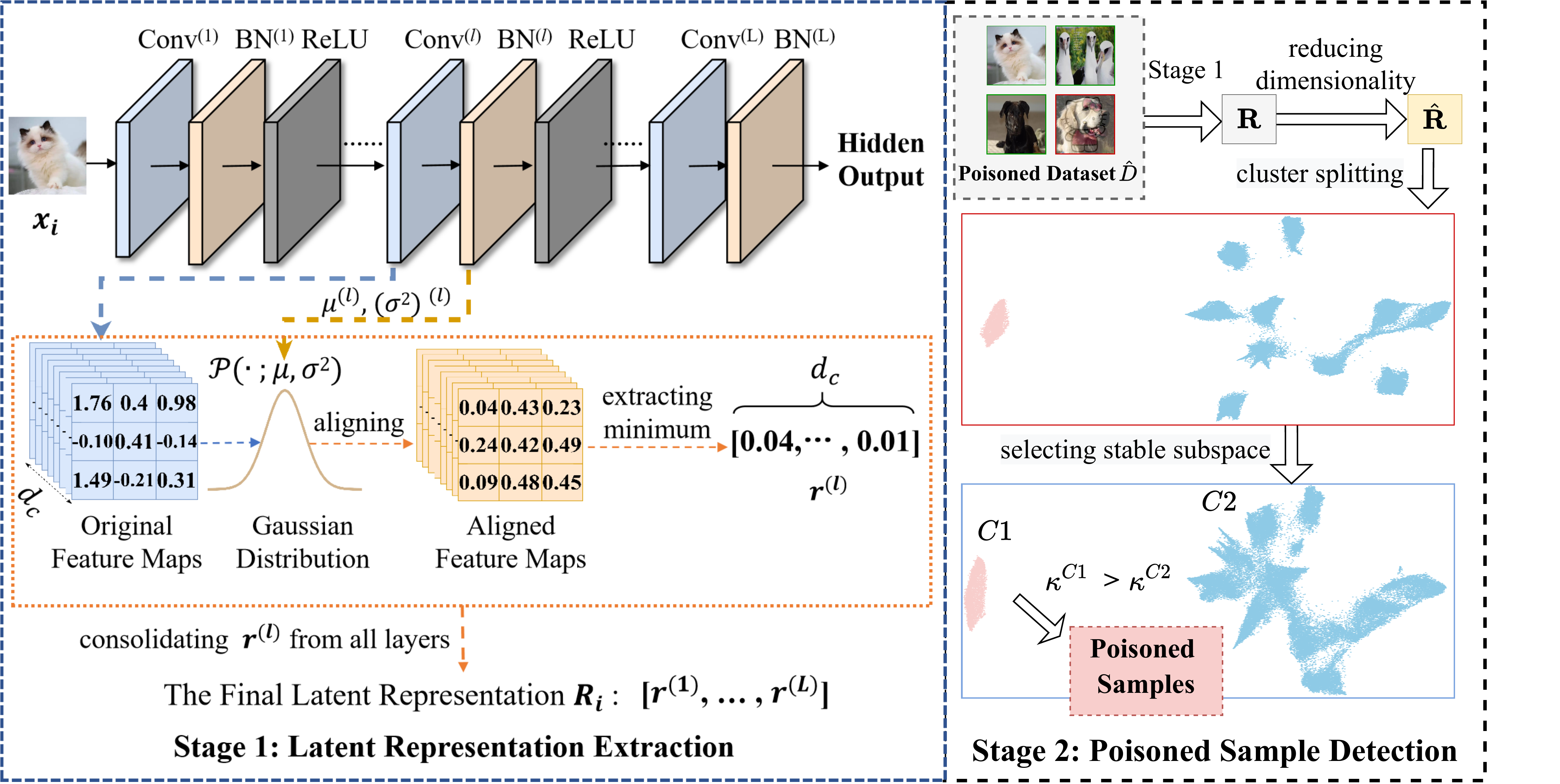}
\vspace{-0.3em}
\caption{The main pipeline of FLARE. \textbf{Stage 1: Latent Representation Extraction}: A backdoored model is trained on a poisoned dataset, and each training sample $x_i$ is forwarded to generate feature maps at hidden layers. The value ranges of each feature map are aligned using the statistics of the corresponding BN layers. An abnormal value is then extracted from each aligned feature map, and these values are aggregated across all hidden layers to form the final representation $\mathbf{R}_i$. \textbf{Stage 2: poisoned samples Detection}: With the representations $\mathbf{R}$ of all training samples, UMAP reduces $\mathbf{R}$ to $\hat{\mathbf{R}}$, followed by density-based clustering to separate samples into two clusters. The cluster with higher stability $\kappa^C$ is identified as poisoned.}
\label{fig:pipeline}
\vspace{-1.5em}
\end{figure*}

\vspace{0.3em} 
\noindent\textbf{Settings.} We hereby also use BadNets and WaNet on CIFAR-10 for our discussions. To evaluate the prediction consistency, we calculate the difference in prediction confidence between the original and perturbed predictions on the initially predicted label for both benign and poisoned samples. Other settings are the same as those used in Section~\ref{sec:local}.

\vspace{0.3em} 
\noindent\textbf{Results.} As shown in Figure~\ref{fig:conf_diff}, for BadNets (A2O), benign and poisoned samples exhibit significantly different confidence difference patterns, making them easily distinguishable. In contrast, under A2A and UT attacks, both benign and poisoned samples display similar confidence difference patterns (\ie, the confidence differences of poisoned samples also approach 1, similar to those of benign samples), indicating that perturbations significantly impact the predictions of poisoned samples. The sensitivity of poisoned samples to perturbations, similar to that of benign samples, indicates that DNNs struggle to overfit to poisoned samples in A2A and UT attacks. Thus, the assumption that backdoor connections are easier to learn than benign ones does not hold for A2A and UT attacks.


\subsection{Revisiting Latent Separability on a Particular Layer}
Previous studies~\cite{chen2018detecting,ma2022beatrix} explored the latent separability between benign and poisoned samples in the feature space. In particular, these works primarily focused on the final hidden layer as a representative intermediate layer. We argue that their success also partly relied on the assumption that backdoor can be easily learned. In this section, we verify it. 

\vspace{0.3em} 
\noindent\textbf{Settings.} We also conduct experiments using BadNets and WaNet on the CIFAR-10 dataset to facilitate our analysis. To assess latent separability, we use t-SNE visualization to examine the latent representations of both benign and poisoned samples across various hidden layers, with particular focus on shallow, middle, and deep layers. We select 2,000 random samples from the training set, including both benign and poisoned ones, with a poisoning rate of 0.1. All other settings are consistent with those described in Section~\ref{sec:local}.

\vspace{0.3em} 
\noindent\textbf{Results.} 
As shown in Figure~\ref{fig:tsne_diff}, poisoned and benign samples do not consistently display separability in specific layers. For example, under the BadNets (A2A) and BadNets (UT) attacks, separability is evident in shallow and middle layers, such as Layer 6 or Layer 10, but diminishes in deeper layers for BadNets (A2A). In contrast, under WaNet (A2A), separability is not evident across most layers. These observations challenge the implicit assumption in existing latent-separability-based purification methods that backdoor triggers are sparsely embedded and primarily affect deeper feature representations, highlighting the necessity of evaluating the broader impact of poisoned samples across multiple layers throughout the model rather than limiting the analysis to a single layer.

\section{The Proposed Method} 
\vspace{-0.6em}
\subsection{Overview}

Motivated by previous findings, we introduce FLARE, a dataset purification method that leverages hidden features across all hidden layers. As illustrated in Figure~\ref{fig:pipeline}, FLARE consists of two main stages: \textbf{(1) Latent Representation Extraction:} For each sample, FLARE constructs a comprehensive latent representation by consolidating the abnormal values from all hidden layers’ feature maps. Specifically, FLARE first aligns the values of all feature maps to a uniform scale using the statistics of Batch Normalization (BN) layer. Then, for each training sample, FLARE extracts abnormally large or small values from each aligned feature map, and aggregates the abnormal values across all hidden layers to construct the final latent representations. \textbf{(2) Poisoned Sample Detection:} FLARE detects poisoned samples through cluster analysis. Specifically, FLARE reduces the dimensionality of each final latent representation and then selects a stable subspace by adaptively excluding representations from the last few hidden layers. FLARE performs cluster analysis in the optimal subspace to split the entire dataset into two clusters and identifies the cluster with higher cluster stability as poisoned.

\subsection{Latent Representation Extraction} 

For each training sample, FLARE forms a comprehensive latent representation by leveraging all hidden features of DNNs. This task encounters two main challenges: \textbf{(C1)} hidden features capture diverse characteristics, leading to significant variability in their values; \textbf{(C2)} hidden features are often high-dimensional and noisy. To tackle these challenges, FLARE employs a two-step process: feature alignment and extracting abnormal features. In the alignment step, FLARE normalizes the feature maps to a uniform scale, based on the statistics of BN layers. Subsequently, FLARE extracts an abnormal value from each feature map as its representative and aggregates them across all hidden layers to form the final latent representation. Their technical details are as follows.



\subsubsection{Feature Alignment}
In this work, we adopt batch normalization (BN) to align all feature maps to a uniform scale. It is mostly because BN transformation was initially proposed to mitigate internal covariate shifts, providing a principled way to stabilize feature distributions. Specifically, for a backdoored DNN model $\mathcal{F}$ consisting of $L$ hidden layers, \ie,
\begin{equation}
     \mathcal{F}= \mathrm{FC}\circ f^{(L)} \circ f^{(L-1)} \circ \cdots f^{(l)} \cdots \circ f^{(2)} \circ f^{(1)},
 \end{equation}
where $f^{(l)}$ denotes the $l$-th hidden layer, consisting of a convolutional layer, a BN layer, and an activation function. Let $\mathbf{a}$ represent the output of a convolutional layer, \ie, $\va \in \mathbb{R}^{d_c \times d_h \times d_w}$, where $d_c$ is the number of feature maps, and $d_h$ and $d_w$ are the height and width of each feature map, respectively. Utilizing the mean $\vmu$ and variance $\sigma$ of the following BN layer, we define a transformation \( \mathcal{P}(\cdot;\vmu, \sigma^2) \) to transform all values in $\va$ to obtain the aligned output $\hat{\va}$ via:

\begin{equation}
    \hat{\va} = \mathcal{P}(\va;\vmu, \sigma^2) = \frac{1}{\sqrt{2\pi(\vsigma^{2})}} \exp\left(-\frac{\va - \vmu}{2\vsigma^{2}}\right).
\end{equation}
In this way, all values in $\hat{\va}$ are from a uniform scale $[0, 1]$.


\subsubsection{Extracting Abnormal Features} As backdoor-related features usually function as dominant features, they tend to induce abnormally large or small activations in the feature map, as partly supported by findings in~\cite{cai2022randomized}. Building on this understanding, we propose to focus on identifying outliers within each feature map, specifically targeting the abnormally small or large values. In particular, all feature maps are normalized and their values adhere to a uniform distribution defined by the statistics of the BN layers after the previous alignment step. At this time, both abnormally large and small values fall outside the central distribution, corresponding to regions with low occurrence probability. This allows us to consistently extract the minimum values from each aligned feature map as indicators of backdoor-related features, without needing to separately select the largest or smallest values. Specifically, we can extract the minimum value (\ie, $r_c^{(l)}$) from the aligned output $\hat{a}^{(l)}$ of the $l$-th hidden layer via:

\begin{equation}
    r_c^{(l)} = \min_{\substack{1 \leq i \leq d_w \\ 1 \leq j \leq d_h}}
 \left( \hat{\va}_{c,i,j}^{(l)} \right),
\end{equation}
where $\hat{\va}^{(l)}\in \mathbb{R}^{d_c \times d_w \times d_h}$ contains $d_c$ feature maps, and $\hat{\va}_{c}^{(l)}$ denotes the \( c \)-th feature map of the $l$-th hidden layer, with \( i \) and \( j \) indexing the width \( d_w \) and height \( d_h \), respectively. The minimums from all $d_c$ feature maps are then consolidated to form the \( l \)-th layer representations, denoted as $\vr^{(l)}$:
\begin{equation}
    \vr^{(l)} = \left[ r_1^{(l)}, r_2^{(l)}, \dots, r_{d_c}^{(l)} \right].
\end{equation}

For a sample $\bm{x}_i$, the minimal values of all hidden layers are consolidated to form the final latent representation $\mathbf{R}_i$:

\begin{equation}
    \mathbf{R}_i = [\vr^{(1)}, \vr^{(2)}, \dots, \vr^{(L)}],
\end{equation}
where $L$ is the number of hidden layers.

\vspace{-0.6em}
\subsection{Poisoned Sample Detection}


After obtaining the final representations of all training samples, FLARE applies dimensionality reduction to reduce the computation consumption and improve clustering efficiency. FLARE then conducts cluster analysis to determine which cluster may contain poisoned samples. Arguably, the most straightforward method is to perform clustering on the entire representation directly. However, benign samples from different classes tend to form multiple clusters due to category-specific features, leading to misidentification. To refine this partitioning, FLARE employs a stable subspace selection algorithm to adaptively exclude representations from the last few hidden layers. This approach allows FLARE to isolate an optimal subspace where benign samples are close together rather than dispersing into multiple small clusters.


\subsubsection{Cluster Splitting}

Let \(\mathbf{R} \in \mathbb{R}^{N \times d}\) represent the latent representations of all \(N\) training samples, with each sample in a \(d\)-dimensional latent space. To enhance clustering efficiency and reduce computational demands, FLARE first performs a dimensionality reduction transformation to obtain:

\begin{equation}
\label{eq:reduce_dim}
    \hat{\mathbf{R}} = \mathcal{T}(\mathbf{R}), \quad \hat{\mathbf{R}} \in \mathbb{R}^{n \times d'}, \quad d' \ll d.
\end{equation}
Here, $\mathcal{T}: \mathbb{R}^d \rightarrow \mathbb{R}^{d'}$ denotes the dimensionality reduction function. In this paper, we choose the uniform manifold approximation and projection (UMAP) algorithm~\cite{mcinnes2018umap} as $\mathcal{T}$ for its ability to reduce high-dimensional data while preserving the topological structure of the original data manifold. FLARE then performs clustering on the reduced representation \(\hat{\mathbf{R}}\): $\mathcal{C} = \mathcal{H}(\hat{\mathbf{R}})$, where \(\mathcal{C} = \{C_1, C_2, \dots, C_k\}\) represents the set of clusters identified by a clustering algorithm \(\mathcal{H}\). In this paper, we use the hierarchical density-based spatial clustering of applications with noise (HDBSCAN) algorithm~\cite{mcinnes2017hdbscan} as $\mathcal{H}$ for its ability to handle clusters of varying shapes and densities. HDBSCAN constructs a hierarchy of clusters represented by a condensed tree $\mathbf{T}$, where clusters are partitioned across varying density levels $\lambda = 1 /d_{core}$, where $d_{core}$ denotes the reachability distance of an object to its `minPts-nearest' neighbor. A higher density level corresponds to a smaller $d_{core}$, indicating that the samples in a cluster are closely packed together. By adjusting the density threshold, HDBSCAN can distinguish between dense poisoned samples and more widely distributed benign samples.



\subsubsection{Stable Subspace Selection} 
Given that poisoned samples typically share the same trigger-related features, they tend to aggregate into a tight cluster. In contrast, benign samples, originating from a more natural distribution, typically form compact clusters within each class while remaining separated across different classes. As a result, poisoned clusters remain stable across varying density levels, while benign clusters tend to fragment or shift as the density threshold increases. This difference makes cluster stability an effective indicator for identifying poisoned samples. Cluster stability measures the ability of a cluster to maintain its integrity as the density level $\lambda$ increases. With an increase in $\lambda$, the reachability distance $d_{core}$ between samples decreases, potentially causing distant samples to be assigned to different clusters. Traditionally, in HDBSCAN, cluster stability is calculated by aggregating the $\lambda$ of each sample within a cluster. However, in poisoned sample detection, where the number of poisoned samples is much smaller than that of benign samples, this approach introduces bias toward the majority class (benign samples). To alleviate this problem, we redefine \textit{cluster stability} to focus on the persistence of a cluster across varying $\lambda$, as follows.
\begin{definition}[Cluster Stability in Detecting Poisoned Samples] \label{def:cluster_stability}
For a cluster \(C\), let \(\lambda_s^C = \min_{x \in C} \lambda_x\) be the density level where \(C\) first appears, and \(\lambda_e^C = \max_{x \in C} \lambda_x\) be the density level before \(C\) divides into sub-clusters. The stability \(\kappa^C\) of cluster \(C\) is defined as \(\kappa^C = \lambda_e^C - \lambda_s^C\).
\end{definition}

While benign samples generally form a single cluster, they often tend to form multiple stable clusters, as illustrated in Figure~\ref{fig:pipeline}. These small but stable benign clusters can significantly interfere with detection performance. To address this issue, FLARE develops a subspace selection strategy to isolate a stable space in which poisoned and benign samples form two distinct clusters. This approach is based on the understanding that models tend to capture semantic information in shallow layers and focus on distinguishing features in deeper layers. Accordingly, FLARE excludes features from the last few hidden layers, obtaining $\mathbf{R}'$ as the final representation that includes only the earlier layers: $\mathbf{R}' = [\mathbf{r}^{(1)}, \mathbf{r}^{(2)}, \dots, \mathbf{r}^{(L-k)}]$.

To determine the optimal number of layers \(k\) to exclude, FLARE designs an adaptive algorithm that dynamically selects a suitable \(k\). The algorithm begins with a model configured with all $L$ hidden layers and progressively removes the last hidden layer of the modified model at each step. At each iteration, FLARE splits the condensed tree $\mathbf{T}$ at the root node to form two primary clusters and assesses the stability $\kappa^{c_{max}}$ of the larger cluster $c_{max}$. If the stability $\kappa^{c_{max}}$ surpasses a predefined threshold \(\xi\), it suggests that the benign samples are cohesively grouped within the current subspace. To further ensure the cluster stability against potential anomalies, FALRE generates a condensed tree at each step and traverses from the root node of $c_{max}$ down to depth $d$. If the stability at any depth exceeds \(\xi\), the current subspace is considered stable. The details of the selection process are provided in Algorithm~\ref{alg:adaptive_layer_selection}.

\subsubsection{Stability-based Detection} 
Within this obtained subspace, FLARE splits the condensed tree $\mathbf{T}$ of the entire poisoned dataset at the root node to form two primary clusters, $C_1$ and $C_2$. FLARE then evaluates the \textit{cluster stability} $\kappa^{C_i}$ for each cluster $C_i$ and identifies the cluster with the higher stability as poisoned. The set of poisoned samples $\mathcal{S}_{p}$ is:

\begin{equation} 
\mathcal{S}_{p} = \{x \mid x \in C_{p}\}, 
\end{equation}
where
\begin{equation}
    C_p = \arg\max\limits_{\{C_1,C_2\}}(\kappa^{C_1}, \kappa^{C_2}).
\end{equation}
The detailed procedure for detecting poisoned training samples is presented in Algorithm~\ref{alg:Stability-based Detection}.


\begin{algorithm}[!t]
\caption{Stable Subspace Selection}
\label{alg:adaptive_layer_selection}
\begin{algorithmic}[1]
\REQUIRE backdoored model $\mathcal{F}$ with $L$ hidden layers; threshold $\xi$; maximum depth $d$.
\ENSURE the stable representation subspace $\mathbf{R}'$.
\FOR{$k = 0$ \textbf{to} $L-1$}
    \STATE extract latent representations $\mathbf{R} \in \mathbb{R}^{N \times d}$ from $\mathcal{F}$;
    \STATE generate reduced representation $\mathbf{R}'$ by excluding the last $k$ layers of $\mathbf{R}$, $ \mathbf{R}' = [\mathbf{r}^1, \mathbf{r}^2, \dots, \mathbf{r}^{L-k}]$;
    \STATE apply dimensionality reduction (\eg, UMAP) to get $\hat{\mathbf{R}}$ as in Eq. 6;
\STATE apply clustering analysis (\eg, HDBSCAN) on $\hat{\mathbf{R}}$ to construct the condensed tree $\mathbf{T}$;
    \STATE initialize queue $Q \leftarrow \{ (n_0,\, 0) \}$ \triangledComment{$n_0$ is the root of $\mathbf{T}$};
    \WHILE{$Q$ is not empty}
        \STATE dequeue $(n,\, h)$ from $Q$;
        \IF{$h \geq d$}
           \STATE \textbf{break}
        \ENDIF
        \STATE $\lambda_n \leftarrow$ the density level of node $n$ in $\mathbf{T}$;
        \IF{node $n$ has children} 
            \STATE choose the child node $c_{\max}$ with the largest size;
            \STATE $\lambda_c \leftarrow$ the density level of node $c_{\max}$ in $\mathbf{T}$;
            \STATE compute the cluster stability $\kappa^{c_{max}} = \lambda_c - \lambda_n$;
            \IF{$\kappa^{c_{max}} > \xi$}
                \STATE \textbf{return} $\mathbf{R}'$;
            \ENDIF
            \STATE enqueue $(c_{\max},\, h + 1)$ into $Q$;
        \ENDIF
    \ENDWHILE
\ENDFOR
\end{algorithmic}
\end{algorithm}

\begin{algorithm}[!t]
\caption{Stability-based Detection}
\label{alg:Stability-based Detection}
\begin{algorithmic}[1]
\REQUIRE the stable representation subspace $\mathbf{R}'$.
\ENSURE the detected poisoned samples $\mathcal{S}_p$.
\STATE apply dimensionality reduction (\eg, UMAP): $\hat{\mathbf{R}} = \mathcal{T}(\mathbf{R}')$ as in Eq. 6;
\STATE cluster $\hat{\mathbf{R}}$ to construct the condensed tree $\mathbf{T}$;
\STATE split $\mathbf{T}$ at the root node into two clusters: $C_1$ and $C_2$;
\STATE compute the stability $\kappa^{C_i}$ for each cluster $C_i$ according to Definition  IV.1;
\STATE identify the poisoned cluster $C_p$ with the higher stability score $\kappa^{C_i}$;
\STATE \textbf{return} the detected poisoned samples $\mathcal{S}_p = \{x \mid x \in C_p\}$;
\end{algorithmic}
\end{algorithm}


\subsection{Post-Detection Strategy 1: Secure Training from Scratch} 

After completing the above detection process, we can remove the detected poisoned samples, denoted as $\hat{\mathcal{D}}_p$, from the training dataset. The remaining samples form a purified dataset, $\hat{\mathcal{D}}_b \triangleq \mathcal{D}- \hat{\mathcal{D}}_p$, which is assumed to contain only benign samples. Defenders can then train a backdoor-free model $\mathcal{M}(\cdot, \theta')$ on $\hat{\mathcal{D}}_b$ using standard training procedures, \ie,

\begin{equation}
\min\limits_{\theta} \sum\limits_{(\vx,y)\in \hat{\gD}_b} \mathcal{L}(\gM(\vx;\theta'),y),
\end{equation}
where $\mathcal{L}(\cdot)$ is the cross-entropy loss function.

\subsection{Post-Detection Strategy 2: Backdoor Removal} 

We can utilize the detected poisoned samples to directly remove backdoors from the model. This process is achieved through a two-step method of unlearning and relearning, ensuring the mitigation of backdoors while preserving the model's performance on benign tasks.

\vspace{0.3em} 
\noindent\textbf{Step 1: Unlearning.} This step aims to eliminate the effect of the trigger by unlearning the identified poisoned samples. Specifically, we aim to maximize the cross-entropy loss for these poisoned samples with respect to their malicious labels. To achieve this, we minimize the negative loss of the backdoored model $\mathcal{F}(\cdot;\theta)$ over the detected poisoned samples $\hat{\mathcal{D}}_p$:
\begin{equation}
    \min\limits_{\theta}\frac{1}{|\hat{\mathcal{D}}_p|} \sum_{(\vx,y) \in \hat{\mathcal{D}}_p} -\mathcal{L}(\mathcal{F}(\vx;\theta),y).
\end{equation}

\vspace{0.3em} 
\noindent\textbf{Step 2: Relearning.}  The unlearning step can effectively mitigate backdoor effects, but it also may result in degraded performance on benign samples. To alleviate this, we perform a relearning step that utilizes the remaining benign samples $\hat{\mathcal{D}}_b \triangleq \mathcal{D}- \hat{\mathcal{D}}_p$. This step fine-tunes the model as follows:
\begin{equation}
    \min\limits_{\theta} \frac{1}{|\hat{\mathcal{D}}_b|} \sum_{(\vx,y) \in \hat{\mathcal{D}}_b} \mathcal{L}(\mathcal{F}(\vx;\theta), y),
\end{equation}
Both unlearning and relearning steps are performed in each epoch to optimize the model’s performance on benign data without reintroducing susceptibility to backdoor triggers.

\section{Experiments}
\label{sec:exp}
\subsection{Main Settings}
\vspace{0.3em} 
\noindent\textbf{Datasets and Models.} We conduct experiments on two classical image classification benchmark datasets, including CIFAR-10~\cite{krizhevsky2009learning} and Tiny-ImageNet~\cite{chrabaszcz2017downsampled}. 
For the primary experiments, we employ ResNet-18~\cite{he2016deep} as the default architecture. 
For ablation studies, we use
VGG19~\cite{simonyan2014very} and MobileNetV2~\cite{sandler2018mobilenetv2}.


\vspace{0.3em} 
\noindent\textbf{Attacks Configurations.} 
To validate the 
generalizability
of our defense, we evaluate it against four representative types of backdoor attacks across three attack modes. These attacks are categorized as follows: \textbf{(1)} sample-agnostic attacks: BadNet~\cite{gu2017badnets}, Blend~\cite{chen2017targeted}, Trojan~\cite{trojan2018}; \textbf{(2)} sample-specific attacks: IAD~\cite{nguyen2020input}, WaNet~\cite{nguyen2021wanet}, ISSBA~\cite{li2021invisible}; \textbf{(3)} clean-label attack: LC~\cite{turner2019label}; 
\textbf{(4)} sparse attack: SIBA~\cite{gao2024backdoor}. The attack modes include the all-to-one (A2O), all-to-all (A2A), and untargeted (UT) paradigms. To comprehensively assess each attack mode, attacks are adapted accordingly. For example, BadNets is evaluated in three formats: BadNets (A2O), BadNets (A2A), and BadNets (UT). These attacks are implemented using the open-source \texttt{BackdoorBox} toolkit~\cite{backdoorbox},
strictly 
following the settings 
outlined in their papers. 
For all 
attacks, we set the target label $y_t$ to 0. Given the complexity of the Tiny-ImageNet dataset, we focus on the most representative attacks, omitting some similar-effect attacks to avoid excessive computational costs. To ensure consistently high attack success rates, we set the poisoning rate to 0.1 and the cover rate to 0.2. 

\begin{table*}[t]
 \vspace{-1.5em}
\centering
\setlength{\tabcolsep}{6.6pt}
 \caption{The detection performance (\%) of FLARE on the CIFAR-10 dataset using the ResNet-18 model. The best results (highest TPRs and lowest FPRs) are in bold, while TPRs $<$ 80\% and FPRs $>$ 20\% are marked in \red{red} as failed cases.}
\vspace{-0.5em}
\label{tab:detection-cifar10}
\begin{tabular}{c|crrrrrrrrrrrr}
\toprule
 \multirow{2}{*}{Attack Mode$\downarrow$}&Defenses$\rightarrow$ & \multicolumn{2}{c}{AC} & \multicolumn{2}{c}{SCALE-UP}  & \multicolumn{2}{c}{MSPC} & \multicolumn{2}{c}{IBD-PSC} & \multicolumn{2}{c}{CT} & \multicolumn{2}{c}{FLARE} \\
\cmidrule(lr){3-4} \cmidrule(lr){5-6} \cmidrule(lr){7-8} \cmidrule(lr){9-10} \cmidrule(lr){11-12} \cmidrule(l){13-14}
& Attacks$\downarrow$     & TPR   & FPR    & TPR   & FPR   & TPR& FPR   & TPR& FPR   & TPR   & FPR   & TPR   & FPR \\ 
\midrule
---&No Poison& ---& 7.31 &---& \red{31.44} & ---& 0.45&---&1.31 &---  & 7.90& ---&  \bf{0.01}\\
\midrule
 \multirow{9}{*}{A2O}&BadNets&99.54& 0.00& \bf{100.00}& 19.44  & 99.06  & 13.14 & \bf{100.00}&5.65  & \bf{100.00}    & 6.74      &  \bf{100.00} & \bf{0.00} \\ 
&Blend & \red{0.00}& 6.08 &  98.54  &  \red{25.76}    & \bf{100.00} &10.05 &\bf{100.00}  & 3.77&  \bf{100.00} &7.28   & \bf{100.00}  & \bf{0.02} \\  
&Trojan & 99.62& 0.09&   \bf{100.00}  & 19.35    &\bf{100.00} &17.89&\bf{100.00} & 7.99& \bf{100.00}& 0.23& \bf{100.00} & \bf{0.00} \\ 
&IAD & 99.46&\bf{0.00}& \bf{100.00} & 15.24    & 98.00  & 11.83 &  \bf{100.00} & 9.59&   \bf{100.00}   & 0.30  &    \bf{100.00} & \bf{0.00} \\ 
&WaNet & 90.36 & \bf{0.00}& \red{26.28}&  \red{30.68}     &    \red{55.12}    &18.59& \bf{99.98} & 8.39& 98.38 & 2.40&   98.14  & \bf{0.00} \\ 
&ISSBA & 99.94& \bf{0.00}& 97.94 &  15.83    & 98.02 &  10.32 &\bf{100.00}  & 2.30 & \bf{100.00}& 4.73& 99.98  & 0.01 \\ 
&SIBA&99.84 &0.07& 92.34&18.06   &\bf{100.00}&10.05&\bf{100.00} &  9.95&\bf{100.00}&0.55 & \bf{100.00}&\bf{0.00}\\
& LC&\red{0.00} &7.40 & \bf{100.00}& \red{31.61}   & \bf{100.00}& 18.47&\bf{100.00} & 0.03& \red{0.00} &  \bf{0.00}&  \bf{100.00} & \bf{0.00} \\
\midrule
\multirow{7}{*}{A2A}&BadNets &\red{0.04} &10.02&  \red{20.38}& 19.89   &\red{11.74} &16.77 & \red{7.56} & 5.82 & \red{4.14} & 2.64 & \bf{100.00} & \bf{0.00} \\ 
&Blend & \red{0.00}&10.02& \red{4.34}&\red{28.00}   &\red{12.86}& 16.76&\red{12.84}&3.42&\red{14.98}&3.30&\bf{99.58}&\bf{0.00} \\ 
&Trojan &\red{0.00} &9.96& \red{28.60}&\red{28.83}   &\red{15.41}& 16.06&\red{26.16}&5.23&\red{15.52}&0.48& \bf{100.00}&\bf{0.00}\\
&WaNet &\red{0.00} &9.96&\red{4.88}&\red{25.41}   &\red{13.06}&16.00 &\red{13.08}&0.52& \red{11.96}& 9.18&\bf{95.76}&\bf{0.00} \\
&IAD &\red{0.00} &9.96& \red{23.12}&14.68   & \red{11.34}& 13.35&\red{35.00}&0.66&\red{5.78}&0.22&\bf{99.98}&\bf{0.00} \\
&ISSBA&\red{0.00} &9.96&\red{33.08}&18.91   &\red{12.46}&14.82&\red{13.08}&0.52&\red{12.46}& 14.82&\bf{99.70}&\bf{0.27} \\
&SIBA &  \red{0.00}& 9.96& \red{26.08}&\red{26.07} &\red{12.10} &11.30 &\red{10.42} & 5.30&\red{0.36} & 3.75 & \bf{99.74} & \bf{0.00}\\
\midrule
\multirow{7}{*}{UT}&BadNets& \red{0.66}& 9.94&  \red{7.60}  & \red{22.99}    & \red{9.42}   & 10.46 &  \red{7.12}& 7.29& \red{0.06}& 0.01& \bf{100.00}  & \bf{0.00} \\ 
&Blend & \red{0.00}&10.03 &\red{16.46}& \red{30.17}&\red{16.23}&15.81&\red{8.82}&10.13&\red{10.48}&14.33& \bf{99.44} & \bf{0.03}\\
&Trojan& \red{0.22}&9.94&\red{10.06}&\red{21.87}&\red{11.06}&8.64&\red{3.02}&0.24&\red{0.24} &1.76 &\bf{100.00} & \bf{0.00}\\
&IAD &\red{0.04} &10.02&\red{9.68}&14.45&\red{8.64}&15.01&\red{8.26}&3.70&\red{2.80}& 4.82& \bf{99.44} & \bf{0.03}\\
&WaNet & \red{0.30}& 9.90&\red{6.04}&\red{26.12}&\red{8.70}&13.32&\red{8.80}&3.61 & \red{6.40}&8.15 &\bf{80.52} & \bf{0.10}\\
 &ISSBA &\red{0.18}&9.53&\red{7.76}&\red{21.37}&\red{9.80}&15.39&\red{8.98}&8.37&\red{2.36}&5.64 & \bf{98.56} & \bf{0.01}\\
 &SIBA&\red{0.24} &9.99 & \red{7.06} & \red{20.89}& \red{8.07}&10.92&\red{6.40} & 6.48 &\red{2.40} &5.04& \bf{97.94} & \bf{0.00}\\
\bottomrule
\end{tabular}
\end{table*}

\begin{table*}[t]
\centering
\setlength{\tabcolsep}{7pt}
\caption{The detection performance (\%) of FLARE on the TinyImageNet dataset using the ResNet-18 model. The best results (highest TPRs and lowest FPRs) are in bold, while TPRs $<$ 80\% and FPRs $>$ 20\% are marked in \red{red} as failed cases.}
\vspace{-0.5em}
\label{tab:detection-tiny}
\begin{tabular}{c|crrrrrrrrrrrr}
\toprule
\multirow{2}{*}{Attack Mode$\downarrow$}&Defenses$\rightarrow$ & \multicolumn{2}{c}{AC} & \multicolumn{2}{c}{SCALE-UP}  & \multicolumn{2}{c}{MSPC} & \multicolumn{2}{c}{IBD-PSC} & \multicolumn{2}{c}{CT} & \multicolumn{2}{c}{FLARE} \\
\cmidrule(lr){3-4} \cmidrule(lr){5-6} \cmidrule(lr){7-8} \cmidrule(lr){9-10} \cmidrule(lr){11-12} \cmidrule(l){13-14}
& Attacks$\downarrow$     & TPR   & FPR    & TPR   & FPR   & TPR& FPR   & TPR& FPR   & TPR   & FPR   & TPR   & FPR \\ 
\midrule
--- & No Poison& --- & 0.49& --- &4.62 & ---& 0.61& ---& 0.05 & ---& 18.65&--- & 2.89\\
\midrule
\multirow{5}{*}{A2O} & BadNets & \red{0.00}&0.15 & 94.61& \red{27.67}& 98.02& \red{40.80}&  98.01& 0.75& \bf{100.00}& 0.65 &\bf{100.00}&  \bf{0.00}\\ 
&Blend &  \red{46.50}&\bf{0.00} &91.68 &4.94 & 95.94 &0.59 &\bf{99.99} &  6.61&94.09&13.96 &99.97&  \bf{0.00} \\ 
&Trojan &\red{0.00} & \bf{0.01}&82.83 & 5.89&\bf{100.00} & 0.65&82.70 & 1.75& \bf{100.00}&11.57 & \bf{100.00}&  \bf{0.01} \\ 
&WaNet & \red{62.18} & \bf{0.00}&  \red{0.00}& 5.25 &\red{0.03}& 0.45 &\bf{96.62} & 0.21&  \red{79.42}& 3.37& 93.68& \bf{0.00}\\ 
\midrule
\multirow{4}{*}{A2A} & BadNets  & \red{0.04}&0.46 & \red{2.58}& 3.81& \red{0.49}& 0.62 &\red{0.15} &0.05 & \red{0.71}& 10.23  &\bf{100.00} &\bf{0.00} \\
&Blend& \red{0.00}&0.50 &\red{0.25} &3.26 & \red{0.20}& 0.10&\red{0.03} &0.26 & \red{0.00} & 1.20 & \bf{99.57} & \bf{0.00}\\
&Trojan  & \red{0.00}&0.49& \red{1.39}&3.85 & \red{0.40}& 0.42&\red{ 0.06} & 0.29&\red{0.00} & 1.20 &\bf{99.99} & \bf{0.01}\\
\midrule
\multirow{4}{*}{UT}&BadNets  &\red{0.45} & 0.35& \red{3.94} &4.47 &  \red{0.61} & 0.63&  \red{0.00}& \bf{0.00}& \red{0.58}& \red{27.54}&\bf{100.00}&  \bf{0.00} \\ 
&Blend &\red{0.54} &\bf{0.00} & \red{1.70}&2.74 & \red{0.00}&  0.40&\red{0.00} &0.21 &  \red{0.04}& 0.30& \bf{100.00} & 0.01 \\ 
&Trojan &\red{9.36} &\bf{0.00} & \red{3.90}&4.60 & \red{0.20}& 0.10&\red{0.00} &0.18 & \red{0.00} &1.00 & \bf{100.00} & \bf{0.00} \\
\bottomrule
\end{tabular}
\vspace{-0.5em}
\end{table*}

\begin{table*}[!t]
\vspace{-1.5em}
\caption{Effectiveness of secure training from scratch (FLARE (P)) on the ``purified" CIFAR-10 dataset. The best results (highest BAs and lowest ASRs) are in bold, while all failed cases (BA drop or ASR $>$ 10\%) are marked in \red{red}.}

\vspace{-0.5em}
\centering 
\setlength{\tabcolsep}{3.4pt} 
\begin{tabular}{c|crr|rr|rrrrrrrrrrrrrr}
\toprule
 \multirow{2}{*}{Attack Mode$\downarrow$}&   Defenses$\rightarrow$& \multicolumn{2}{c|}{Benign Model} & \multicolumn{2}{c|}{No Defense} & 
 \multicolumn{2}{c}{AC} & \multicolumn{2}{c}{SCALE-UP } &\multicolumn{2}{c}{MSPC}  &\multicolumn{2}{c}{IBD-PSC} &\multicolumn{2}{c}{CT} & \multicolumn{2}{c}{FLARE (P)}  \\ 
\cmidrule(lr){3-4} \cmidrule(lr){5-6} \cmidrule(lr){7-8} \cmidrule(lr){9-10} \cmidrule(lr){11-12} \cmidrule(lr){13-14} \cmidrule(lr){15-16} \cmidrule(l){17-18}
  &  Attacks$\downarrow$    & BA   & ASR   & BA   & ASR    & BA   & ASR   & BA& ASR   & BA& ASR   & BA   & ASR  & BA   & ASR & BA   & ASR \\ 
\midrule
\multirow{7}{*}{A2O} &BadNets& 94.16& 0.61& 92.94&100.00 &93.58 &0.81 & 90.04&\bf{0.29} &91.63&  0.61&92.48& 0.46&  92.73 &0.43 &\bf{93.96}&0.63  \\
 & Blend & 94.16&0.38 & 93.65 &100.00 &93.35& \bf{0.00} &89.14& \red{90.74}&84.41& \bf{0.00}&92.91& 0.35&93.36& 0.85&\bf{93.41}& 0.19 \\
 &Trojan  & 94.16&1.61 & 93.04 &100.00   & 93.45&\red{28.24}& 90.10&\bf{0.45} &91.61 & \red{14.06} &  90.63  & 0.46 &\bf{93.49}  &  1.86 &93.43& 1.93\\
 &IAD & 94.16& 5.26&93.80& 99.99 & 93.44 & \red{12.87}  & 88.01& \bf{0.06} &84.78& \red{96.91}&  88.05&  0.21  &93.54&5.89 & \bf{94.05}& 5.86\\
 &WaNet & 94.16&0.71 &92.53  &   97.83  &92.88&0.92 & 89.70& \red{96.44}& 90.13 & \red{94.34} &90.80& \bf{0.15}   & 92.86  &0.95& \bf{94.10}& 0.73\\
 &ISSBA &94.16 & 0.60&93.50 & 100.00 & \bf{93.48}& 0.76   &87.75&0.26 &  84.40& \bf{0.19} & 93.05 &  0.61 &93.09&0.61 & 93.26& 0.76\\
 & SIBA & 94.16& \red{25.10}& 94.33& 98.68 & 93.16&\red{18.82} & 88.76& 6.02 &84.83 &4.88 & 85.88&\bf{0.04} & 92.96& \red{16.86} & \bf{94.15}& \red{23.91}\\
 &LC & 94.16& 1.11& 84.24 & 100.00 & 82.16 &\red{99.93}  &83.01&\bf{0.00} &  83.26&0.15&\bf{84.66} &\bf{0.00} & 84.51&\red{100.00}&84.55&  \bf{0.00} \\ 
  \midrule
\multirow{7}{*}{A2A} &BadNets  & 94.16& 0.48&92.74& 83.40  &82.78 &\red{64.56}& 88.43&\red{62.91}&87.59 & \red{78.19} &  90.46 & \red{62.85}& 91.89& \red{61.86} & \bf{93.53}& \bf{0.80}\\
 &Blend &94.16 & 8.09&93.08 &85.20 & 83.79& \red{85.40}&90.21 &\red{84.25} & 90.01&\red{83.20} &93.10 & \red{76.35}&\bf{93.40} & \red{82.84}& 93.24 & \bf{7.25}\\
 &Trojan  &94.16 & 1.36& 93.85& 91.90 &84.14 &\red{90.59} &90.55& \red{86.35}& 90.43 &\red{89.13} & 92.64& \red{89.14}&93.19& \red{89.35}& \bf{93.26} & \bf{1.73}\\
 &IAD & 94.16&1.69 & 93.48 &91.01 & 83.98&\red{88.65} &92.53 &\red{84.09} & 92.50&\red{87.34} & \bf{93.50}& \red{86.01}& 93.31& \red{88.56}& 93.39 &\bf{1.86} \\
 &WaNet &94.16& 0.71& 92.89 & 87.86 & 83.34 &\red{85.40} &\red{82.10}& \red{82.80}& 90.69 & \red{82.99}& 84.10& \red{82.00} & 92.33& \red{85.63}&\bf{92.79}& \bf{1.07} \\
 &ISSBA  &94.16& 0.53&93.66 &93.15 &84.14 & \red{91.48}&92.26 &\red{90.14} & 90.45 &\red{89.46} & \bf{93.44}& \red{91.01}& 92.20&\red{84.00} & 93.30 &  \bf{4.27}\\
 &SIBA  &94.16 &2.45 &93.74 & 55.59&83.93   & \red{50.10}& 91.78& \red{41.59}&  93.08& \red{56.86}&92.81 &\red{53.11} &\bf{93.40}&\red{52.06}&  93.37 & \bf{3.10}\\
 \midrule
\multirow{7}{*}{UT} &BadNets  &94.16 & 0.50&91.15 &  92.50& 81.70&  \red{92.43}& 85.21&  \red{92.62}& 85.95&  \red{92.10} &89.53&  \red{91.05}&90.89&  \red{90.50} &\bf{93.51}&  \bf{0.56}\\
 &Blend  &94.16& \red{69.55}& 93.41 & 80.82&\red{83.36} &\red{84.32} &88.69&\red{88.09} & 85.39&\red{85.70} &85.59&\red{87.59} & 92.15& \red{87.17}& \bf{93.30}& \red{\bf{63.59}}\\
 &Trojan  &94.16& 5.81&91.39& 91.97& 83.19&\red{88.46} & 89.74&\red{87.56} & 82.40& \red{88.93}&92.55 &\red{87.02} &92.53 &\red{88.48}&  \bf{93.36}& \bf{6.12} \\
 &IAD  &94.16 & \red{21.01}&92.06 & 90.81& 83.10&\red{88.00} & 90.58&\red{89.13} & 85.20&\red{87.60} &92.56 &\red{87.64} & 92.45& \red{86.99}&  \bf{93.40} & \bf{\red{19.96}}\\
 &WaNet& 94.16& 2.47&92.08 &  75.55 & 82.99&\red{72.63} &  84.22& \red{76.10}& 90.56&\red{74.35} & \red{81.02}&\red{73.40} &\bf{91.05}&\red{75.21} &90.05& \bf{1.82}\\
 &ISSBA &94.16&0.24 &91.94  & 80.65& 89.35&\red{86.96} &89.83 & \red{80.77}&83.11 &\red{88.62} &92.14 &\red{87.28} &92.47 & \red{86.26}&  \bf{92.83}& \bf{0.16}\\
 &SIBA  &94.16& \red{51.00}& 92.23& 90.66& 83.39&\red{88.01} &90.39 &\red{91.81} &84.20&\red{90.84} &92.25 & \red{89.04}& 92.71 & \red{91.95} & \bf{93.25} & \red{\bf{52.15}}\\
\bottomrule
\end{tabular}
\label{tab:defense_scratch}
\end{table*}

\begin{table*}[!t]
\caption{Effectiveness of secure training from scratch (FLARE (P)) on the ``purified'' Tiny-ImageNet dataset. The best results (highest BAs and lowest ASRs) are in bold, while all failed cases (BA drop or ASR $>$ 10\%) are marked in \red{red}.}

\vspace{-0.5em}
\centering 
\setlength{\tabcolsep}{3.4pt}
\begin{tabular}{c|crr|rr|rrrrrrrrrrrrrr}
\toprule
\multirow{2}{*}{Attack Mode$\downarrow$}&    Defenses$\rightarrow$& \multicolumn{2}{c|}{Benign Model}& \multicolumn{2}{c|}{No Defense} & 
 \multicolumn{2}{c}{AC} & \multicolumn{2}{c}{SCALE-UP } &\multicolumn{2}{c}{MSPC}  &\multicolumn{2}{c}{IBD-PSC} &\multicolumn{2}{c}{CT} & \multicolumn{2}{c}{FLARE (P)}  \\ 
\cmidrule(lr){3-4} \cmidrule(lr){5-6} \cmidrule(lr){7-8} \cmidrule(lr){9-10} \cmidrule(lr){11-12} \cmidrule(lr){13-14} \cmidrule(lr){15-16} \cmidrule(l){17-18}
 &    Attacks$\downarrow$     & BA   & ASR    & BA   & ASR   & BA& ASR   & BA& ASR   & BA   & ASR  & BA   & ASR & BA   & ASR & BA   & ASR\\ 
\midrule
\multirow{5}{*}{A2O} & BadNets& 67.75& 0.01&   62.10&96.52 &\red{42.00} & \red{97.25} &65.17& 2.57&59.76 & 0.41 &\bf{68.25} & 0.13& 66.69&1.96&67.88&\bf{0.02}\\
&Blend&  67.75& 0.07&66.91 & 100.00 & 68.41& \red{80.05}& \bf{70.86}&\red{99.37}& 68.96&\red{99.45}&63.92&2.09& 67.97&\red{100.00}&66.50 &\bf{0.15}\\
&Trojan & 67.75& 0.00& 66.06& 100.00 &  65.00& \red{100.00}& 70.22&\red{99.55}& \bf{70.75} & 0.69 & 70.15& \red{99.57}&68.82&\bf{0.00}& 70.18& \bf{0.00}\\
&WaNet & 67.75& 0.03&62.12 & 99.42& 68.02 & \red{97.67}&66.52 &\red{98.29}& 66.27&\red{98.97}&68.45& 2.52&65.61 &\red{68.56}& \bf{68.79} & \bf{0.05}\\
\midrule
\multirow{3}{*}{A2A}&BadNets  & 67.75& 1.15&68.01 &   22.93 & 69.35& \red{13.84} & 64.37 &\red{10.73} & 68.33&\red{20.00}& \bf{69.73}& \red{13.87}&68.30 &\red{19.20}& 65.53& \bf{1.07}\\
&Blend & 67.75& 0.77&69.54 &  48.67 &  69.44& \red{35.43} & 65.13&\red{32.80} &67.00 &\red{34.27} &66.34 &\red{32.11} & 65.49 &\red{36.20} & \bf{70.48} &\bf{0.73}\\
&Trojan &  67.75& 0.65& 68.20 & 29.26&  \bf{69.47} & \red{16.12}& 68.99& \red{14.97} &65.33  &\red{28.10} & 64.69& \red{24.46}&65.20 &\red{28.08} &68.79 & \bf{0.73}\\
\midrule
\multirow{2}{*}{UT}&BadNets & 67.75 & 2.77 & 66.36&  91.20 &  68.36&  \red{91.10}&62.62&  \red{92.05} &69.71&  \red{91.10}&68.64&  \red{91.00}&60.21 &  \red{92.10}& \bf{70.74} & \bf{2.29}\\
&Blend & 67.75& \red{58.38}& 70.01&  99.26&  \bf{70.79}&  \red{98.13}& 70.28 &  \red{98.64}&66.79 & \red{98.60}&68.39 &\red{99.11} & 68.20&\red{98.10} & 69.08 &\bf{\red{49.54}}\\
\bottomrule
\end{tabular}
\label{tab:defense_scratch_tiny}
\vspace{-1em}
\end{table*}

\begin{figure*}[!t]
	\centering
 \begin{minipage}{0.90\linewidth}
    \begin{minipage}{0.196\linewidth}
            \includegraphics[width=1\linewidth]{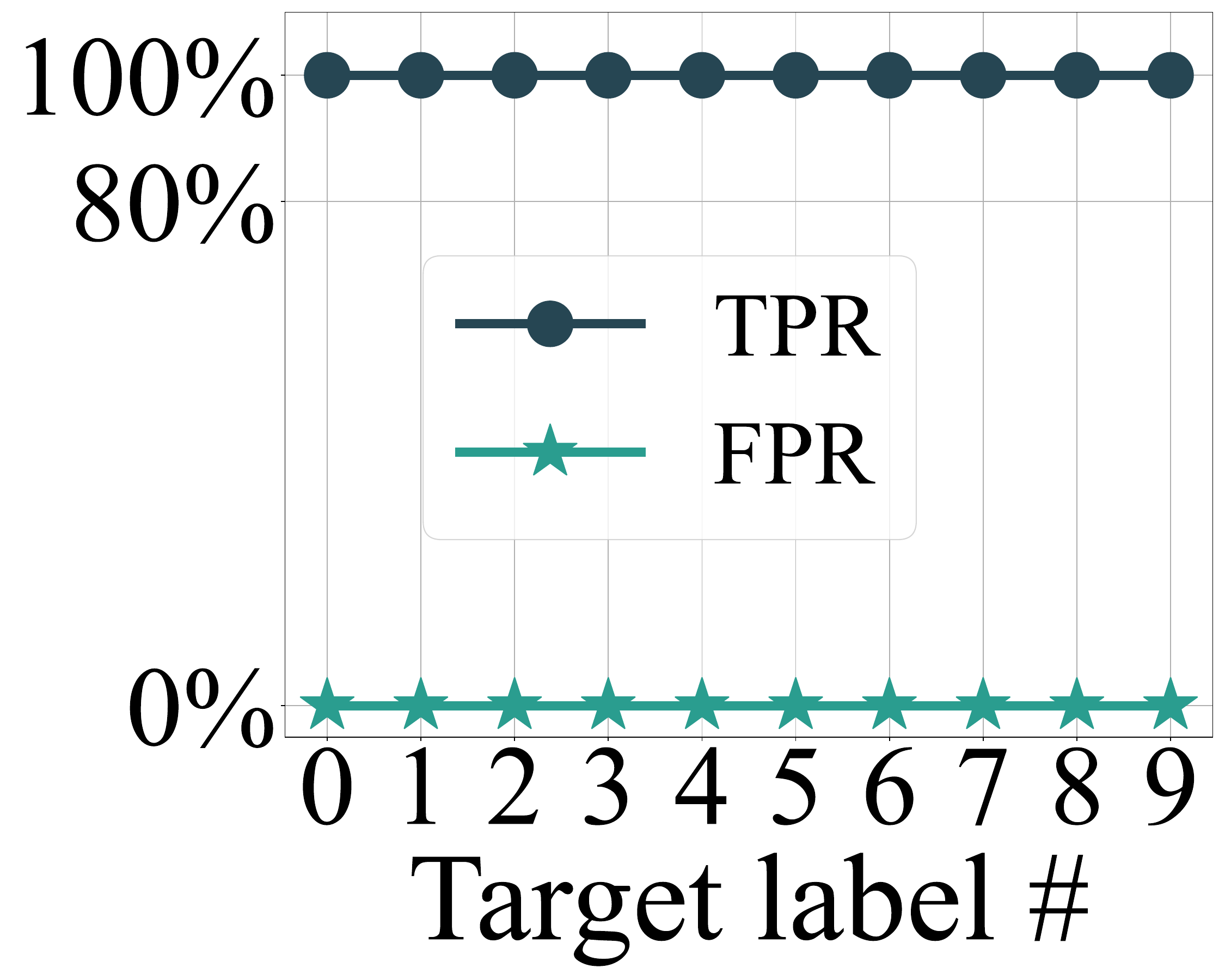}
            \centerline{(a) BadNets (A2O)}
    \end{minipage}
     \begin{minipage}{0.196\linewidth}
            \includegraphics[width=1\linewidth]{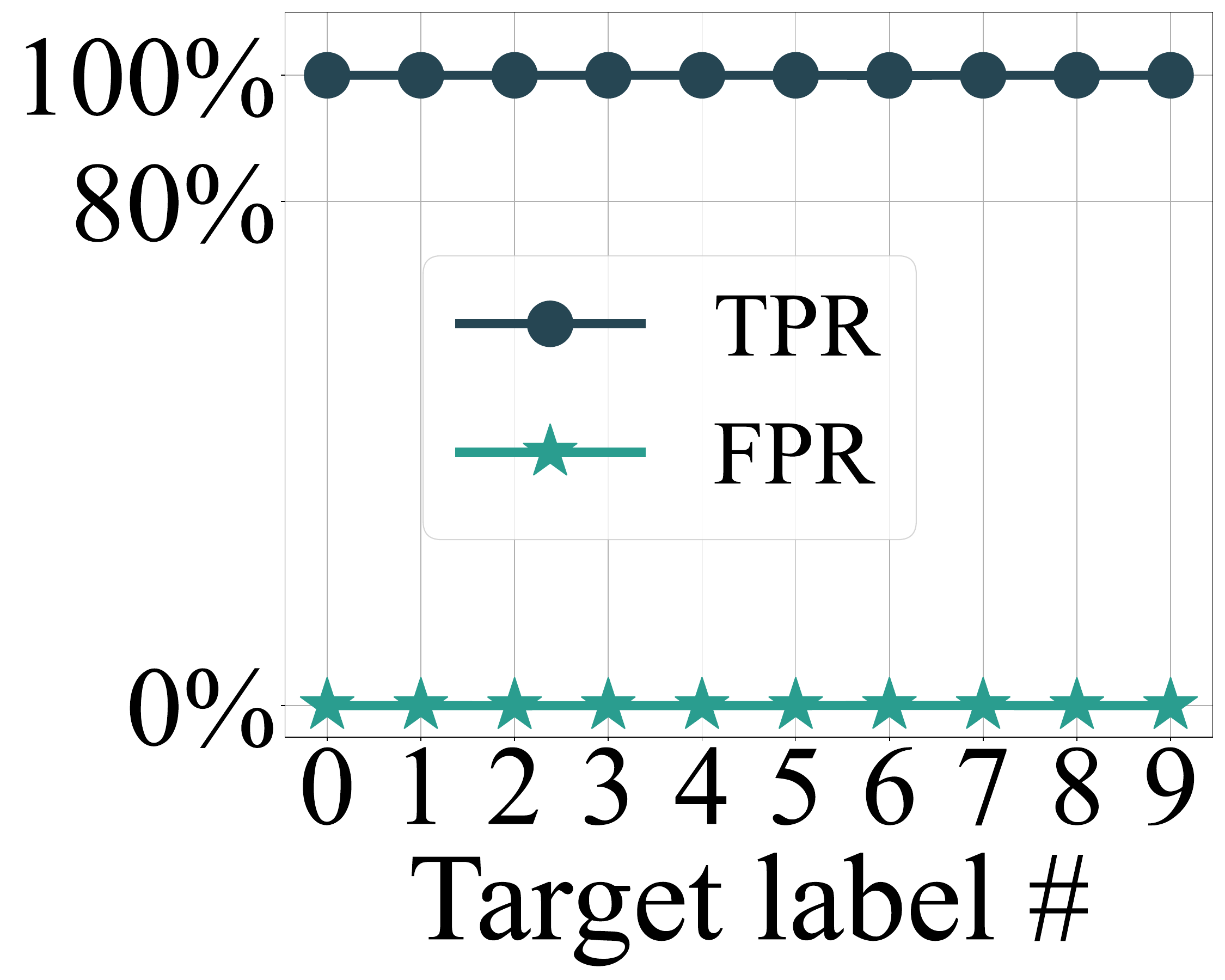}
            \centerline{(b) Blend (A2O)}
    \end{minipage}
    \begin{minipage}{0.196\linewidth}
            \includegraphics[width=1\linewidth]{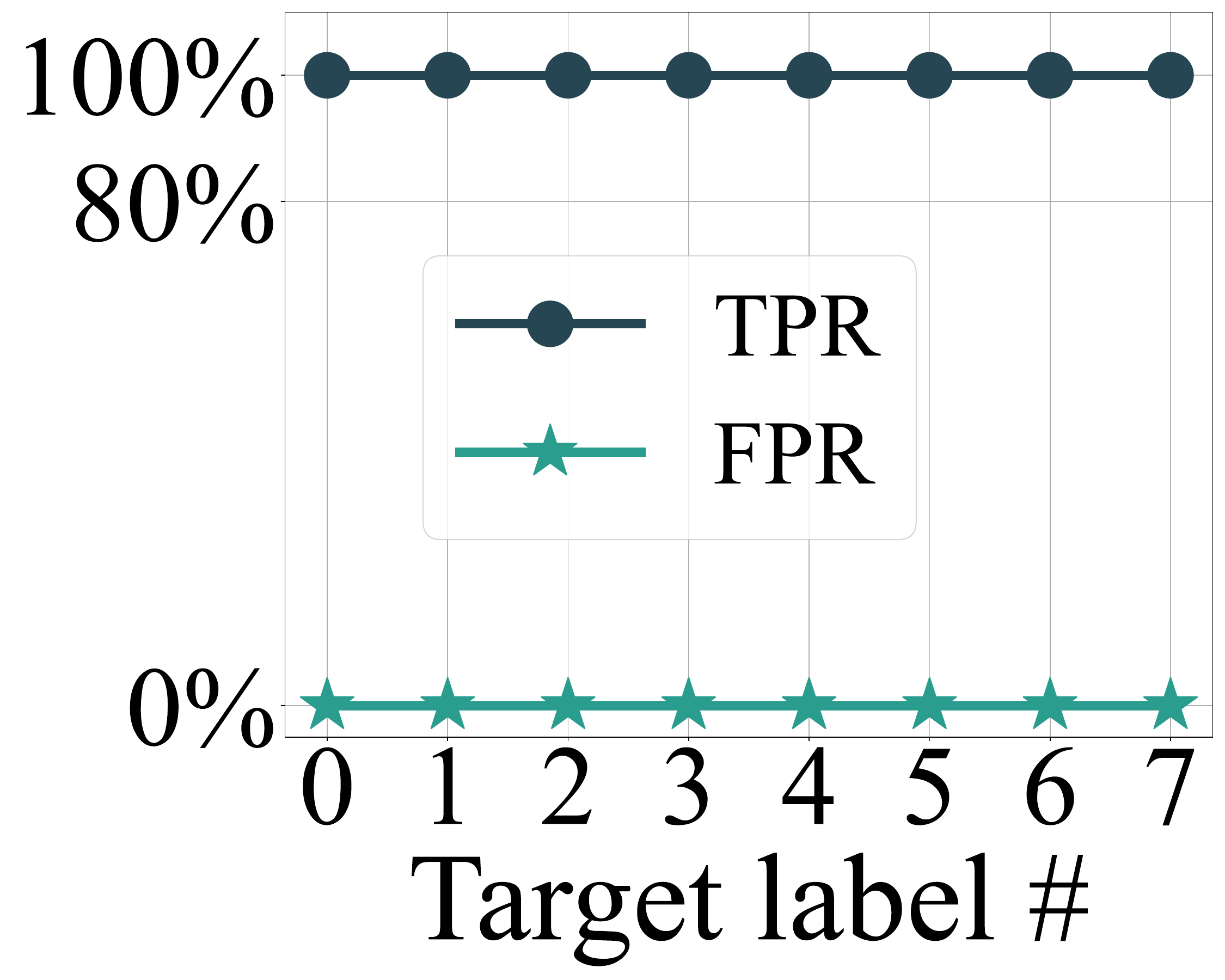}
            \centerline{(c) LC }
    \end{minipage}
    \begin{minipage}{0.196\linewidth}
            \includegraphics[width=1\linewidth]{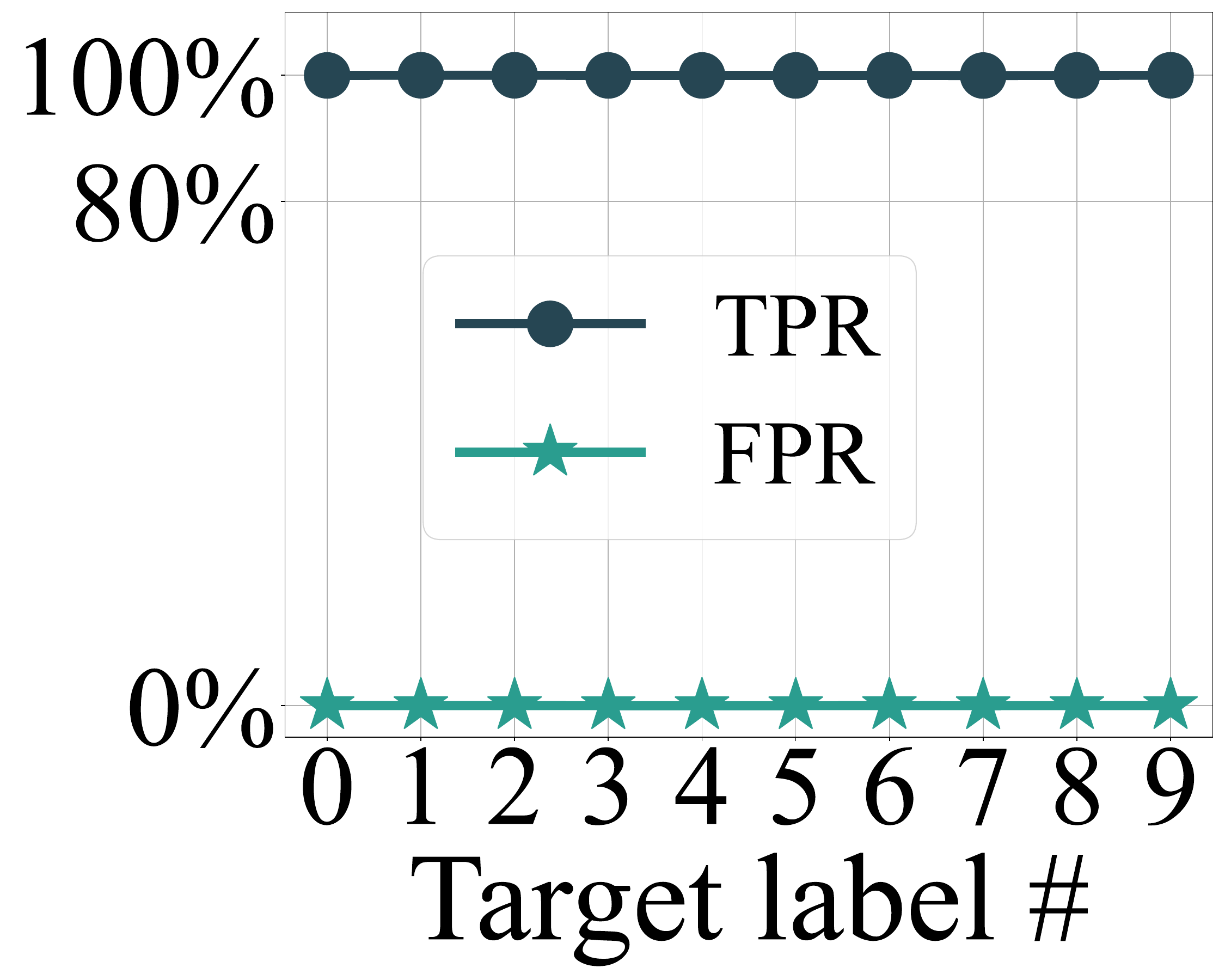}
            \centerline{(d) ISSBA (A2O)}
    \end{minipage}
    \begin{minipage}{0.196\linewidth}
            \includegraphics[width=1\linewidth]{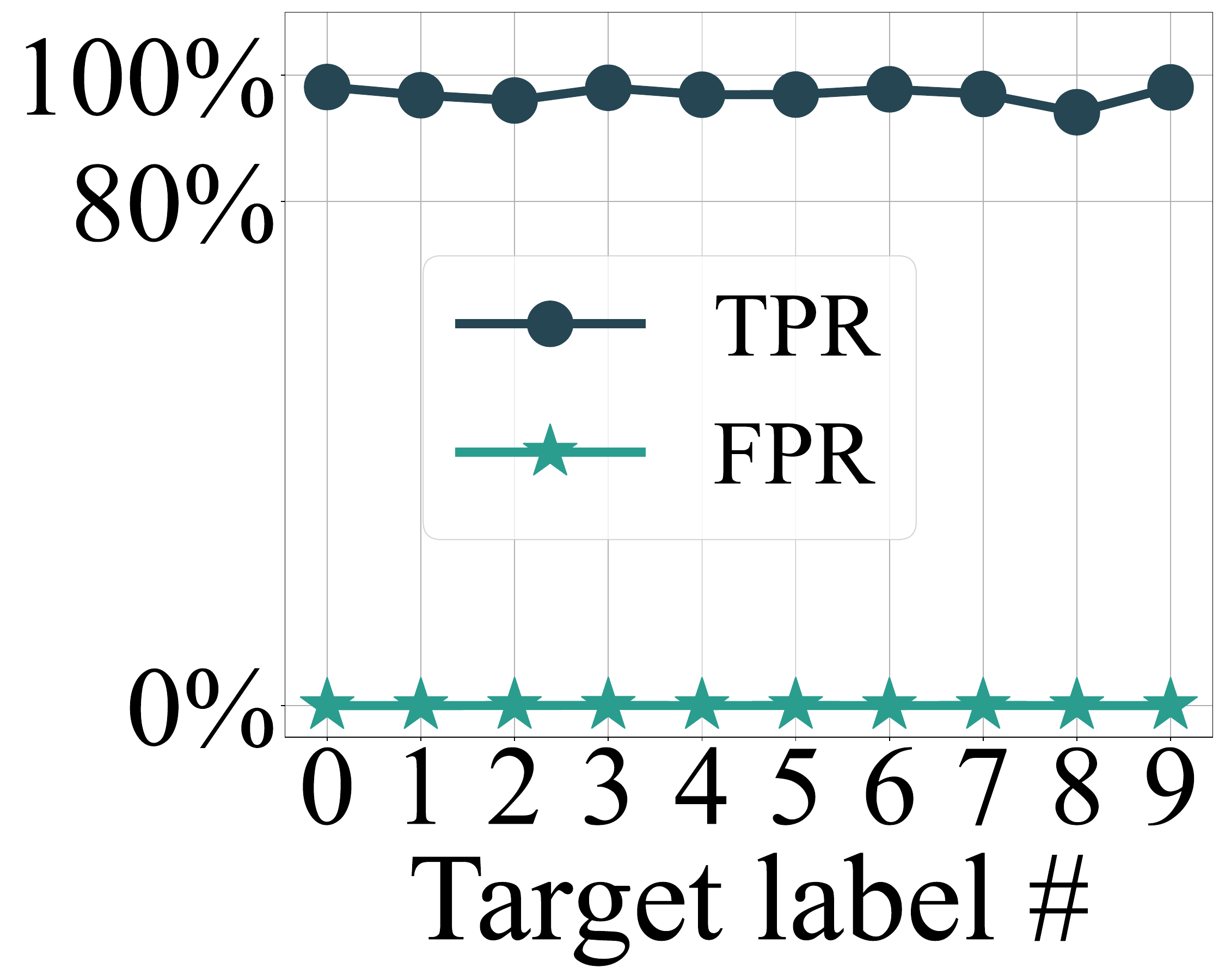}
            \centerline{(e) WaNet (A2O)}
    \end{minipage}
\end{minipage}
 \caption{Performance of our defense across different target labels of CIFAR-10. The consistently high TPR and low FPR across various target labels indicate stable and effective defense performance.}
 \label{fig:targets}
  \vspace{-1em}
\end{figure*}

\vspace{0.3em} 
\noindent\textbf{Evaluation Metrics.} 
We 
measure the 
detection performance using true positive rate (TPR) and false positive rate (FPR). TPR reflects the proportion of correctly identified poisoned samples, while FPR represents the proportion of benign samples incorrectly classified as poisoned. Higher TPR and lower FPR indicate more effective detection. In addition to the detection metrics, we also assess 
benign accuracy (BA) and attack success rate (ASR) after purification. 
Higher BA and lower ASR indicate that the model's primary functionality is preserved, and backdoor effects are effectively neutralized.

\subsection{Detection Performance}
\label{sec:detection}
We mainly assess the effectiveness of FLARE against various backdoor attacks on the CIFAR-10 and Tiny-ImageNet datasets. The results, detailed in Table~\ref{tab:detection-cifar10} and Table~\ref{tab:detection-tiny}, demonstrate that FLARE consistently achieves promising performance across different attack scenarios,

with the TPRs reaching 100\% (or near 100\%)
while maintaining FPRs near 0\%. The results demonstrate a significant improvement over baseline defenses, which fail in A2A and UT attack scenarios (highlighted in red). This failure is primarily due to their implicit assumptions that backdoor correlations are easier to learn than benign ones, which does not hold in more complex attack configurations. Additionally, FLARE maintains an FPR close to zero on datasets composed entirely of benign samples, indicating that it has minimal false positives.

\subsection{Effectiveness of Secure Training from Scratch}
To assess the effectiveness of FLARE (P), which retrains models on a purified dataset with detected poisoned samples removed, we evaluated the BAs and ASRs of the retrained models on the CIFAR-10 and Tiny-ImageNet datasets. In particular, we define `failed cases' as a BA drop of more than 10\% compared to the original BA of the backdoored model without defense or an ASR greater than 10\%, in line with the evaluation criteria used in existing defenses~\cite{xu2024towards,chen2025refine}. As shown in Table~\ref{tab:defense_scratch} and Table~\ref{tab:defense_scratch_tiny}, the ASRs of retrained models approach to nearly 0\% in almost all cases. In contrast, all baseline methods fail in most cases, especially under all-to-all and untargeted attacks. In particular, we also observe some outliers with a high ASR, \eg, models under Blend (UT) and SIBA (UT) on CIFAR-10. This is mainly because even adding triggers of these attacks to a backdoor-free benign model may result in some ASR, instead of due to the failure of our method.

\subsection{Effectiveness of Backdoor Removal}
To assess the effectiveness of FLARE (R), which directly removes backdoors by unlearning the detected poisoned samples, we also evaluated the BAs and ASRs of the purified models on the CIFAR-10 and Tiny-ImageNet datasets. As shown in Table~\ref{tab:defense_unlearn} and Table~\ref{tab:defense_unlearn_tiny}, FLARE (R) significantly reduces the ASRs of different attacks to nearly 0\%, while the impact on BA remains minimal, generally around 1\%. Compared to existing advanced defenses, FLARE (R) consistently proves effective across various attacks and datasets.

\subsection{Ablation Study}
We conduct ablation studies to assess the impact of five specific factors on the effectiveness of FLARE: \textbf{(1)} target label for A2O attacks, \textbf{(2)} model architectures, \textbf{(3)} the module for selecting a stable subspace, \textbf{(4)} the hyper-parameters of the lambda threshold $\xi$ and maximum depth $d$, \textbf{(5)} the granularity in feature extraction. 


\begin{table}[t]
\centering
\setlength{\tabcolsep}{3pt}
\caption{Detection performance of FLARE on the CIFAR-10 dataset with different model architectures.}
\vspace{-0.5em}
\label{tab:detection-models}
\begin{tabular}{ccrrrr}
\toprule
Attack Mode$\downarrow$&  Models$\rightarrow$ & \multicolumn{2}{c}{VGG-19}  & \multicolumn{2}{c}{MobileNet\_v2}\\
\cmidrule(lr){3-4} \cmidrule(l){5-6} 
& Attacks$\downarrow$     & TPR (\%) & FPR (\%)  & TPR (\%) & FPR (\%) \\ 
\midrule
\multirow{3}{*}{A2O} & BadNets  & 100.00 & 0.00 & 100.00 & 0.00 \\
&LC & 100.00 & 0.00& 100.00 & 0.00\\
&ISSBA &99.98& 0.00& 99.90 & 0.00\\
\midrule
A2A & BadNets &100.00 & 0.00& 100.00 & 0.00\\
\midrule
UT & BadNets &100.00 & 9.86&100.00 & 0.00\\
\bottomrule
\end{tabular}
\vspace{-1.2em}
\end{table}

\vspace{0.3em} 
\noindent\textbf{Impact of Target Labels.} In our main experimetns, the target label for all A2O attacks is initially set to 0. To further assess the effectiveness of FLARE against variations in the target label, we test five representative A2O attacks, including patch-based, clean-label, and sample-specific backdoor attacks, targeting each of the ten labels in the CIFAR-10 dataset. The TPRs and FPRs are displayed in Figure~\ref{fig:targets}. As shown, FLARE consistently performs well across different attacks and target labels, with TPRs generally at 100\% and FPRs near 0\%. The results demonstrate that our defense is effective against a range of A2O backdoor attacks and target labels.

\vspace{0.3em} 
\noindent\textbf{Impact of Model Architectures.} In our main results, the model architecture is initially set to ResNet-18. To further assess the robustness of FLARE, we also evaluate it with VGG-19 and MobileNet v2 architectures. The results, shown in Table~\ref{tab:detection-models}, indicate that FLARE consistently achieves high TPRs, approaching 100\%, while maintaining FPRs near 0\%.

\vspace{0.3em} 
\noindent\textbf{Impact of the Stable Subspace Selection Module.}  Some sophisticated attacks, such as WaNet, employ regularization samples that exhibit similar trigger patterns to actual poisoned images while maintaining correct labels. This poses a significant challenge for detection mechanisms, making it increasingly difficult to differentiate between benign and poisoned samples. To tackle this challenge, we developed a module designed to enhance the compactness of benign clusters. We assessed the effectiveness of this module by evaluating FLARE's detection performance with and without its integration. The results demonstrate that FLARE, with this module, achieves a False Positive Rate (FPR) of 0.00\%, a significant improvement compared to the 20.15\% FPR observed without the module. This indicates that the subspace selection module effectively improves detection performance.

\begin{table*}[!t]
\vspace{-1.5em}
\caption{Comparisons of FLARE (R) with SOTA backdoor-removal defenses on CIFAR-10. The best results (highest BAs and lowest ASRs) are in bold, while all failed cases (BA drop or ASR $>$ 10\%) are marked in \red{red}.}
\vspace{-0.5em}
\label{tab:defense_unlearn}
\setlength{\tabcolsep}{3.4pt}
\centering 
\begin{tabular}{c|crr|rrrrrrrrrrrrrrrr}
\toprule
 \multirow{2}{*}{Attack Mode$\downarrow$}  & Defenses$\rightarrow$ & \multicolumn{2}{c|}{No Defense} &
 \multicolumn{2}{c}{FP} & \multicolumn{2}{c}{NAD } &\multicolumn{2}{c}{AMW}  &\multicolumn{2}{c}{ABL} &\multicolumn{2}{c}{SEAM} & \multicolumn{2}{c}{BTI-DBF}&  \multicolumn{2}{c}{FLARE (R)}  \\ 
 \cmidrule(lr){3-4} \cmidrule(lr){5-6} \cmidrule(lr){7-8} \cmidrule(lr){9-10} \cmidrule(lr){11-12} \cmidrule(lr){13-14} \cmidrule(lr){15-16} \cmidrule(l){17-18}
  &  Attacks$\downarrow$     & BA   & ASR    & BA   & ASR   & BA& ASR   & BA& ASR   & BA   & ASR  & BA   & ASR & BA   & ASR & BA   & ASR\\ 
\midrule
\multirow{9}{*}{A2O}&BadNets & 92.94 & 100.00&\red{82.34} &0.84 & 85.63& 2.38&89.41&4.83 & 92.90& 0.37&   89.60& 4.05&  92.08 &0.58 &  \bf{92.96}& \bf{0.36}\\
 &Blend  & 93.65 & 100.00 &\red{83.56} & \red{44.11}& 84.83& 1.29&87.38& 2.20& 91.27& \red{79.29}&  89.78&3.85  & 90.48&0.60 & \bf{91.69}&\bf{0.36}\\
& Trojan  & 93.04& 100.00 &83.53 &3.57  &86.44 & 4.77&87.55& 2.78& \bf{92.85} &1.60 & 90.05&2.46 &90.08& 1.89&92.20& \bf{0.83}\\
 &IAD &93.80 &99.99  &\red{83.44}& \red{78.70}& 86.89& 6.28& 87.00& 3.25& 91.25&\red{64.92} & 89.19& \red{12.96} & 91.34& 2.13&\bf{92.30}&\bf{1.39}\\
 &WaNet & 92.53 &97.83 &89.55&\red{22.32} &87.89&\bf{1.86}  & 85.54& 2.20& \bf{92.41}& \red{11.77}&  88.98& 2.40& 90.01 & \red{13.46} &91.61&2.29\\
& ISSBA& 93.50& 100.00& 83.38 &\bf{0.00} & 86.85& 7.61& 87.86& 1.06& \bf{92.46}&\red{85.07}  &89.20 &8.44  & 90.63 & 0.85&92.15& 0.58\\
& SIBA &94.33& 98.68& \red{78.31}&\red{90.62} & 87.16&  \red{15.71}& 88.25& \red{11.26} &90.66 & \red{19.60}&  87.44& \red{22.17}&89.34&\red{24.94} & \bf{92.21}& \bf{2.74}\\
&  LC&84.24& 100.00&83.81 & \bf{0.00}& 87.08&  \red{12.06}& 86.56& 3.86&82.97&\bf{0.00} &  88.31 & 7.99 & \bf{89.29}& 7.03& 83.78&5.25\\ 
 \midrule
\multirow{7}{*}{A2A} &BadNets  &92.74& 83.40&\red{82.43}&  5.48&86.26& 1.76&86.84& 1.47&86.21 & \red{11.98} &  88.84&  1.17 & 90.40& 1.57  & \bf{93.09}&\bf{0.15} \\
&Blend  & 93.08&85.20 &\red{82.03} & 9.05&86.37& 6.19& 88.53& 4.44&90.95 &\red{12.04} &  87.74&  \bf{2.81}&90.46&\red{23.34}& \bf{91.51} & 6.58\\
&Trojan  & 93.85&91.90 & \red{83.18} & 7.84&86.59 &3.30  &88.24 &1.90 &  \bf{93.60}& \red{12.11}& 88.98&1.50 &  91.74 &  9.07& 93.09& \bf{0.49}\\
& WaNet &93.93& 89.28&\red{82.23}& 8.17&84.66& 1.91&86.25& 1.49& \bf{92.41} &\red{11.77}&   88.13&  1.47& 91.08& 3.10 &92.27& \bf{0.66}\\
& IAD &  93.48&91.01 &\red{81.91}&5.64 &87.01 & 5.86& 88.36& 1.80& \bf{93.00}& \red{12.50}& 90.11 & 1.81 & 91.75& 2.38 &92.39&\bf{0.60}\\
& ISSBA &93.66&93.65 & \red{79.81}& 5.05 & \red{79.60}& 2.31&88.46& 1.34& 92.48& \red{12.59} &  88.83& 0.92&\bf{92.16}&  0.80&92.09& \bf{0.10}\\
&SIBA &93.74& 55.59&\red{83.64}& 6.43&\red{81.09} & 4.66& 88.31& 3.99& \bf{92.94}&  \red{11.82}&88.34 &  7.01 &90.94& 5.61&92.38 &\bf{2.43}\\
\bottomrule
\end{tabular}

\end{table*}

\begin{table*}[!t]
\caption{Comparisons of FLARE (R) with SOTA backdoor-removal defenses on Tiny-ImageNet (\%). The best results (highest BAs and lowest ASRs) are in bold, while all failed cases (BA drop or ASR $>$ 10\%) are marked in \red{red}.}
\vspace{-0.5em}
\label{tab:defense_unlearn_tiny}
\centering 
\setlength{\tabcolsep}{4.8pt}
\begin{tabular}{c|crr|rrrrrrrrrrrrrrrr}
\toprule
\multirow{2}{*}{Attack Mode$\downarrow$}  &    Defenses$\rightarrow$ & \multicolumn{2}{c|}{No Defense} &
 \multicolumn{2}{c}{FP} & \multicolumn{2}{c}{NAD} &\multicolumn{2}{c}{ABL} &\multicolumn{2}{c}{SEAM} & \multicolumn{2}{c}{BTI-DBF}&  \multicolumn{2}{c}{FLARE (R)}  \\ 
\cmidrule(lr){3-4} \cmidrule(lr){5-6} \cmidrule(lr){7-8} \cmidrule(lr){9-10} \cmidrule(lr){11-12} \cmidrule(lr){13-14} \cmidrule(l){15-16}
  &   Attacks$\downarrow$     & BA   & ASR    & BA   & ASR   & BA& ASR   & BA& ASR   & BA   & ASR  & BA   & ASR & BA   & ASR\\ 
\midrule
\multirow{5}{*}{A2O}& BadNets &  54.76 & 96.52 & \red{37.84}& \red{95.65}  &46.76& 0.24 & \red{41.00}&\bf{0.00}  & 47.23& 0.10 &  52.57 &  4.15&\bf{55.70} &0.01\\
& Blend  &66.91& 100.00 & \red{52.16}&\red{100.00}  &\bf{67.38}&  10.00&\red{39.03}&\bf{0.00}  & \red{45.53}& 1.68 & \red{53.11}& 2.99 & 63.82&1.20\\
& Trojan  &66.06& 100.00 & 58.51 & \red{100.00} &59.12&5.92 & \red{39.09} &\bf{0.00} & \red{48.39}& \bf{0.00} & \red{54.84}&  \red{99.93}& \bf{65.10} & \bf{0.00}\\
& WaNet  &  62.12& 99.42 &\red{24.47}& \red{99.90} & 53.47& 0.49&\red{42.12}&\bf{0.00}&\red{44.55}&0.36& \red{51.09}& 1.05 & \bf{61.20} & 1.10\\
\midrule
\multirow{3}{*}{A2A}& BadNets& 68.01& 22.93  &\bf{67.61} & 0.43 &61.21&4.45 &\red{44.35} & \red{21.66}& 58.48& 0.22 & 58.22 &8.87 & 67.11 & \bf{0.10}\\
& Blend &  69.54  &48.67 & \bf{69.39}& 1.04  &  62.84&3.71 & \red{37.85}&\red{10.96}&60.33 & \bf{0.24}&\red{52.01}& 1.91& 69.04& 1.90\\
& Trojan &68.20 & 29.26 & \bf{68.12} & 0.76 &  61.01 & \red{18.01} &\red{43.49} &\red{12.08}& 58.60&0.33 & \red{48.95} &7.67 & 67.20 & \bf{0.30}\\
\bottomrule
\end{tabular}
\vspace{-1.4em}
\end{table*}

\begin{table}[!t]
\centering
\caption{Detection performance of FLARE against various backdoor attacks under different threshold $\xi$.}
\vspace{-0.5em}
\label{tab:dff_lambda}
\setlength{\tabcolsep}{4.5pt}
\begin{tabular}{clrrrrr} 
\toprule
 $\xi$& Metrics & \makecell[c]{BadNets\\ (A2O)} & LC & ISSBA & \makecell[c]{BadNets\\ (A2A)} & \makecell[c]{BadNets\\ (UT)}\\
\midrule
\multirow{2}{*}{0.01}&  TPR (\%) & 100.00 & 100.00& 99.98&100.00&100.00\\
&  FPR (\%) & 0.00 &0.01 &0.01&0.00& 0.00\\
\midrule
\multirow{2}{*}{0.02}&  TPR (\%) & 100.00 &100.00&99.98 &100.00&100.00\\
&  FPR (\%) & 0.00 &0.00& 0.01&0.00& 0.00\\
\midrule
\multirow{2}{*}{0.03}&  TPR (\%) & 100.00 &100.00& 99.98&100.00&100.00\\
&  FPR (\%) & 0.00 &0.00& 0.01&0.00& 0.00\\
\midrule
\multirow{2}{*}{0.04}&  TPR (\%) & 100.00 &100.00& 99.98&100.00&100.00\\
&  FPR (\%) & 0.00 &33.89& 0.01&0.00& 0.00\\
\midrule
\multirow{2}{*}{0.05}&  TPR (\%) & 100.00 &100.00 & 99.98&100.00&100.00\\
&  FPR (\%) & 0.00& 33.89&0.01&40.67& 0.00\\
\bottomrule
\end{tabular}
\vspace{-1.5em}
\end{table}




\vspace{0.3em} 
\noindent\textbf{Impact of the Hyper-parameters.} In our search for a stable subspace, we employ two hyperparameters: the lambda threshold $\xi$ and the maximum depth $d$. To assess FLARE’s robustness, we vary $\xi$ from 0.01 to 0.05 across five representative attacks. As shown in Table~\ref{tab:dff_lambda}, while higher $\xi$ values slightly increase TPR, setting $\xi$ between 0.01 and 0.03 consistently achieves optimal results, with TPRs at 100\% and FPRs close to 0\% across all attacks. We also examine $d$ from 1 to 5, observing consistent performance across this range, with TPRs and FPRs remaining close to 100\% and 0\%, respectively. Given the uniform results, we omit detailed data for these variations and set $\xi$ to 0.02 and $d$ to 3 by default.

\vspace{0.3em} 
\noindent\textbf{Impact of Granularity Selection in Feature Extraction.} In our main experiments, FLARE extracts the minimum abnormal value at the channel level. To evaluate the impact of granularity selection on detection performance, we conduct ablation studies comparing the detection performance of layer-wise FLARE (dubbed `FLARE-L') and channel-wise FLARE (dubbed `FLARE-C'). Specifically, FLARE-L extracts the minimum abnormal value from all channels in each layer. In contrast, FLARE-C operates at the channel level, where it extracts a minimum abnormal value from each channel separately and then aggregates these values from all channels to form the feature representation of each layer. We evaluate the performance of both FLARE-L and FLARE-C on representative backdoor attacks, including BadNets (under A2O, A2A, and UT), as well as clean-label attack (\ie, LC) and sample-specific attack (\ie, ISSBA). This experimental setup is consistent with the configurations used in other ablation studies. As shown in Table \ref{tab:detection-Granularity}, FLARE-C consistently outperforms FLARE-L in terms of both TPR and FPR across all attacks. The superior performance of FLARE-C can be attributed to its ability to leverage more fine-grained feature representations, allowing it to capture more subtle differences between benign and poisoned samples, leading to better detection performance. These results verify the effectiveness of our method again.

\begin{table}[t]
\centering
 \caption{The detection performance (\%) of FLARE-L and FLARE-C on the CIFAR-10 dataset.}
\vspace{-0.5em}
\label{tab:detection-Granularity}
\begin{tabular}{c|crrrrrrrrrrrr}
\toprule
 \multirow{2}{*}{Attack Mode$\downarrow$}&Defenses$\rightarrow$ & \multicolumn{2}{c}{FLARE-L} & \multicolumn{2}{c}{FLARE-C}\\
\cmidrule(lr){3-4} \cmidrule(lr){5-6} \cmidrule(lr){7-8} 
& Attacks$\downarrow$      & TPR   & FPR   & TPR   & FPR \\ 
\midrule
---&No Poison& ---& 1.70 & ---&  \bf{0.01}\\
\midrule
 \multirow{3}{*}{A2O}&BadNets& 94.90 & 0.00 & \bf{100.00} & \bf{0.00} \\ 
&LC & 100.00 & 0.00& 100.00 & 0.00\\
&ISSBA & 88.32 & 0.13 & \bf{99.98}  & \bf{0.01} \\ 
\midrule
\multirow{1}{*}{A2A}&BadNets & 89.56&0.00 & \bf{100.00} & \bf{0.00} \\ 
\midrule
\multirow{1}{*}{UT}&BadNets& 89.96 &0.00 & \bf{100.00}  & \bf{0.00} \\ 
\bottomrule
\end{tabular}
\vspace{-1.5em}
\end{table}

\begin{figure*}[!t]
  \vspace{-1.5em}
	\centering
 \begin{minipage}{0.90\linewidth}
    \begin{minipage}{0.190\linewidth}
            \includegraphics[width=1\linewidth]{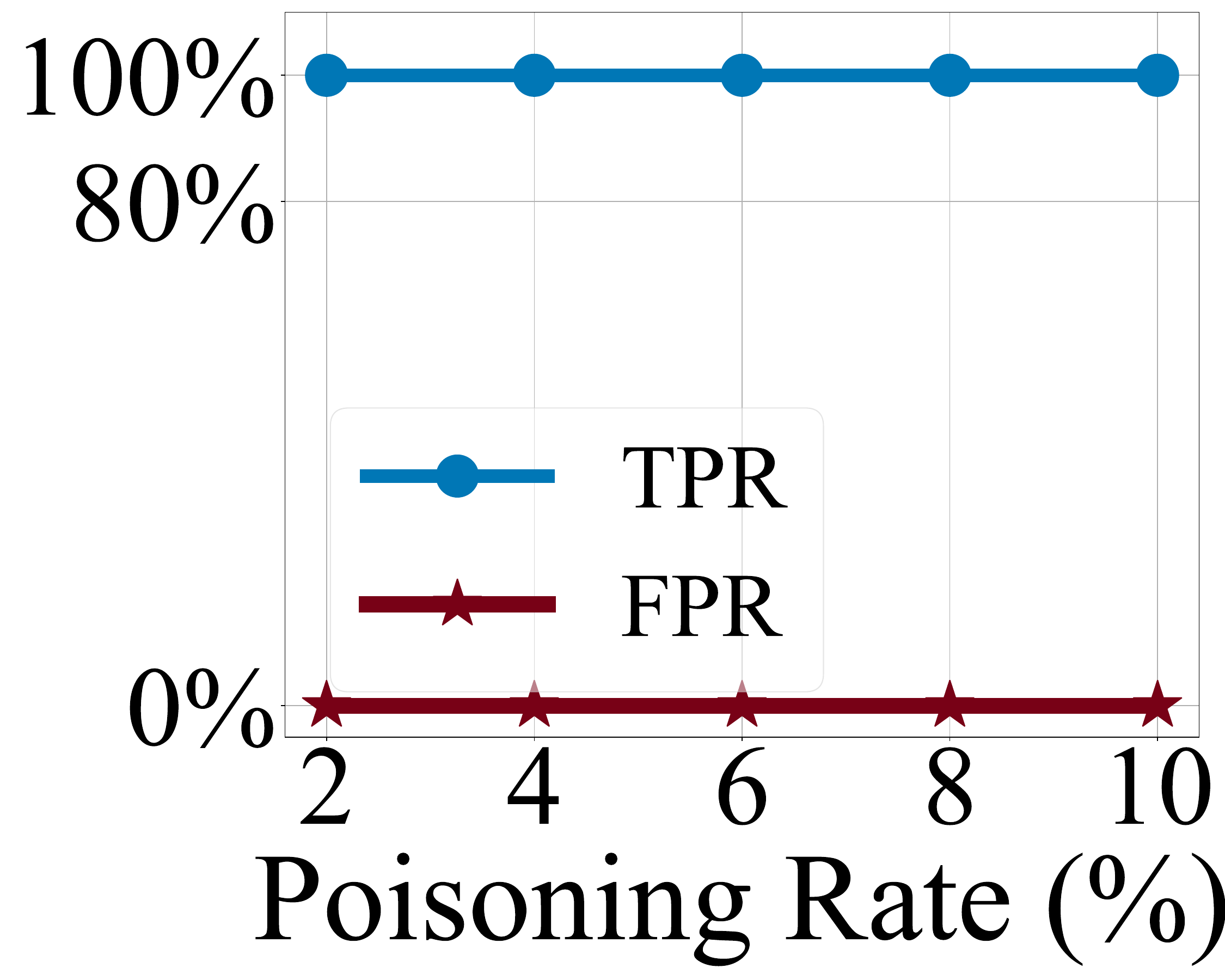}
            \centerline{(a) BadNets (A2O)}
    \end{minipage}
    \begin{minipage}{0.190\linewidth}
            \includegraphics[width=1\linewidth]{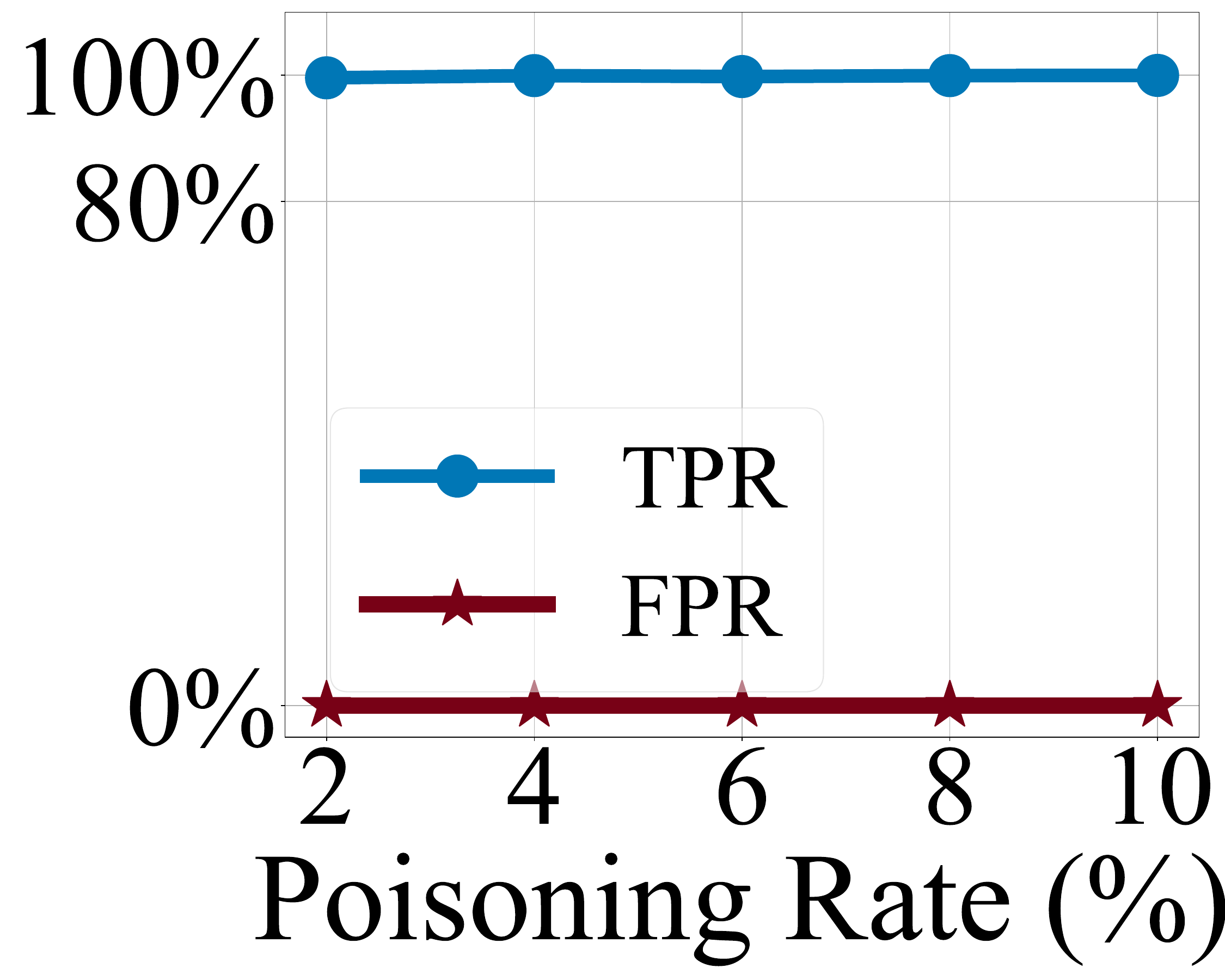}
            \centerline{(b) ISSBA (A2O)}
    \end{minipage}
    \begin{minipage}{0.190\linewidth}
            \includegraphics[width=1\linewidth]{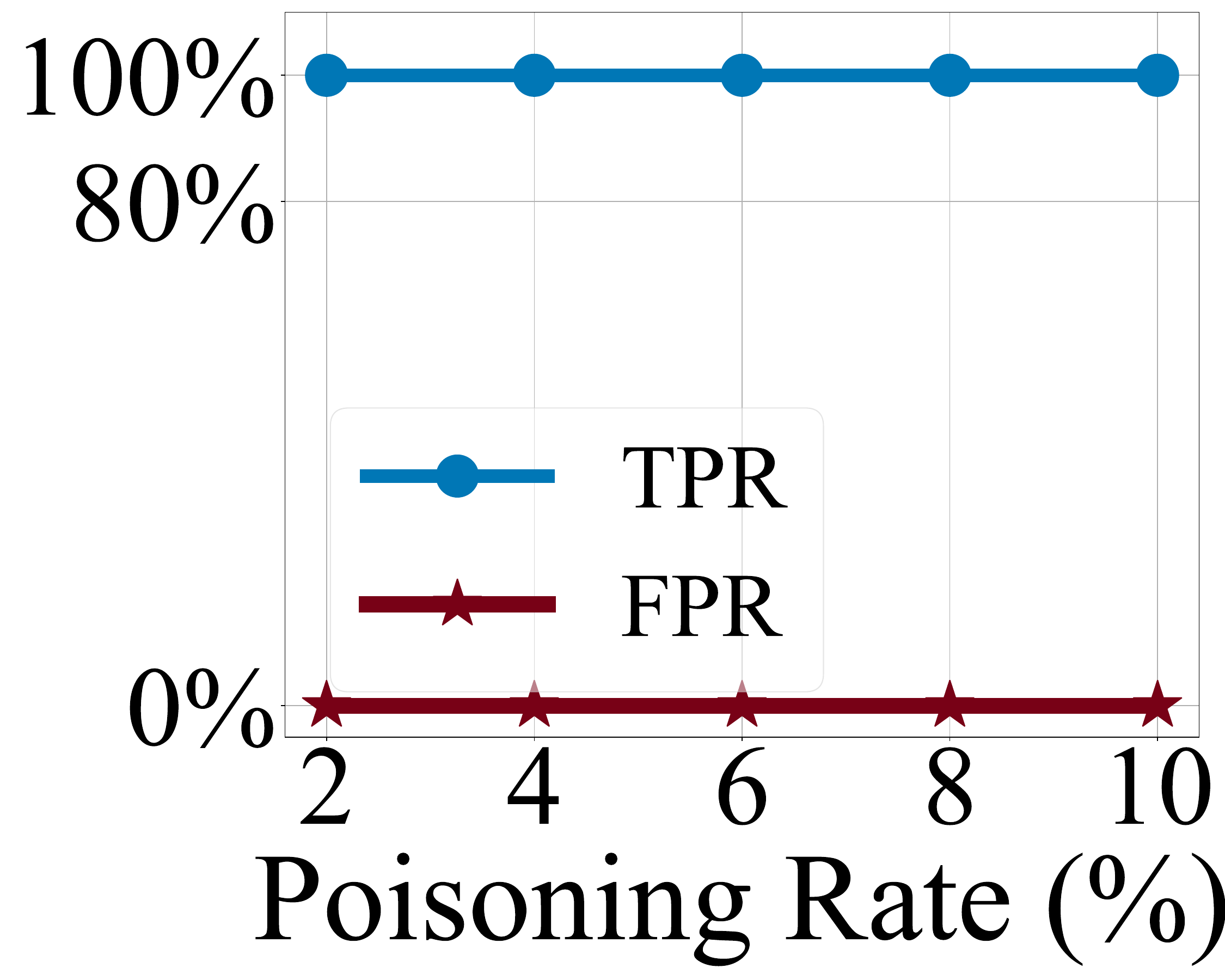}
            \centerline{(c) LC}
    \end{minipage}
     \begin{minipage}{0.190\linewidth}
            \includegraphics[width=1\linewidth]{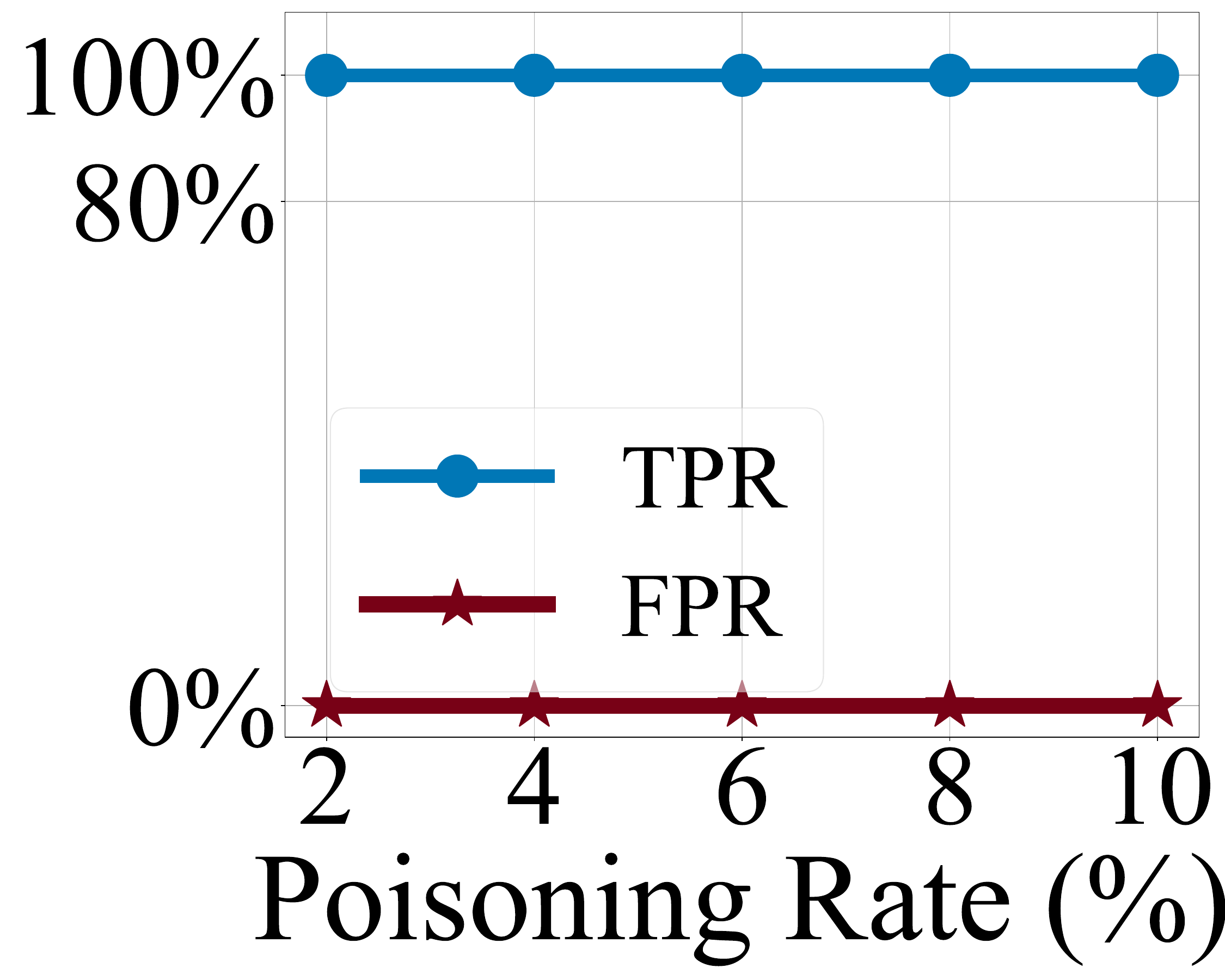}
            \centerline{(d) BadNets (A2A)}
    \end{minipage}
    \begin{minipage}{0.190\linewidth}
            \includegraphics[width=1\linewidth]{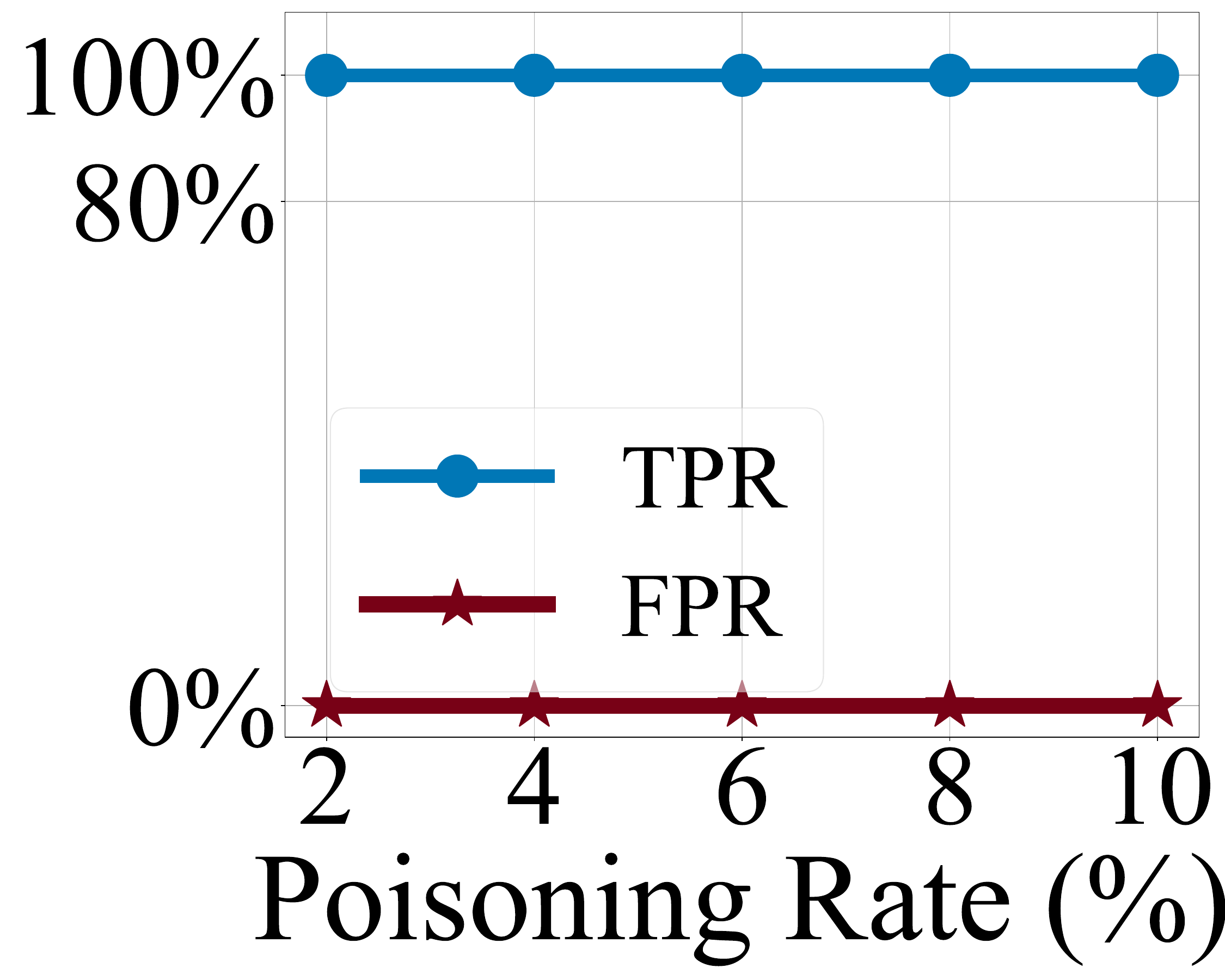}
            \centerline{(e) BadNets (UT)}
    \end{minipage}
\end{minipage}
  \vspace{-0.3em}
 \caption{The impact of the poisoning rate on CIFAR-10. The consistently high TPR and low FPR across various target labels indicate stable and effective defense performance.}
 \label{fig:poisoningratio}
  \vspace{-1.3em}
\end{figure*}



\subsection{Resistance to Potential Adaptive Attacks}

FLARE operates on the assumption that backdoored models learn distinct latent representations for poisoned and benign samples. However, studies in~\cite{hayase2020spectre} indicate that the separations may diminish at low poisoning rates. We assess the effectiveness of FLARE against five representative attacks: BadNets (A2O), ISSBA (A2O), LC, BadNets (A2A), and BadNets (UT). These attacks are conducted on the CIFAR-10 dataset with poisoning rates ($\rho$) ranging from 0.02 to 0.1, ensuring ASRs exceed 80\%. The results in Figure~\ref{fig:poisoningratio} show that FLARE maintains high effectiveness across all poisoning rates, with TPRs close to 100\% and FPRs near 0\%.

We further evaluate the robustness of FLARE against adaptive attacks under the worst-case scenario, where adversaries deliberately reduce the latent separation. The Ada-Patch attack~\cite{qi2023revisiting} designs regularization samples (poisoned images with ground-truth labels) to reduce separation, posing a significant challenge for our detection. We test FLARE against Ada-Patch on the CIFAR-10 dataset with a poisoning rate of 0.1, consistent with our primary experiments. FLARE achieved a TPR of 92.10\% and an FPR of 1.05\%, demonstrating robustness even under this advanced adaptive attack. We argue that the effectiveness mostly stems from its ability to detect distributed anomalies across all hidden layers to accumulate subtle differences, which enhances its detection capabilities.

\section{Potential Limitations and Future Work}

Firstly, our method requires additional time compared to standard model training to some extent. Specifically, it involves a two-stage process: training a model on the suspicious dataset and then performing clustering on the features extracted from the model’s hidden layers to identify (potential) poisoned samples. Arguably, model training is a necessary step in all machine learning workflows, regardless of the presence of defense mechanisms. As such, the training time is not considered an additional cost. Consequently, the only additional overhead introduced by our defense is the time required for the second step. Specifically, in our experiments, training a ResNet-18 model on CIFAR-10 for 100 epochs took 1,365 seconds, whereas feature extraction and clustering took only 180 seconds (\ie, our additional overhead is 180 seconds).

Secondly, our FLARE focuses on detecting poisoned samples but does not address the reverse engineering of backdoor triggers. This task is challenging to a large extent due to the stealthiness of triggers and the lack of prior knowledge about poisoned samples. We will explore techniques to trace back and recover the triggers simultaneously in our future work.

Thirdly, our FLARE mainly focuses on image classification tasks. As multi-modal data (\eg, images, speech, and text) becomes more prevalent, we will explore extending FLARE to cross-modal scenarios to develop integrated defense strategies that combine data from multiple sources to effectively detect and prevent cross-modal backdoor attacks.




\section{Conclusion}


In this paper, we revealed a critical limitation of existing purification methods and proposed a universal approach (\ie, FLARE) to filter out poisoned training samples. Existing methods assumed that backdoor connections between triggers and target labels are easier to learn. However, this assumption did not always hold, particularly in all-to-all and untargeted backdoor attacks. We observed that the latent separation between benign and poisoned samples varies across multiple layers rather than on a particular layer. These findings motivated us to leverage abnormal features from all hidden layers to construct comprehensive representations for cluster analysis. Besides, to further improve separation, we developed an adaptive subspace selection algorithm to dynamically isolate an optimal space for dividing an entire dataset into two clusters. We conducted 22 backdoor attacks on benchmark datasets to comprehensively verify the effectiveness of FALRE. 


\small
\bibliography{main}
\bibliographystyle{unsrt}


\end{document}